\documentclass[conference]{IEEEtran}
\IEEEoverridecommandlockouts
\usepackage{cite}
\usepackage{amsmath,amssymb,amsfonts}

\newtheorem{definition}{Definition}

\usepackage[noend]{algpseudocode}
\usepackage{algorithm}
\usepackage{textcomp}
\usepackage{xcolor}

\usepackage{graphicx}
\usepackage{subcaption}

\usepackage{booktabs} 
\usepackage[rightcaption]{sidecap}
\usepackage{bold-extra}
\usepackage{amsmath,amsfonts,amssymb}

\usepackage{multirow}
\usepackage{xcolor}
\usepackage{xspace}

\usepackage{paralist}

\usepackage{grffile}

\usepackage{epstopdf}

\usepackage{boxedminipage}

\usepackage{tabularx}
\usepackage{array}
\usepackage{tabularx}
\usepackage{array}
\usepackage{hyperref}
\usepackage{textcase}
\usepackage{caption}

\usepackage{stfloats}
\usepackage[capitalise]{cleveref}

\def\BibTeX{{\rm B\kern-.05em{\sc i\kern-.025em b}\kern-.08emT\kern-.1667em\lower.7ex\hbox{E}\kern-.125emX}}
    
\begin{document}

%
%

\title{Characterizing and Utilizing the Interplay Between Core and Truss Decompositions
\thanks{A shorter version of this paper is accepted by 2020 \textit{IEEE} International Conference on Big Data.}}

\author{\IEEEauthorblockN{Penghang Liu}
\IEEEauthorblockA{\textit{University at Buffalo}\\
Buffalo, USA \\
penghang@buffalo.edu}
\and
\IEEEauthorblockN{Ahmet Erdem Sar{\i}y\"{u}ce}
\IEEEauthorblockA{\textit{University at Buffalo}\\
Buffalo, USA \\
erdem@buffalo.edu}
}

\date{}
\maketitle

\begin{abstract}
Finding the dense regions in a graph is an important problem in network analysis. Core decomposition and truss decomposition address this problem from two different perspectives. The former is a vertex-driven approach that assigns density indicators for vertices whereas the latter is an edge-driven technique that put density quantifiers on edges.
Despite the algorithmic similarity between these two approaches, it is not clear how core and truss decompositions in a network are related.
In this work,  we introduce the vertex interplay (VI) and edge interplay (EI) plots to characterize the interplay between core and truss decompositions.
Based on our observations, we devise \textsc{Core-TrussDD}, an anomaly detection algorithm to identify the discrepancies between core and truss decompositions.
We analyze a large and diverse set of real-world networks, and demonstrate how our approaches can be effective tools to characterize the patterns and anomalies in the networks.
Through VI and EI plots, we observe distinct behaviors for graphs from different domains, and identify two anomalous behaviors driven by specific real-world structures. 
Our algorithm provides an efficient solution to retrieve the outliers in the networks, which correspond to the two anomalous behaviors.
We believe that investigating the interplay between core and truss decompositions is important and can yield surprising insights regarding the dense subgraph structure of real-world networks.\\



\end{abstract}


\section{Introduction}
Dense subgraphs in real-world networks contain significant and unusual information.
There are many application domains where dense subgraphs are useful. 
A few use cases are finding  price value motifs in the financial networks~\cite{Du09}, locating  spam link farms in webs~\cite{Kumar99,Gibson05,Dourisboure07}, detecting  DNA motifs in biological networks~\cite{Fratkin06}, and identifying the news stories from microblogging streams in real-time~\cite{Angel12}.
Dense regions are also used to improve the efficiency of computation heavy tasks like distance query computation~\cite{Jin09} and materialized per-user view creation~\cite{Gionis13}.
Core and truss decompositions are effective models to find dense regions with hierarchical relationships.
In the core decomposition~\cite{Seidman83,MaBe83}, vertices are assigned density pointers, called core numbers, that indicate the cohesiveness in the neighborhood.
Likewise, truss decomposition~\cite{Saito06,Cohen08,Verma12,Zhang12} yields truss numbers for edges which can be interpreted as the link strength.


Given the wide application space, core and truss decompositions are unified and extended by new models for different types of graphs~\cite{Sariyuce15, Sariyuce-TWEB17, Li2018}.
However, the relationship between the core and truss decompositions and its impact on the graph structure have been overlooked.
By definition, a large truss number of an edge implies large core numbers on its endpoints.
But, it is not clear what aspects of the graph structure are covered by each decomposition.
Understanding the interplay between those measures can enable more effective network analysis.

\begin{figure}[!b]
\centering
\vspace{-4ex}
\includegraphics[width=0.8\linewidth]{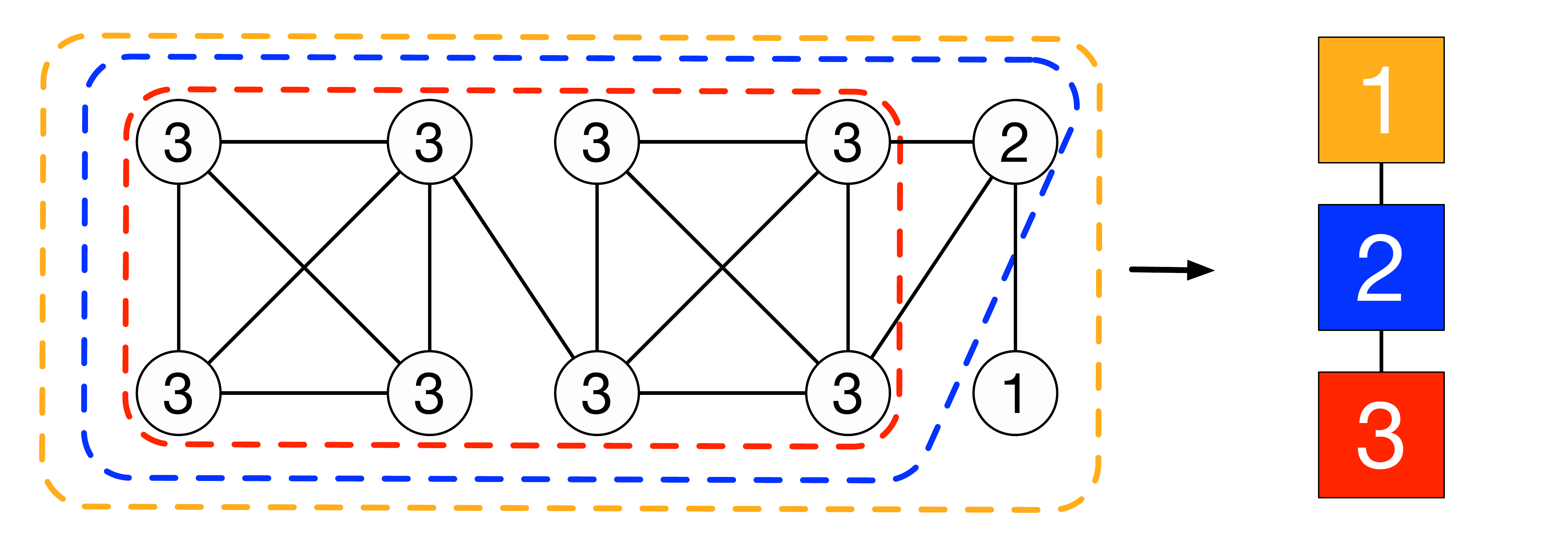}
\includegraphics[width=0.8\linewidth]{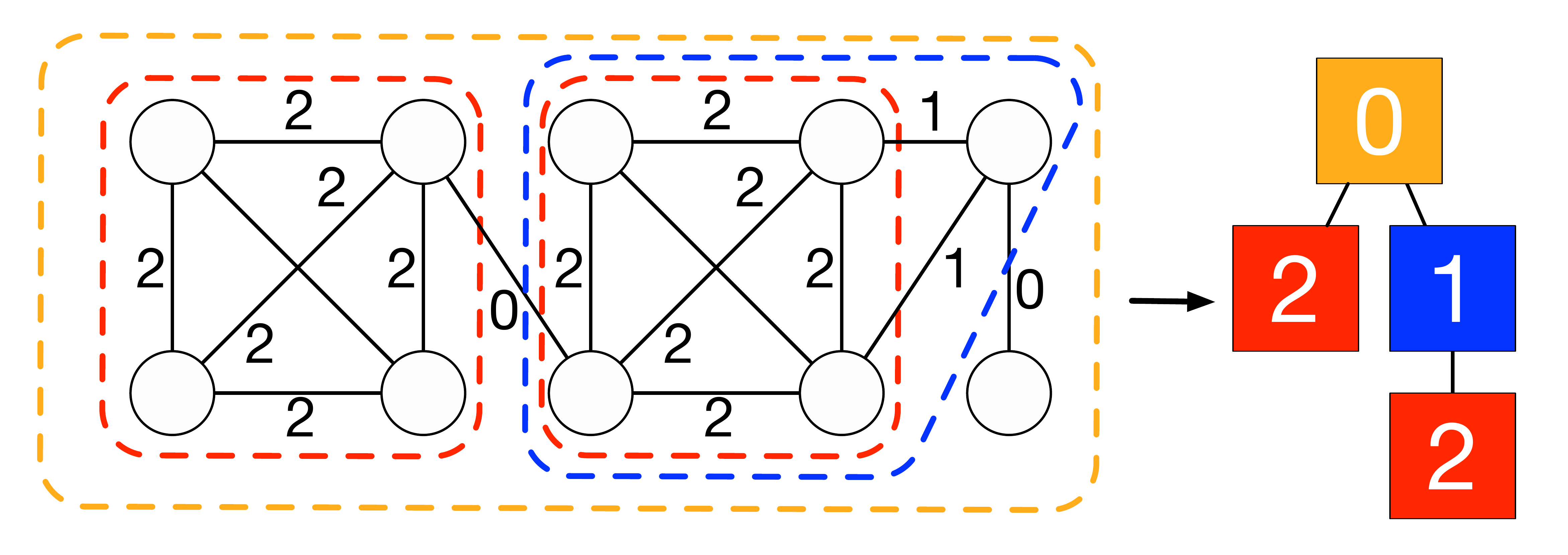}
\vspace{-2ex}
\caption{\small \textbf{Examples for core (top) and truss (bottom) decompositions. At the top, core numbers are shown for each node and red, blue, and orange regions show the 3-, 2-, and 1-cores. They form a hierarchy by containment as denoted; 1-core contains 2-core and 2-core subsumes 3-core.
For the same graph, trusses and truss numbers of edges are presented at the bottom. The entire graph is a 0-truss and the five nodes on the right form a 1-truss. There are two 2-trusses and one of them is a subset of the 1-truss, as denoted by the tree hierarchy.}}
\label{fig:toys}
\end{figure}

In this work, we investigate the interplay between core and truss decompositions in real-world networks and random graphs.
Our first contribution is the vertex interplay (VI) and edge interplay (EI) plots, which analyze the structure of dense subgraphs from two different perspectives. Then, we propose the \textsc{Core-TrussDD} algorithm to identify the anomalies with respect to the core-truss interplay.
The VI plot investigates the core-truss interplay by checking the core numbers of vertices and the truss numbers of their adjacent edges. The EI plot explores the interplay based on the truss numbers of edges and the core numbers of their two endpoints.
We use several real-world networks from various domains and examine the similarities and differences in the core-truss interplay.
Our analysis on VI and EI plots gives interesting results for the core-truss interplay behavior in real-world networks.
We also show that those behaviors cannot be captured when the edges are rewired with a realistic random graph model.
Our algorithm provides an efficient solution to retrieve the outliers in the networks, which reveal the anomalous behaviors observed in the VI plots.

The rest of this paper is organized as follows. \cref{sec:back} gives the background for core and truss decomposition and random graph models. \cref{sec:data} describes the datasets used in the paper. In \cref{sec:vertex} and \ref{sec:edge}, we characterize the core-truss interplay using VI and EI plots. Based on our observations, we propose our anomaly detection algorithm in \cref{sec:alg}. After reviewing related work in \cref{sec:related}, we draw conclusions in \cref{sec:conc}.
Codes for reproducing the results and figures are available at \texttt{https://github.com/penghangliu/Core-Truss.}

\section{Background}\label{sec:back}
Our study explores real-world networks which can be represented as a simple undirected unweighted graph $G = (V,E)$, where $V$ is the set of vertices and $E$ is the set of edges. We represent the neighborhood of a vertex $u$ (the set of nodes connected to $u$) as $N(u)$. 

\subsection{Core decomposition}

The $k$-core subgraph is introduced by Seidman~\cite{Seidman83} for social networks analysis, and also by Matula and Beck~\cite{MaBe83} for clustering and graph coloring.
\textbf{The $\mathbf{\textit{k}}$\,-core is a connected, maximal subgraph such that every node in the subgraph has degree of at least $\mathbf{\textit{k}}$ within the subgraph.}
The core number of a node is the highest value of $k$ such that the node belongs to a $k$-core. We denote the core number of a vertex $u$ by $K(u)$. The maximum core number of all vertices in the graph is defined as the \textit{(core) degeneracy}~\cite{ErHa66}.

Core decomposition is the process of finding the core numbers of nodes, which are used to locate all the $k$-core subgraphs.
Batagelj and Zaversnik introduced an efficient iterative peeling algorithm that uses a bucket data structure to find the core numbers of nodes in $O(|E|)$ time~\cite{BaZa03}.
Starting from the node with the minimum degree, peeling algorithm assigns the degree of a node as its core number and remove it from the graph --- thus decrementing the neighbors' degrees if larger. This process continues until the graph is empty.
$k$-core subgraph (for any $k$) is found by a traversal that visits all reachable nodes whose core numbers are at least $k$. 
The nested structure of $k$-cores reveals a hierarchy by containment.
\autoref{fig:toys} (top) presents an example for the core numbers,  $k$-core subgraphs, and hierarchy.

\subsection{Truss Decomposition}\label{truss intro}

Higher-order structures, also known as motifs or graphlets, have been used to locate dense regions that cannot be detected otherwise with edge-centric methods~\cite{Benson16, Tsourakakis15}.
Finding the frequency and distribution of triangles and other small motifs in real-world networks is a simple yet effective approach used in data analysis~\cite{Jha15, Pinar16, Ahmed15, Rossi17}.
The truss decomposition is inspired by the $k$-core and considers the edges and the triangles they participate in~\cite{Cohen08, Saito06, Verma12, Zhang12}. 
\textbf{$\mathbf{\textit{k}}$\,-truss is a connected, maximal subgraph such that every edge in the subgraph participates in at least $\mathbf{\textit{k}}$ triangles within the subgraph}.
The truss number of an edge is the highest $k$ such that the edge is part of a $k$-truss. In the following, we denote the truss number of an edge $(u,v)$ by $T(u,v)$. The maximum truss number of all edges in the graph is defined as the \textit{truss degeneracy}.

Similar to the core decomposition, finding the truss numbers has two phases; 1) Counting the triangles that each edge participates in, 2) Peeling those counts by choosing the edge with the minimum count, assigning that as truss number, and decrementing triangle counts of neighbors. This requires $O(\sum_{v \in V}{d(v)^2})$ time.
 $k$-trusses also exhibit a hierarchy by containment.
\autoref{fig:toys} (bottom) presents the truss numbers of edges, $k$-truss subgraphs, and their hierarchy on a toy graph.

\noindent \textbf{Relationship between $k$-clique, and $k$-core, $k$-truss}. The $k$-clique is a fully connected subgraph that contains $k$ nodes, and each node is connected to the other $k-1$ nodes. For any edge $(u,v)$ in a $k$-clique, both $u$ and $v$ are connected to the other $k-2$ nodes, thus every edge in the $k$-clique participates in at least $k-2$ triangles within the clique. This shows that $k$-clique is also a $(k-1)$-core and a $(k-2)$-truss.

\subsection{Random Graph Models}
Random graph models are commonly used as the null model for analyzing real-world networks. 
The Erd\H{o}s-R\'{e}nyi model (ER), configuration model, and  Block Two-Level Erd\H{o}s-R\'{e}nyi (BTER) model~\cite{Sesh12} are three random graph models that simulate real-world networks at different levels of authenticity. 
The ER model rewires the edges randomly while keeping the graph size.
The configuration model generates random graphs from the given degree sequence $\overrightarrow{d}$, where $d(u) \in \overrightarrow{d}$ is the degree of vertex $u$. 
In the BTER model, the degree distribution and the clustering coefficient per degree distribution are preserved to the best extent. 
A graph is generated through two phases in the BTER model. 
In the first phase, vertices are clustered into communities and ER model is applied to generate edges within the same community. In the second phase edges between communities are generated regarding the size of the communities. 

\begin{table*}[!b]
\vspace{-2ex}
\caption{\bf \small Statistics of real-world networks and random graphs. The last four columns show the core degeneracy and truss degeneracy numbers. In each group, \textit{Exact} shows the core and truss degeneracy numbers of real-world graphs, and \textit{BTER} presents the same numbers (on average) for the random graphs generated with the BTER model.
BTER model shows good capability of approximating the core degeneracy, but it does not perform well in capturing the truss degeneracy.}
\renewcommand{\tabcolsep}{1ex}
\centering
\begin{tabular}{|c|c|r|r|r|r|r|r|}
\hline
\multirow{2}{*}{Category}   & \multirow{2}{*}{Name}  & \multicolumn{1}{c|}{\multirow{2}{*}{$|V|$}} & \multicolumn{1}{c|}{\multirow{2}{*}{$|E|$}} & \multicolumn{2}{c|}{Core degeneracy}& \multicolumn{2}{c|}{Truss degeneracy} \\
& & & & \textit{Exact} &  \textit{BTER} & \textit{Exact} & \textit{BTER}\\ \hline
\multirow{8}{*}{Social} 	&	\texttt{Hamster}	&	1.86K	&	12.5K	&	\textbf{	20	}&	 17 &	\textbf{	7	}&	 5 \\	
		&	\texttt{Email}	&	36.7K	&	184K	&	\textbf{	43	}&47&	\textbf{	20	}&27	\\	
			&	\texttt{YouTube}	&	1.13M	&	2.99M	&	\textbf{	51	}&78&	\textbf{	17	}&48	\\	
	&	\texttt{Catster}	&	150K	&	5.45M	&	\textbf{	419	}&312&	\textbf{	205	}&164	\\		
		&	\texttt{Dogster}	&	427K	&	8.54M	&	\textbf{	248	}&300&	\textbf{	91	}&192	\\	
		&	\texttt{Flickr}	&	1.72M	&	15.6M	&	\textbf{	568	}&310&	\textbf{	276	}&110	\\
	&	\texttt{LiveJournal}	&	4.00M	&	34.7M	&	\textbf{	360	}&33&	\textbf{	350	}&6	\\	
	&	\texttt{Orkut}	&	3.07M	&	117M	&	\textbf{	253	}&80&	\textbf{	76	}&40	\\	
	&	\texttt{Email-Eu-core}	&	1.00K	&	25.5K	&	\textbf{	34}&36 &	\textbf{	21	}&11	\\	
\hline
\multirow{5}{*}{Autonomous sys.}	&	\texttt{As-733}	&	6.47K	&	12.6K	&	\textbf{	12	}&13&	\textbf{	8	}&11	\\	
	&	\texttt{Oregon-2}	&	10.9K	&	31.2K	&	\textbf{	31	}&22&	\textbf{	23	}&16	\\	
	&	\texttt{Caida}	&	26.5K	&	53.4K	&	\textbf{	22	}&28&	\textbf{	14	}&23	\\	
	&	\texttt{Gnutella}	&	62.6K	&	148K	&	\textbf{	6	}&6&	\textbf{	2	}&1\\	
	&	\texttt{Skitter}	&	1.70M	&	11.1M	&	\textbf{	111	}&191&	\textbf{	66	}&146\\	\hline
\multirow{5}{*}{Citation}		&	\texttt{DBLP}	&	12.6K	&	49.6K	&	\textbf{	12	}&13&	\textbf{	7	}&5	\\	
		&	\texttt{Cora}	&	23.2K	&	89.2K	&	\textbf{	13	}&9&	\textbf{	9	}&3\\	
	&	\texttt{HepTh}	&	27.7K	&	352K	&	\textbf{	37	}&31&	\textbf{	28	}&14	\\	
	&	\texttt{CiteSeer}	&	384K	&	1.74M	&	\textbf{	15	}&14&	\textbf{	11	}&3	\\	
	&	\texttt{Patent}	&	3.78M	&	16.5M	&	\textbf{	64	}&9&	\textbf{	34	}&1	\\	\hline
\multirow{4}{*}{Collaboration}		
	&	\texttt{DBLP\_pp}	&	8.41K	&	22.9K	&	\textbf{	44	}&16&	\textbf{	43	}&3	\\	
&	\texttt{DBLP\_dbs}	&	8.10K	&	23.0K	&	\textbf{	35	}&10&	\textbf{	34	}&2	\\	
	&	\texttt{DBLP\_dm}	&	16.4K	&	33.9K	&	\textbf{	24	}&7&	\textbf{	23	}&1	\\	\hline
\multirow{5}{*}{Web}		&	\texttt{Blogs}	&	1.22K	&	16.7K	&	\textbf{	36	}&38&	\textbf{	23	}&12\\	
	&	\texttt{NotreDame}	&	326K	&	1.09M	&	\textbf{	155	}&144&	\textbf{	153	}&47	\\	
	&	\texttt{Stanford}	&	282K	&	1.99M	&	\textbf{	71	}&123&	\textbf{	60	}&93	\\	
	&	\texttt{Google}	&	876K	&	4.32M	&	\textbf{	44	}&86&	\textbf{	42	}&72	\\
	&	\texttt{BerkStan}	&	685K	&	6.65M	&	\textbf{	201	}&258&	\textbf{	199	}&178	\\		\hline
\end{tabular}

\label{tab:data}
\end{table*}

\section{Datasets}\label{sec:data}
In order to explore patterns in various types of real-world networks, we cover datasets from five different categories: social, autonomous systems, citation, collaboration, and web hyperlink networks. The datasets are obtained from SNAP~\cite{snap}, DBLP~\cite{dblp}, and Konect~\cite{konect}.
Various statistics are summarized in~\cref{tab:data}.


\noindent \textbf{Social networks.} Our dataset includes \texttt{Catster}, \texttt{Dogster}, \texttt{Flickr}, \texttt{Hamster}, \texttt{LiveJournal}, \texttt{YouTube}, and \texttt{Orkut} friendship networks that are formed among the members of the social networking websites. In addition, we consider the \texttt{Email} network among the employees of Enron Corporation. We also include \texttt{Email-Eu-core}, a network representing the emails between members of a large European research institution.

\noindent \textbf{Autonomous systems.} In this category, we have the five router networks, \texttt{As-733}, \texttt{Caida}, \texttt{Oregon-2}, and \texttt{Skitter}, and the \texttt{Gnutella} graph, which is the network among the hosts of Gnutella website.

\noindent \textbf{Citation Networks.}  Here we have five networks including \texttt{Cora}, \texttt{CiteSeer} network extracted from the CiteSeer library, \texttt{DBLP} network among the papers in DBLP website, \texttt{HepTh} network among the publications in the arXiv's High Energy Physics Theory section, and \texttt{Patent} citation network among the patents registered with the United States Patent and Trademark Office.

\noindent \textbf{Collaboration Networks.} Here we consider three co-authorship networks from DBLP. \texttt{DBLP-dm} includes the authors and papers from top data mining conferences (SIGKDD, WWW, WSDM, ICDM, and SDM) in last ten years. \texttt{DBLP-dbs} is similarly constructed for the top database venues (VLDB, SIGMOD, and ICDE) and \texttt{DBLP-pp} is likewise for the top parallel processing conferences (IPDPS, HPDC, SC, and ICS).

\noindent \textbf{Web Networks.} We include five hyperlink networks that are constructed among websites (edge directions are ignored). \texttt{Stanford}, \texttt{NotreDame}, and \texttt{BerkStan} networks are formed in the domain of Stanford University, University of Notre Dame, and UC Berkeley and Stanford. \texttt{Blogs} contains the front-page hyperlinks between blogs in the context of the 2004 US election and \texttt{Google} is another hyperlink network released in 2002 by Google as a part of the Google Programming Contest.

We also consider the random graph models to validate our findings in real-world networks. For each real-world network, we generate 10 corresponding random graphs using BTER model and consider the average numbers for core and truss degeneracy, as shown in~\cref{tab:data}. We also generate random graphs with ER and configuration models, but the results are not shown as they fail to provide any close approximates for core and truss degeneracy.
The core degeneracy of the real-world networks ranges from 6 to 568, and the truss degeneracy ranges from 2 to 350.
The degeneracy of social networks are commonly higher than the other types.
Core degeneracy can be approximated very well by the BTER model for most datasets.
However, the BTER model is less successful in capturing the truss degeneracy.

\begin{figure*}[!t]
\centering
\vspace{-3ex}
\hspace{-2ex}
\begin{subfigure}{0.33\textwidth}
\includegraphics[width=\linewidth]{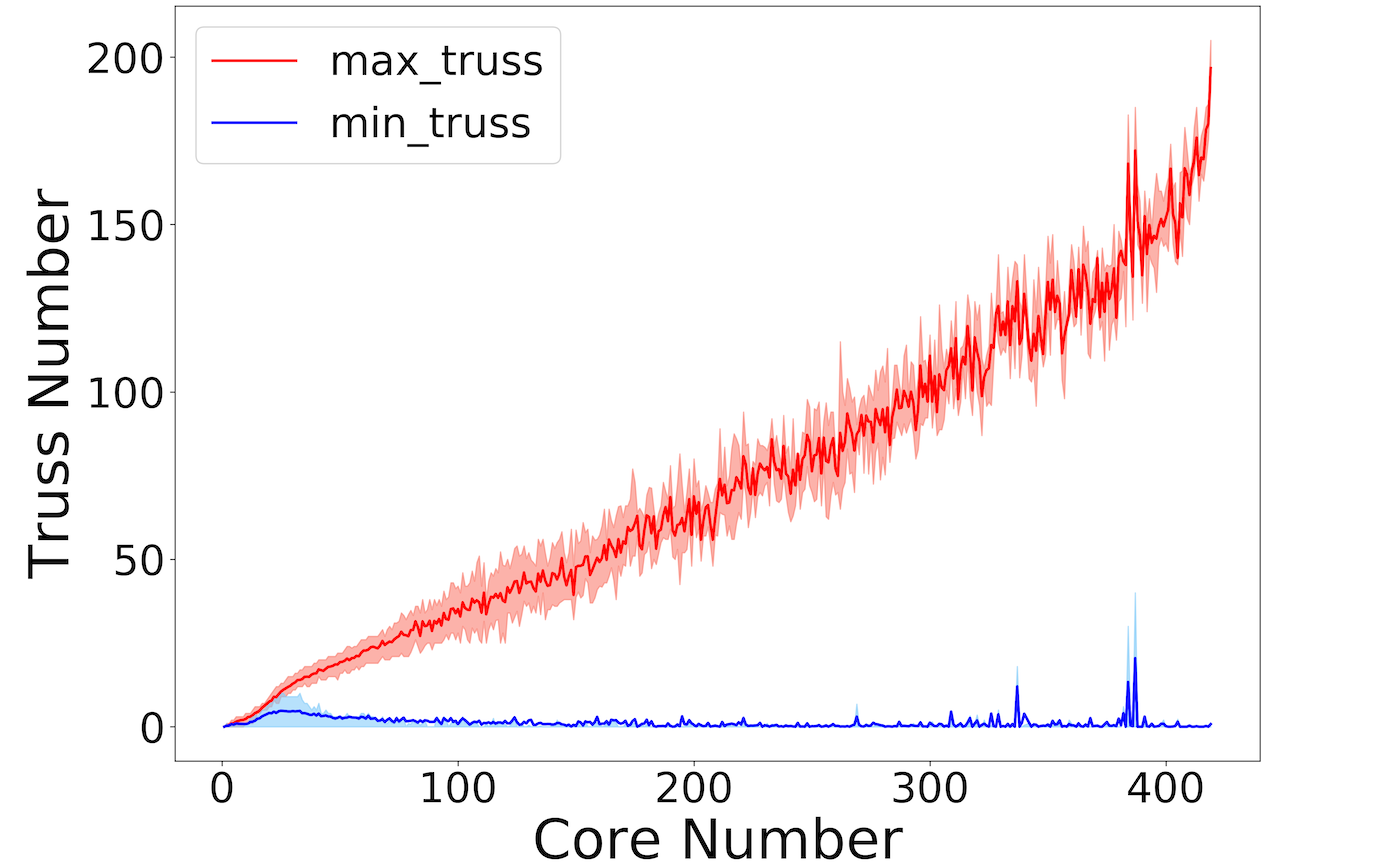}
\vspace{-1.5\baselineskip}
\caption{\small \tt Catster}
\label{fig:catsterVI}
\end{subfigure}
\hspace{-4ex}
\begin{subfigure}{0.33\textwidth}
\includegraphics[width=\linewidth]{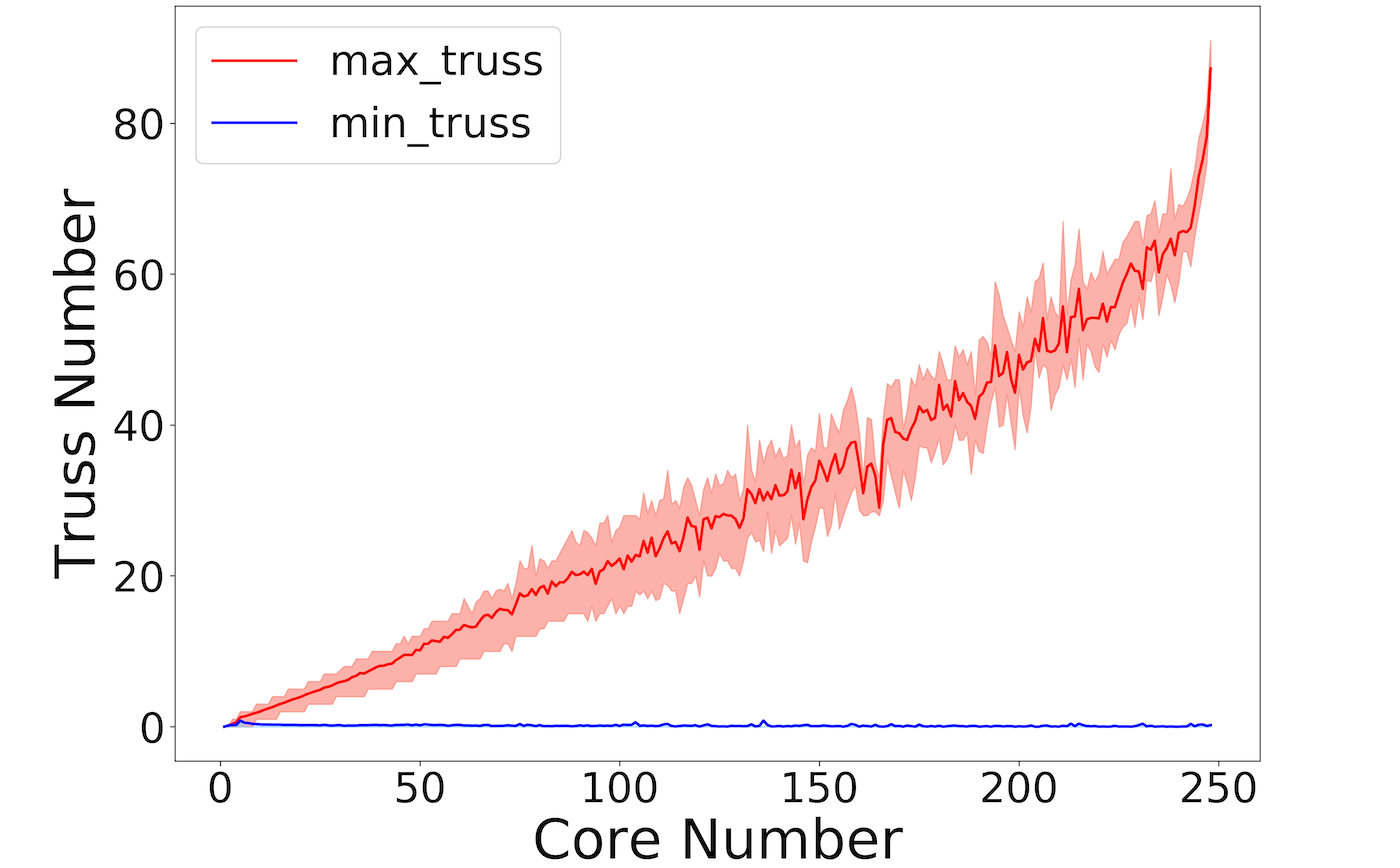}
\vspace{-1.5\baselineskip}
\caption{\small \tt Dogster}
\label{fig:dogsterVI}
\end{subfigure}
\hspace{-4ex}
\begin{subfigure}{0.33\textwidth}
\includegraphics[width=\linewidth]{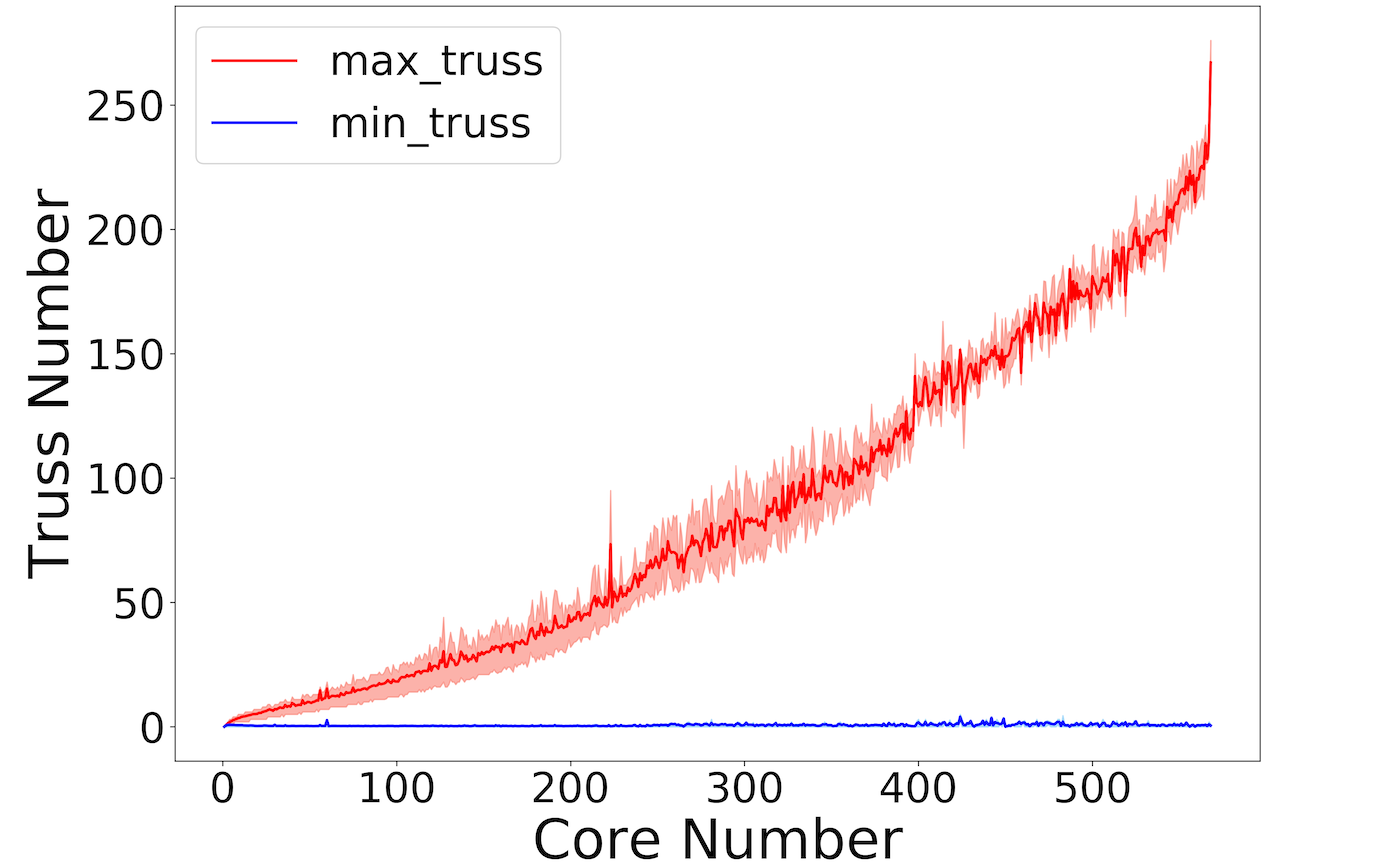}
\vspace{-1.5\baselineskip}
\caption{\small \tt Flickr}
\label{fig:flickrVI}
\end{subfigure}
\hspace{-4ex}

\hspace{-2ex}
\begin{subfigure}{0.33\textwidth}
\includegraphics[width=\linewidth]{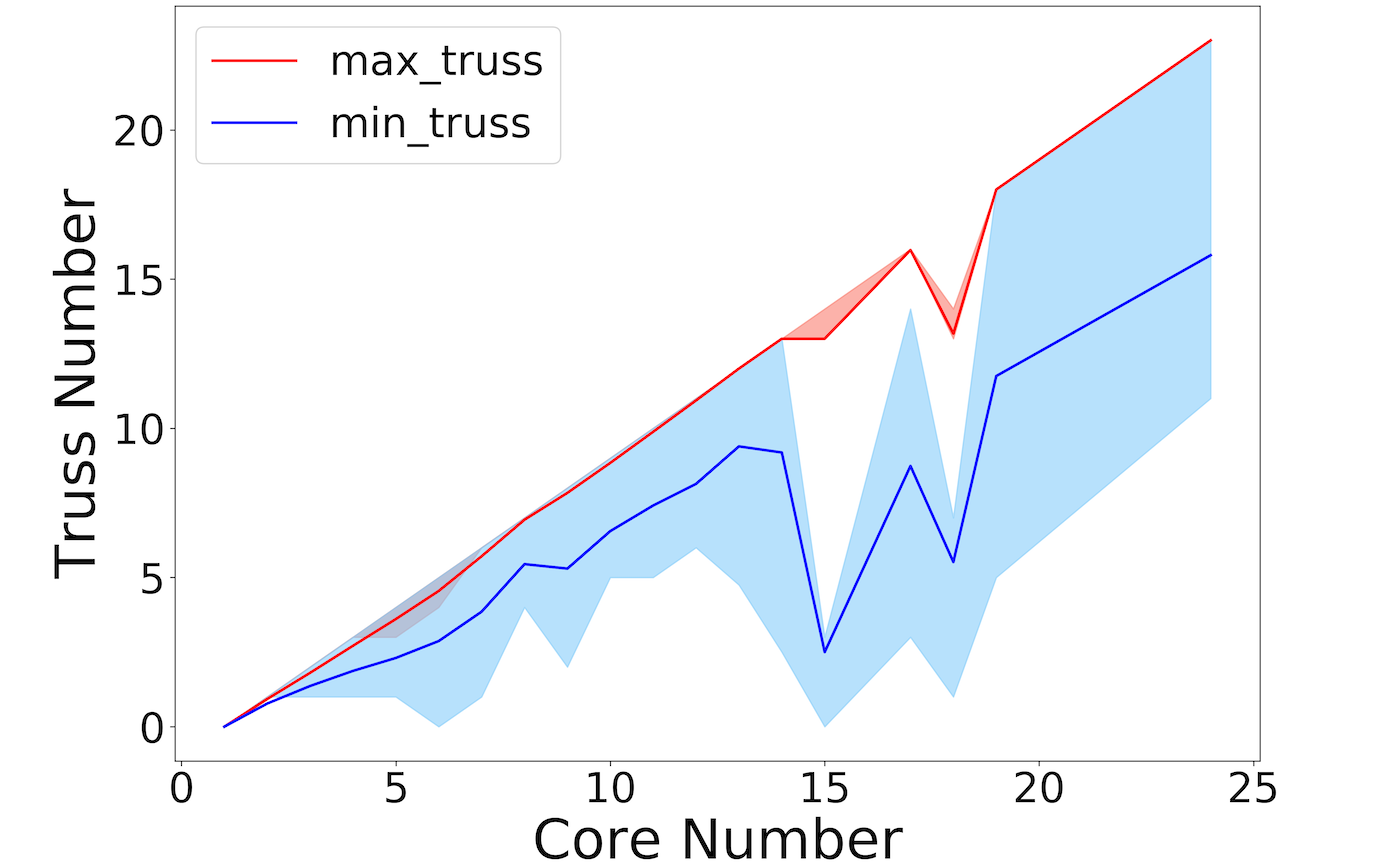}
\vspace{-1.5\baselineskip}
\caption{\small \tt DBLP-dm}
\label{fig:dmVI}
\end{subfigure}
\hspace{-4ex}
\begin{subfigure}{0.33\textwidth}
\includegraphics[width=\linewidth]{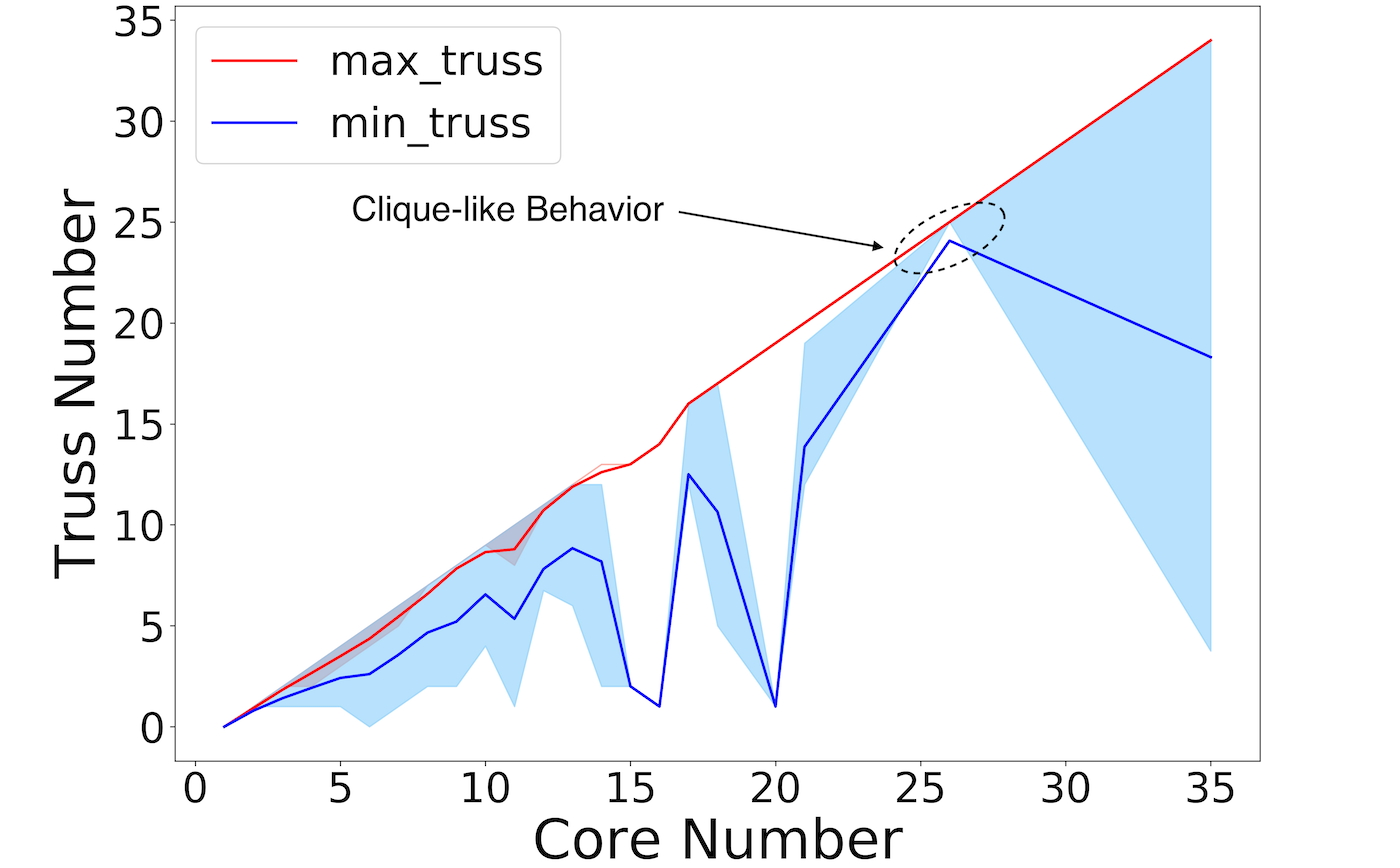}
\vspace{-1.5\baselineskip}
\caption{\small \tt DBLP-dbs}
\label{fig:dbsVI}
\end{subfigure}
\hspace{-4ex}
\begin{subfigure}{0.33\textwidth}
\includegraphics[width=\linewidth]{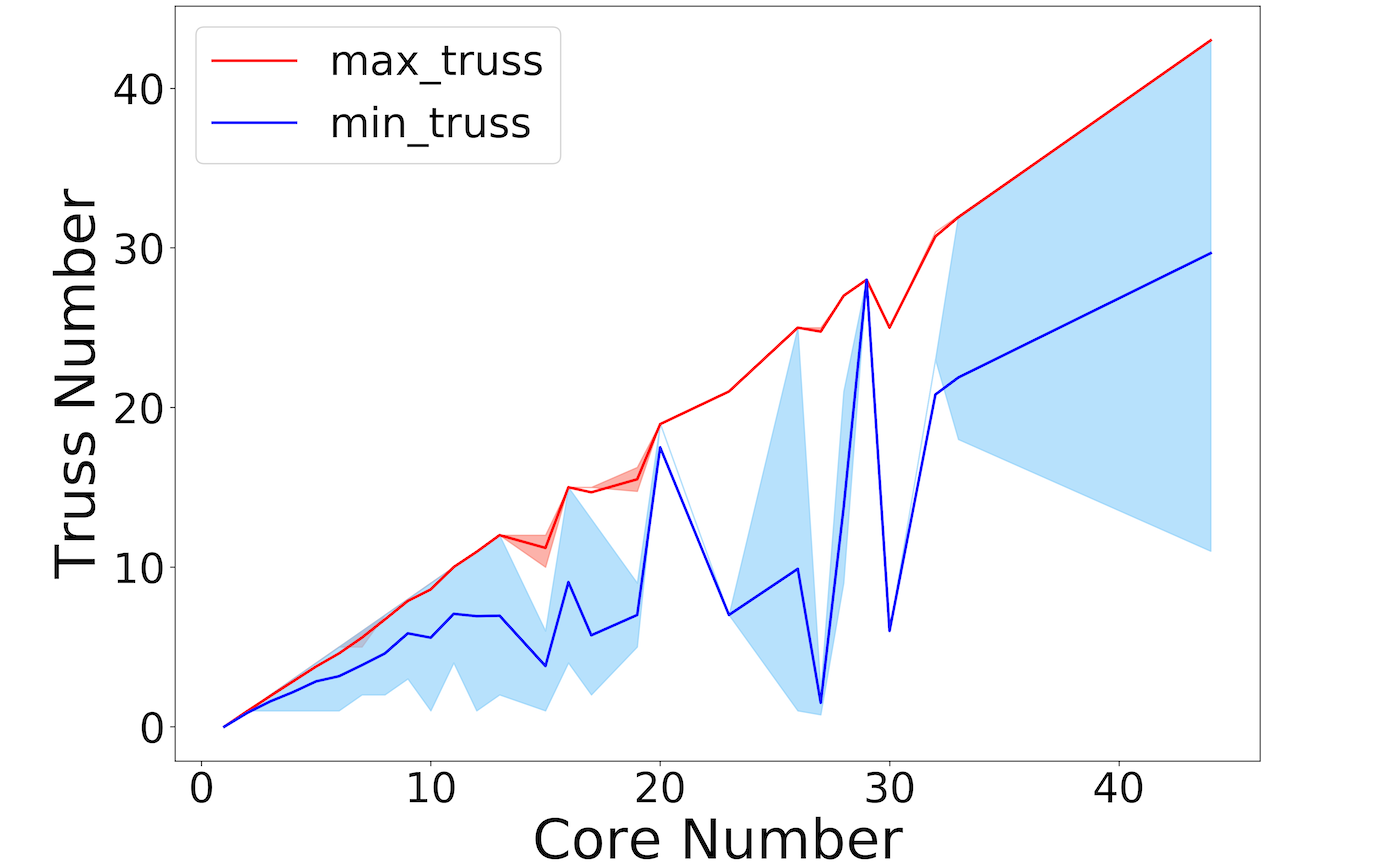}
\vspace{-1.5\baselineskip}
\caption{\small \tt DBLP-pp}
\label{fig:ppVI}
\end{subfigure}
\hspace{-4ex}

\hspace{-2ex}
\begin{subfigure}{0.33\textwidth}
\includegraphics[width=\linewidth]{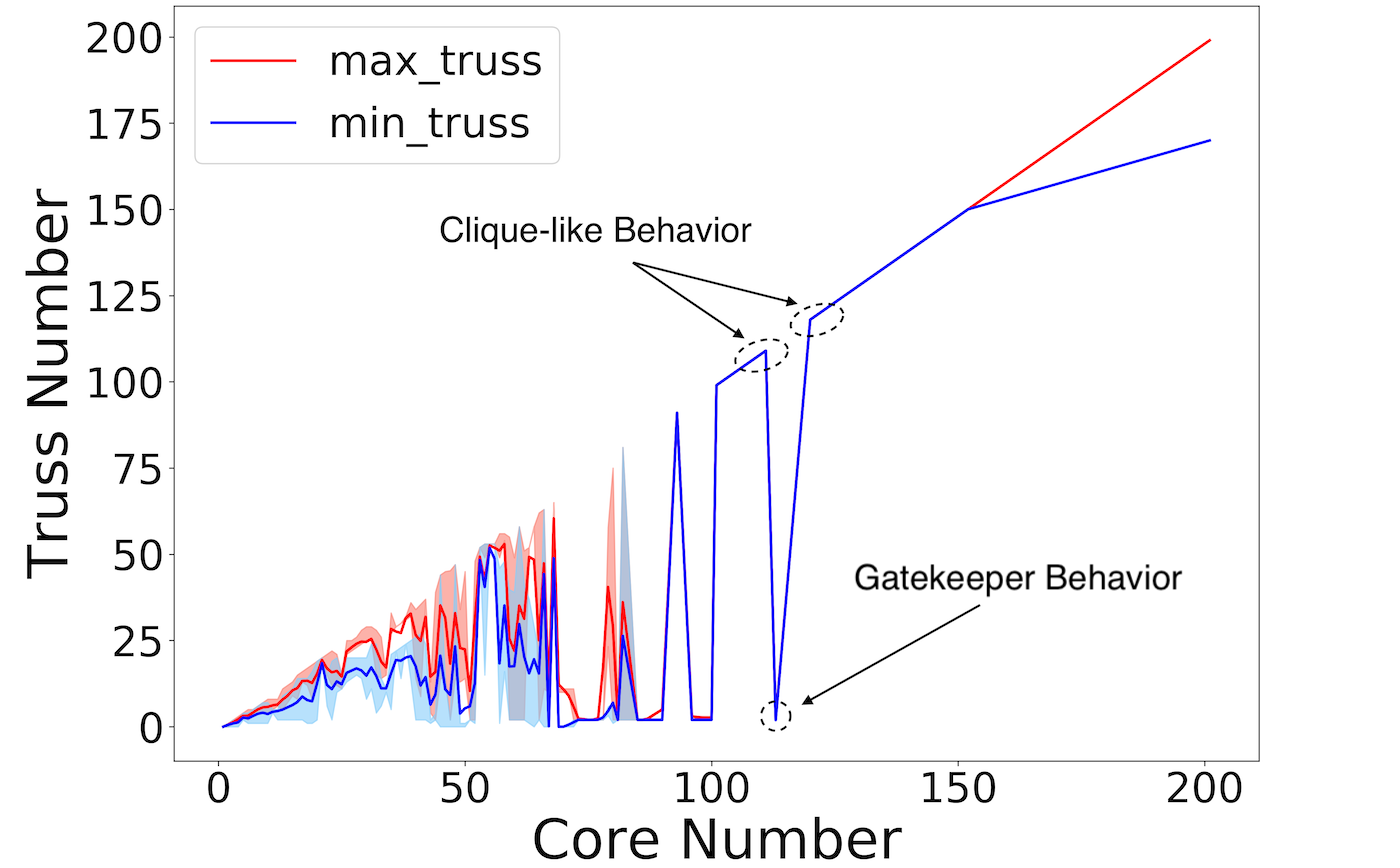}
\vspace{-1.5\baselineskip}
\caption{\small \tt BerkStan}
\label{fig:berkstanVI}
\end{subfigure}
\hspace{-4ex}
\begin{subfigure}{0.33\textwidth}
\includegraphics[width=\linewidth]{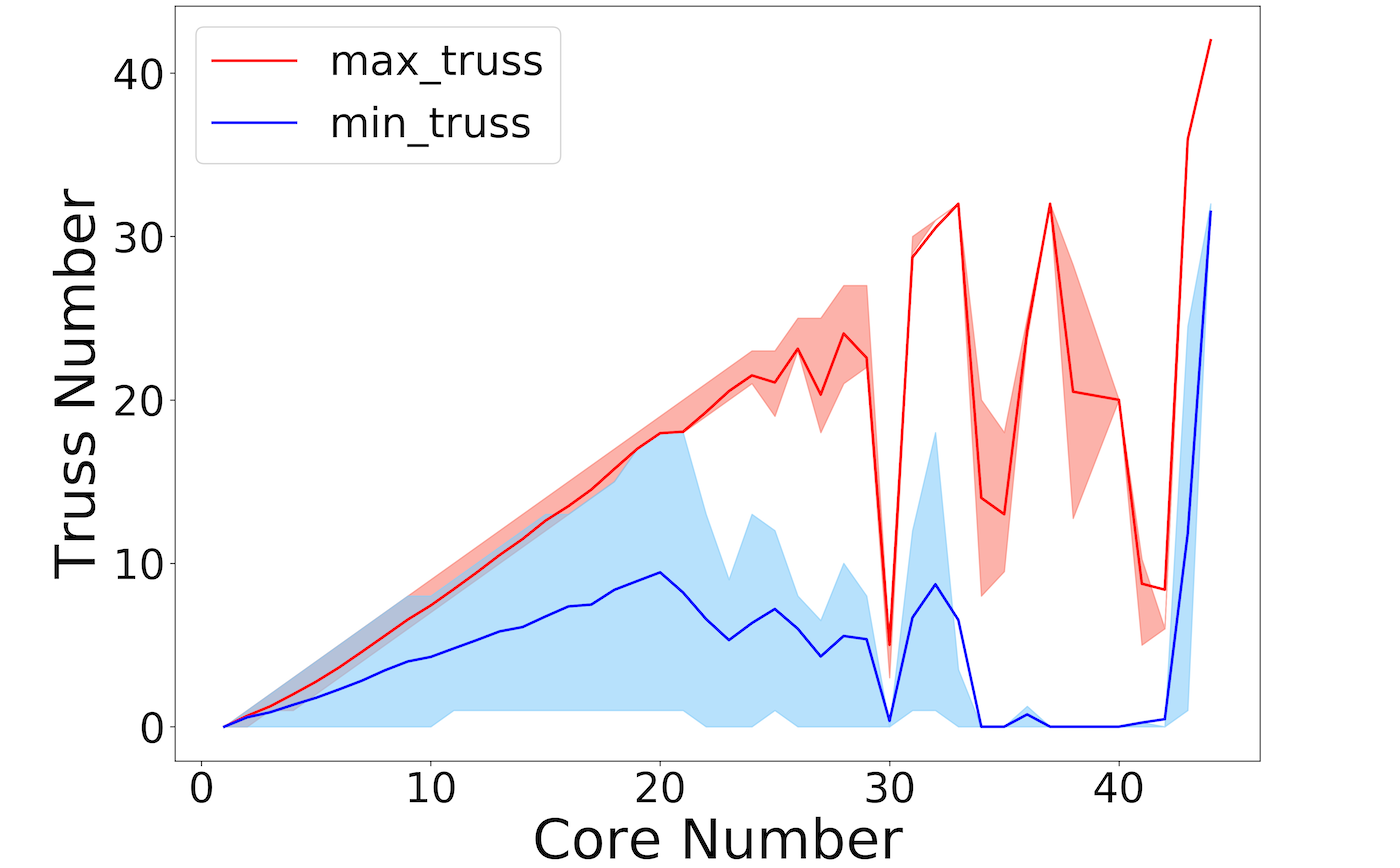}
\vspace{-1.5\baselineskip}
\caption{\small \tt Google}
\label{fig:googleVI}
\end{subfigure}
\hspace{-4ex}
\begin{subfigure}{0.33\textwidth}
\includegraphics[width=\linewidth]{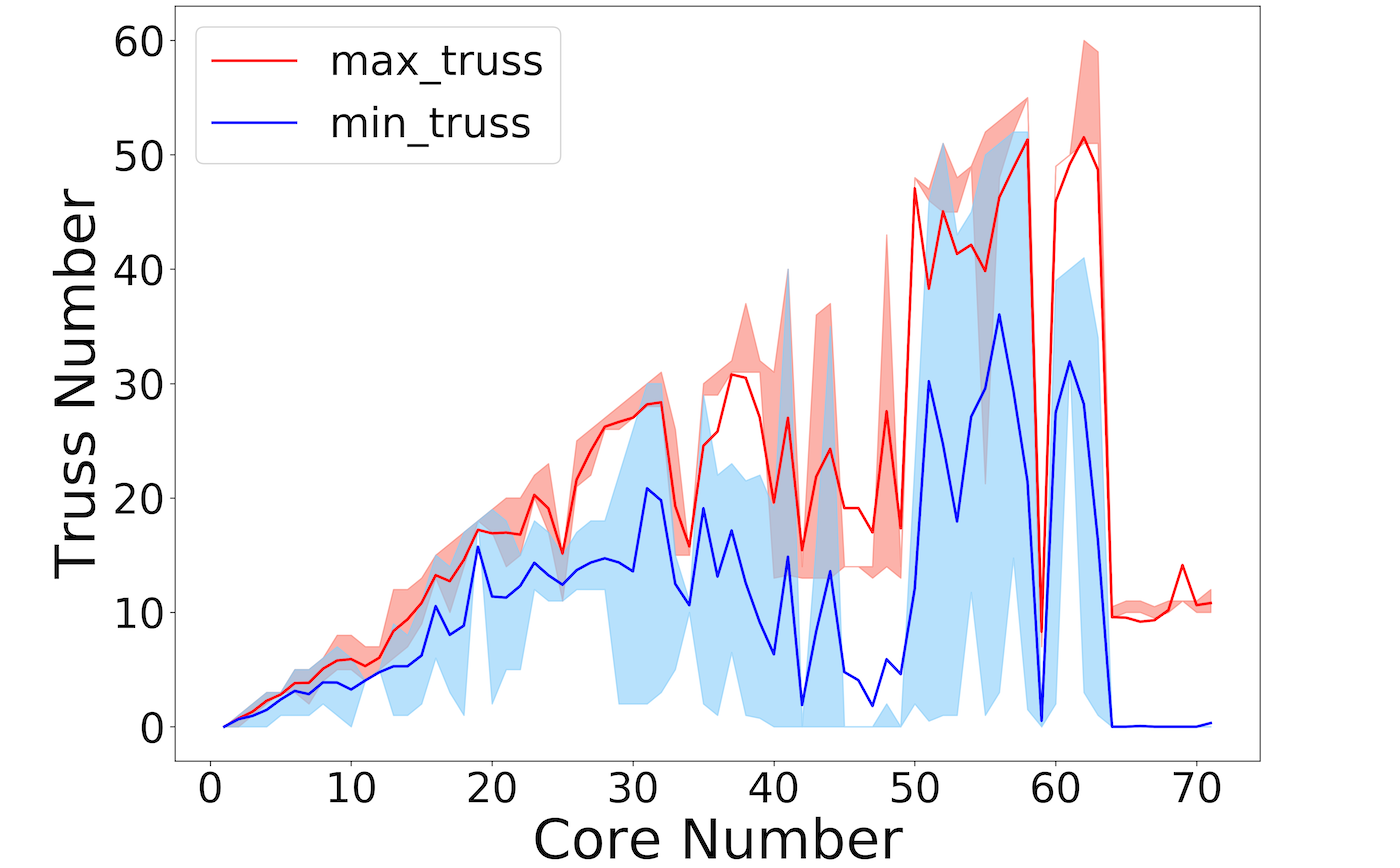}
\vspace{-1.5\baselineskip}
\caption{\small \tt Stanford}
\label{fig:stanfordVI}
\end{subfigure}
\hspace{-4ex}
\vspace{-0.5\baselineskip}
\caption{\small \bf Vertex interplay (VI) plots for some real-world graphs (all are available in 
Figures \ref{fig:VI_ext_1} and \ref{fig:VI_ext_2}). 
For each vertex with a particular core number, the maximum and minimum of the truss numbers of surrounding edges are shown.
Average and interquartile ranges are computed over the vertices with the same core number.
Social networks in Figures~\ref{fig:catsterVI},~\ref{fig:dogsterVI},~\ref{fig:flickrVI} exhibit a common (and expected) behavior, where maximum truss numbers of surrounding edges is larger for vertices with large core numbers whereas the minimum truss numbers of those stay low for all the core numbers.
Collaboration networks, shown in Figures~\ref{fig:dmVI},~\ref{fig:dbsVI},~\ref{fig:ppVI}, present a different picture; vertices with large core numbers are only surrounded by edges with large truss numbers (increasing minimum truss numbers).
Lastly, web networks' characteristic behaviors (presented in Figures~\ref{fig:berkstanVI},~\ref{fig:googleVI},~\ref{fig:stanfordVI}) are unlike others, where the maximum and minimum truss numbers demonstrate non-monotonic behavior since some vertices with large core numbers serve as a gatekeeper (as in \cref{fig:hub-like}) among dense regions (thus surrounded with low truss numbers.)}
\vspace{-2ex}
\label{fig:VI}
\end{figure*}

\begin{figure}[!b]
\vspace{-3ex}
\begin{subfigure}[t]{0.23\textwidth}
\includegraphics[width=\linewidth]{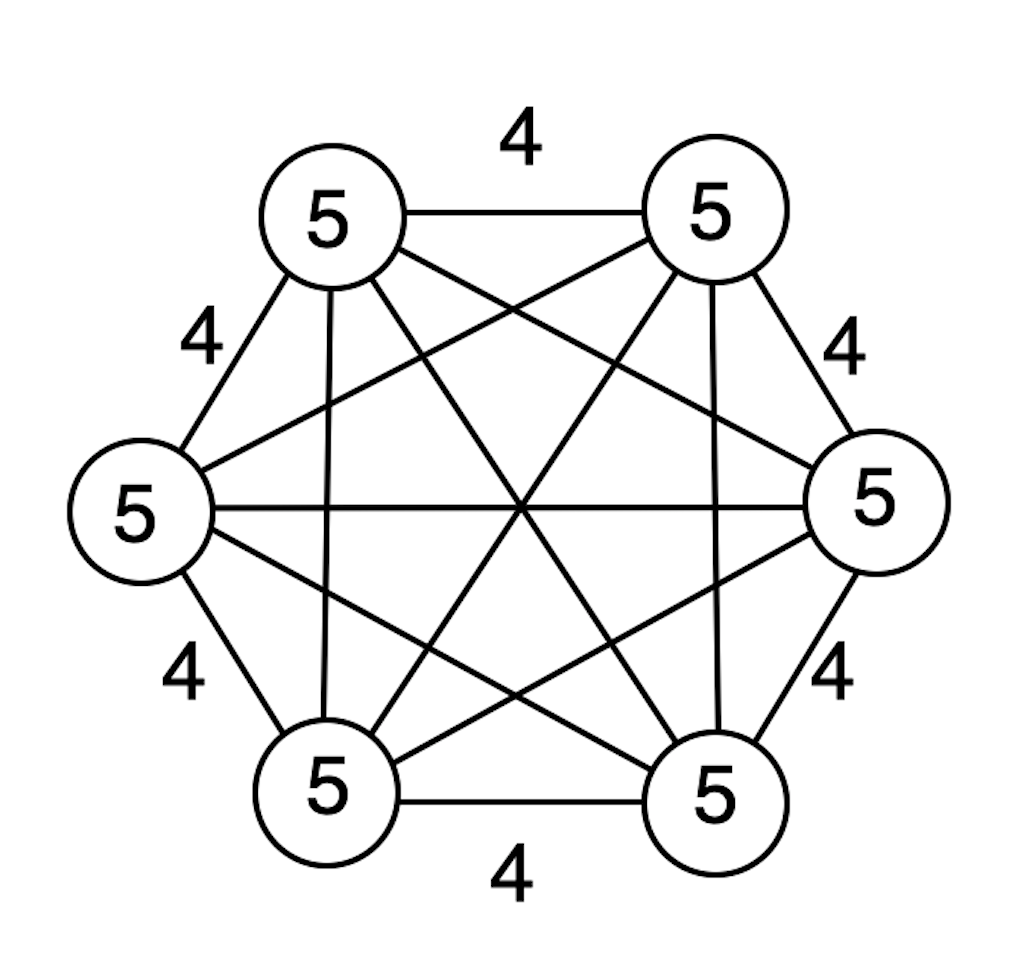}
\vspace{-4ex}
\caption{\small Clique-like structure}
\label{fig:clique-like}
\end{subfigure}
\begin{subfigure}[t]{0.25\textwidth}
\includegraphics[width=\linewidth]{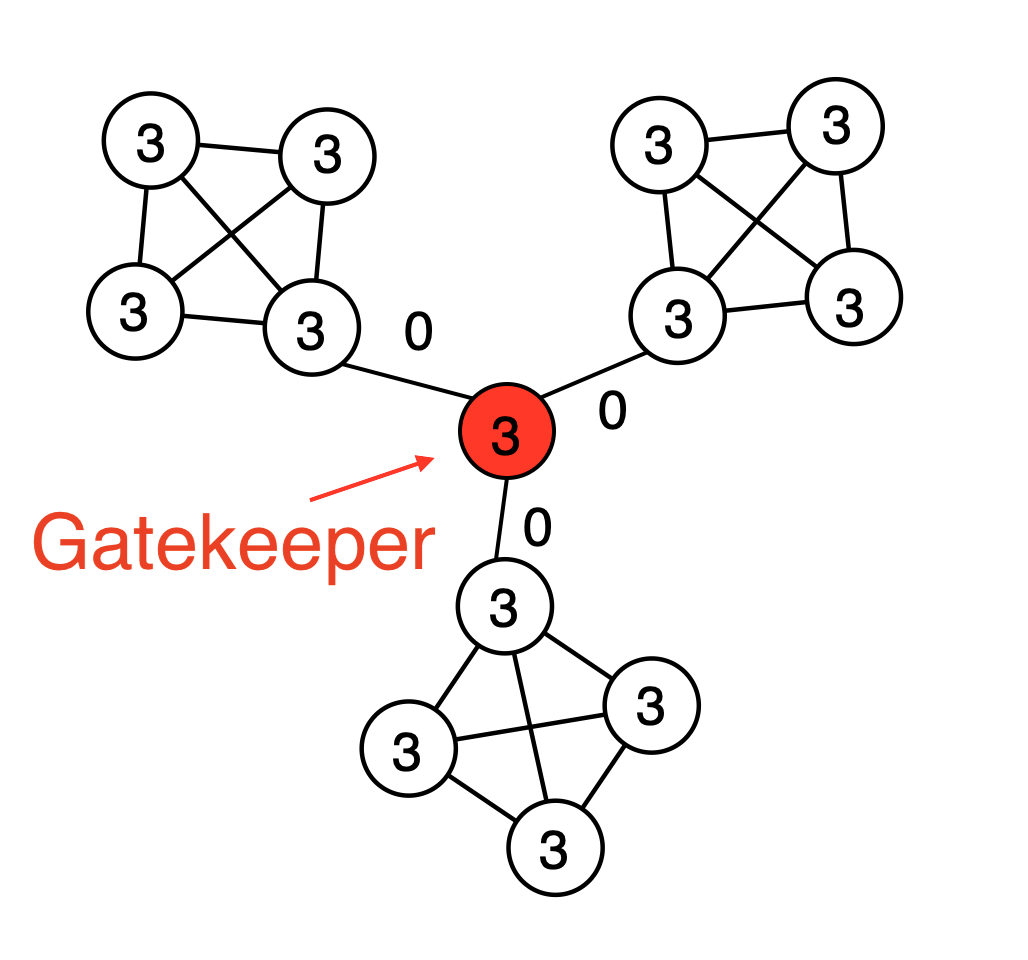}
\vspace{-4ex}
\caption{\small Gatekeeper structure}
\label{fig:hub-like}
\end{subfigure}

\caption{\small \bf Examples of clique-like and gatekeeper structure behaviors. In the 6-clique, the core numbers of vertices are 5 and the truss numbers of edges are 4. The gatekeeper (red) has three neighbors that are isolated from each other. Although the core number (3) shows that the gatekeeper belongs to a cohesive subgraph, its neighborhood structure is not cohesive, as truss numbers are zero.}
\label{fig:clique-hub}
\end{figure}

\section{Vertex Based Analysis}\label{sec:vertex} 

%
\begin{figure*}[!t]
\vspace{-3ex}
\hspace{-2ex}
\begin{subfigure}{0.36\textwidth}
\includegraphics[width=\linewidth]{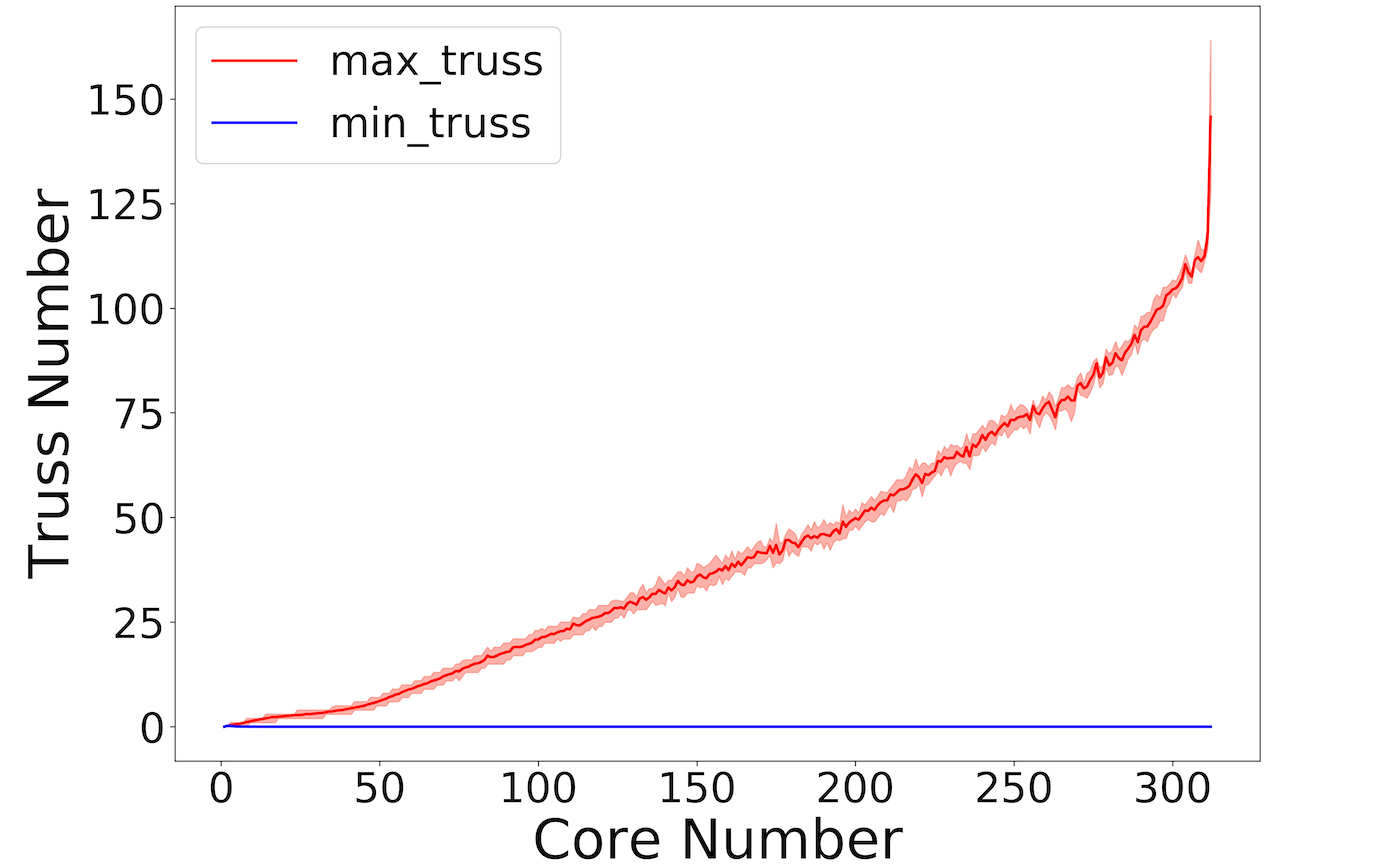}
\vspace{-1.5\baselineskip}
\caption{\small \texttt{Catster} by BTER}
\label{fig:catsterBTER}
\end{subfigure}
\hspace{-4ex}
\begin{subfigure}{0.36\textwidth}
\includegraphics[width=\linewidth]{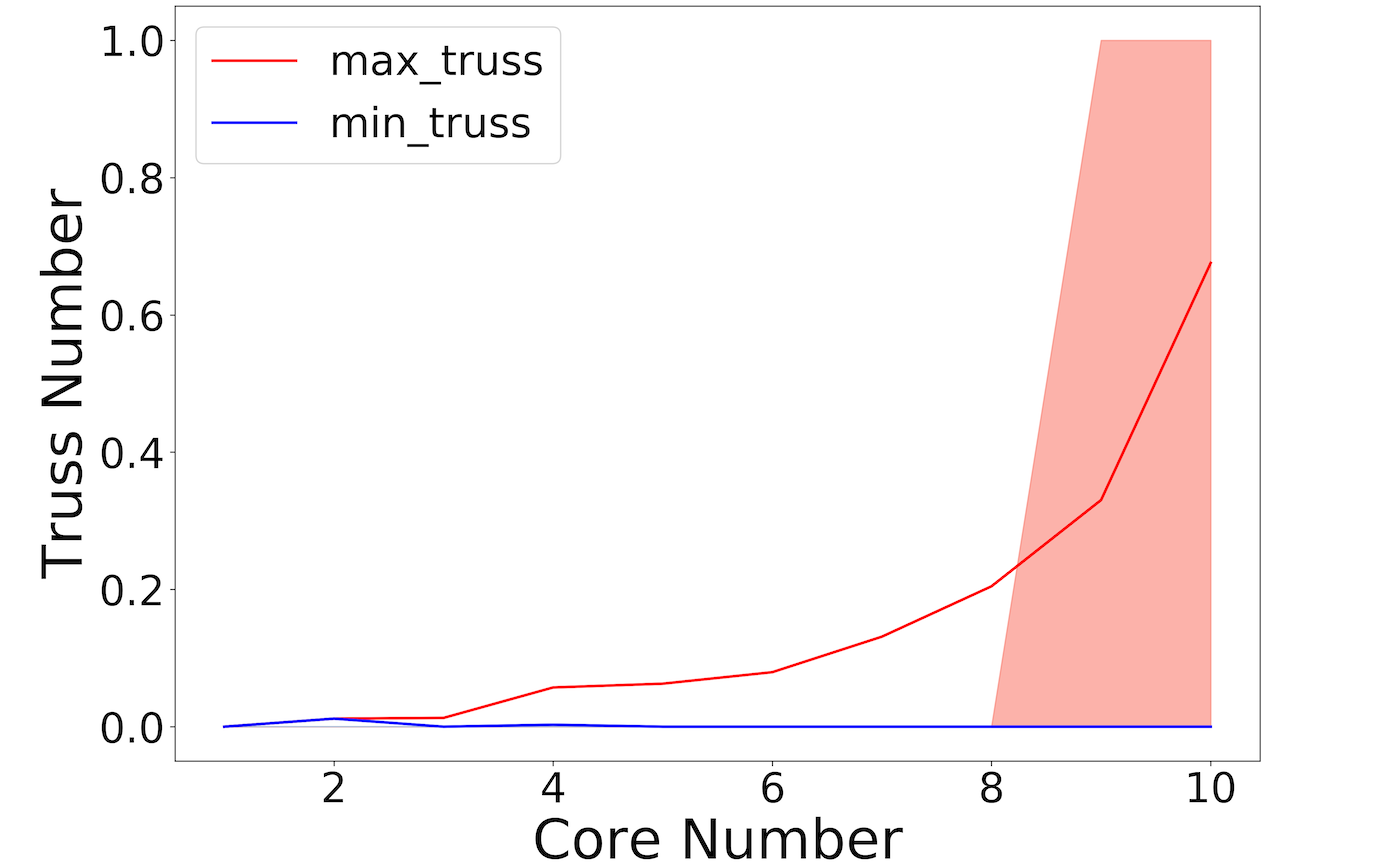}
\vspace{-1.5\baselineskip}
\caption{\small \texttt{DBLP-dbs} by BTER}
\label{fig:dbsBTER}
\end{subfigure}
\hspace{-4ex}
\begin{subfigure}{0.36\textwidth}
\includegraphics[width=\linewidth]{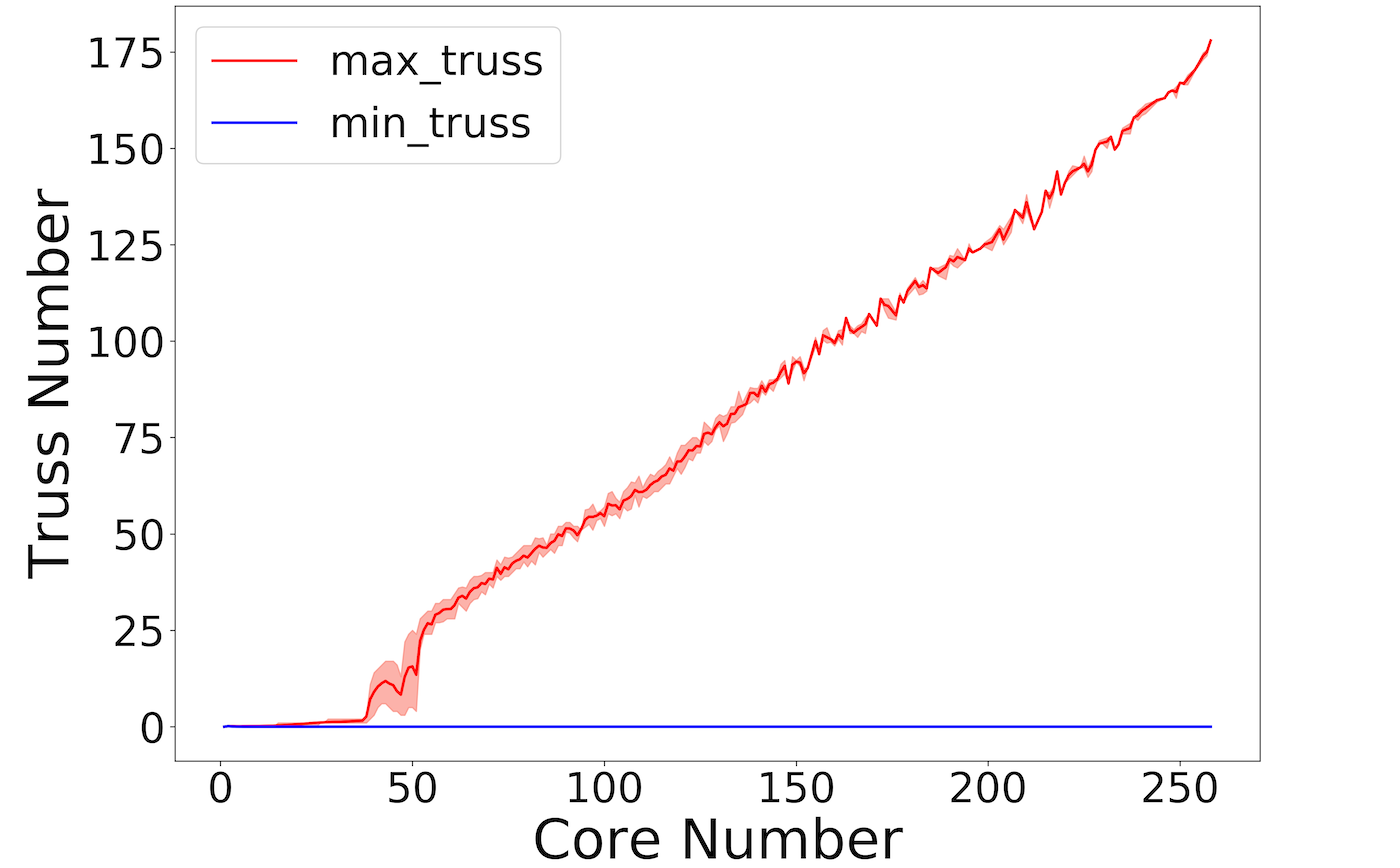}
\vspace{-1.5\baselineskip}
\caption{\small \texttt{BerkStan} by BTER}
\label{fig:berkstanBTER}
\end{subfigure}
\hspace{-4ex}
\vspace{-0.5\baselineskip}
\caption{\small \bf Vertex interplay (VI) plots for random graphs generated from three real-world datasets.
For each vertex with a particular core number, the maximum and minimum of the truss numbers of surrounding edges are shown.
Average and interquartile ranges are computed over the vertices with same core number.
\cref{fig:catsterBTER}, \cref{fig:dbsBTER}, and \cref{fig:berkstanBTER} give the VI plots of BTER model for \texttt{Catster}, \texttt{DBLP-dbs}, and \texttt{BerkStan}.
BTER model can approximate the most general behavior of core-truss patterns to some extent, as we observed in \texttt{Catster}. However, it fails to preserve the interesting behavior in \texttt{DBLP-dbs} and \texttt{BerkStan}, i.e., the clique-like and gatekeeper structures.}
\label{fig:randomVI}
\vspace{-4ex}
\end{figure*}

The core number of a vertex quantifies the cohesiveness around it.
Truss numbers serve the same purpose for the edges.
By definition, we can say that a vertex with a low (high) core number should be connected to some vertices with low (high) core numbers.
It is not clear though, how the core number of a vertex is related to the truss numbers of its edges.
In this section, we analyze the interplay of core and truss numbers from the perspective of a vertex.
We introduce the {\bf vertex interplay (VI) plot} to demonstrate the spectrum of edges around vertices with a particular core number.
We show that the VI plots of networks from the same domain present a consistent behavior and the ones from different domains show a variety.
We first investigate the VI plots of real-world networks, and then compare our findings to the VI plots of the corresponding randomized networks.

\subsection{Real-world Networks}
\begin{figure*}[!p]
\vfill
\vspace{-3ex}
\hspace{-2ex}
\begin{subfigure}{0.35\textwidth}
\includegraphics[width=\linewidth]{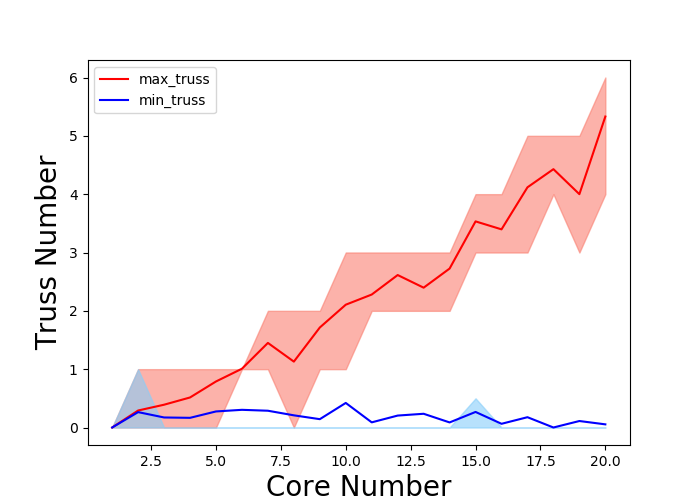}
\vspace{-1.5\baselineskip}
\caption{\tt Hamster}
\end{subfigure}
\hspace{-4ex}
\begin{subfigure}{0.35\textwidth}
\includegraphics[width=\linewidth]{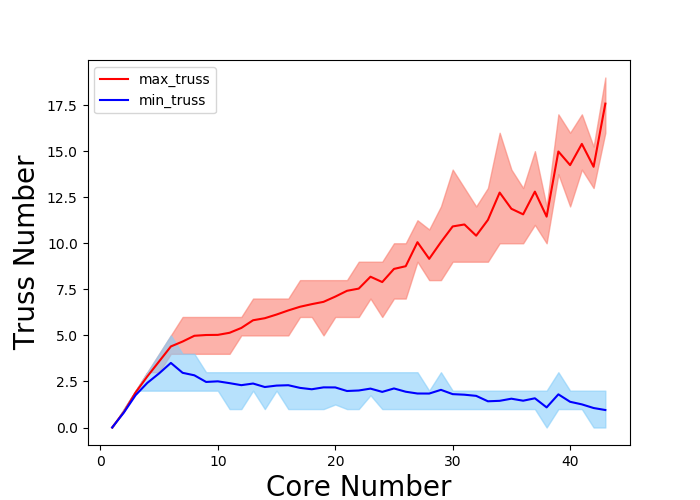}
\vspace{-1.5\baselineskip}
\caption{\tt Email}
\end{subfigure}
\hspace{-4ex}
\begin{subfigure}{0.35\textwidth}
\includegraphics[width=\linewidth]{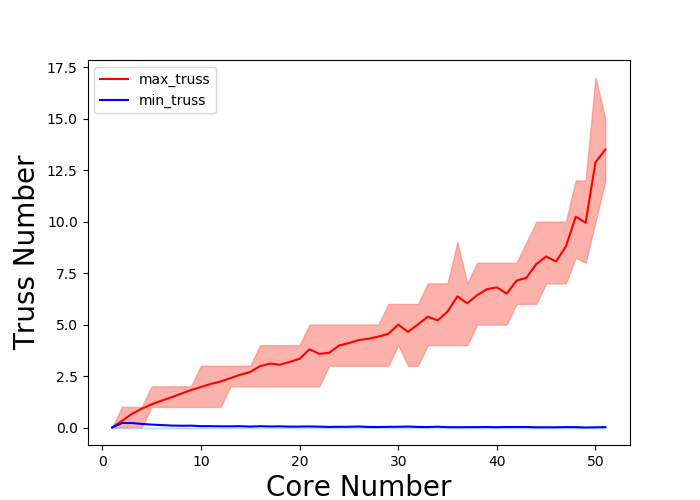}
\vspace{-1.5\baselineskip}
\caption{\tt YouTube}
\end{subfigure}
\hspace{-4ex}

\hspace{-2ex}
\begin{subfigure}{0.35\textwidth}
\includegraphics[width=\linewidth]{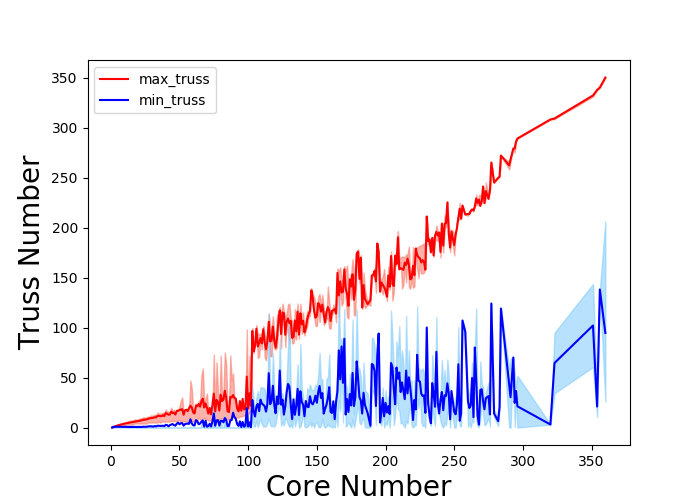}
\vspace{-1.5\baselineskip}
\caption{\tt LiveJournal}
\end{subfigure}
\hspace{-4ex}
\begin{subfigure}{0.35\textwidth}
\includegraphics[width=\linewidth]{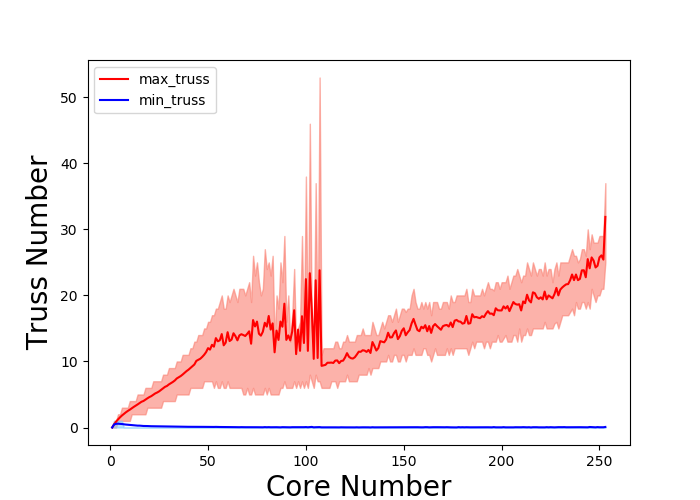}
\vspace{-1.5\baselineskip}
\caption{\tt Orkut}
\end{subfigure}
\hspace{-4ex}
\begin{subfigure}{0.35\textwidth}
\includegraphics[width=\linewidth]{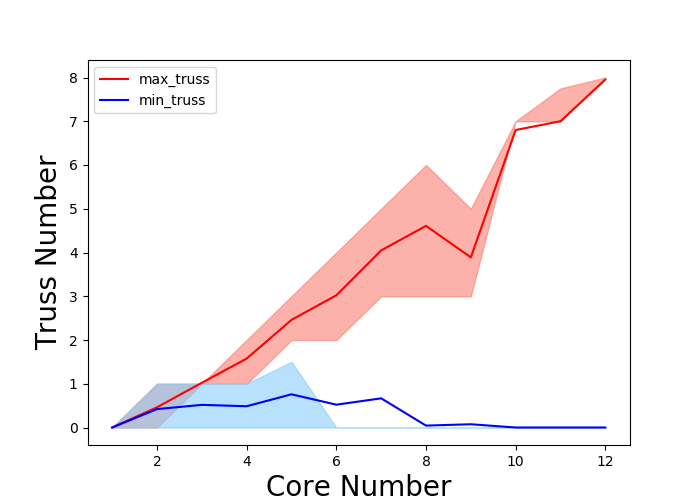}
\vspace{-1.5\baselineskip}
\caption{\tt AS-733}
\end{subfigure}
\hspace{-4ex}

\hspace{-2ex}
\begin{subfigure}{0.35\textwidth}
\includegraphics[width=\linewidth]{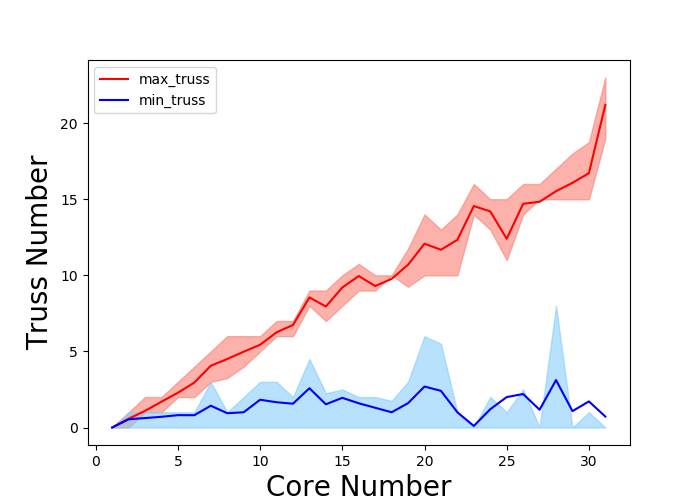}
\vspace{-1.5\baselineskip}
\caption{\tt Oregon-2}
\end{subfigure}
\hspace{-4ex}
\begin{subfigure}{0.35\textwidth}
\includegraphics[width=\linewidth]{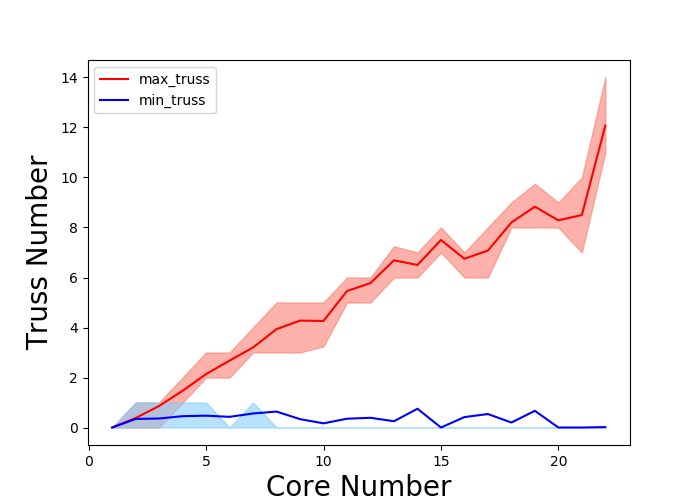}
\vspace{-1.5\baselineskip}
\caption{\tt Caida}
\end{subfigure}
\hspace{-4ex}
\begin{subfigure}{0.35\textwidth}
\includegraphics[width=\linewidth]{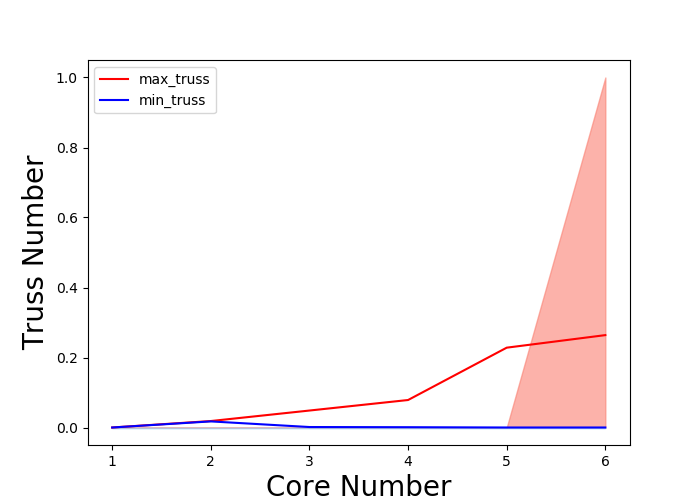}
\vspace{-1.5\baselineskip}
\caption{\tt Gnutella}
\end{subfigure}
\hspace{-4ex}
\vspace{-0.5\baselineskip}
\caption{\bf Vertex interplay (VI) plots for other real-world graphs (listed but not shown in the paper)(part I).
For each vertex with a particular core number, the maximum and minimum of the truss numbers of surrounding edges are shown.
Average and interquartile ranges are computed over the vertices with same core number.
}
\label{fig:VI_ext_1}
\vfill
\end{figure*}

\begin{center} 
\begin{figure*}[h]
\vspace{-3ex}
\hspace{-2ex}
\begin{subfigure}{0.36\textwidth}
\includegraphics[width=\linewidth]{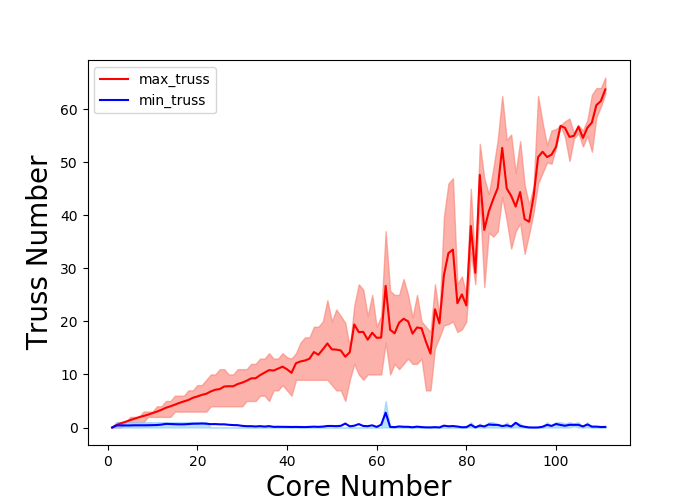}
\vspace{-1.5\baselineskip}
\caption{\tt Skitter}
\end{subfigure}
\hspace{-4ex}
\begin{subfigure}{0.36\textwidth}
\includegraphics[width=\linewidth]{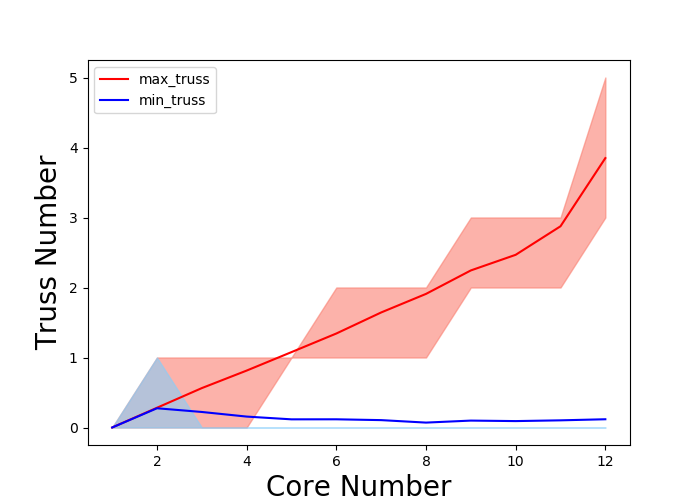}
\vspace{-1.5\baselineskip}
\caption{\tt DBLP}
\end{subfigure}
\hspace{-4ex}
\begin{subfigure}{0.36\textwidth}
\includegraphics[width=\linewidth]{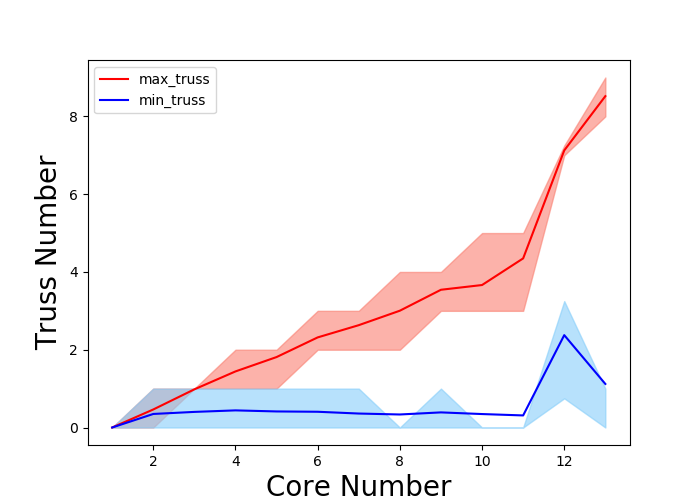}
\vspace{-1.5\baselineskip}
\caption{\tt Cora}
\end{subfigure}
\hspace{-4ex}

\hspace{-2ex}
\begin{subfigure}{0.36\textwidth}
\includegraphics[width=\linewidth]{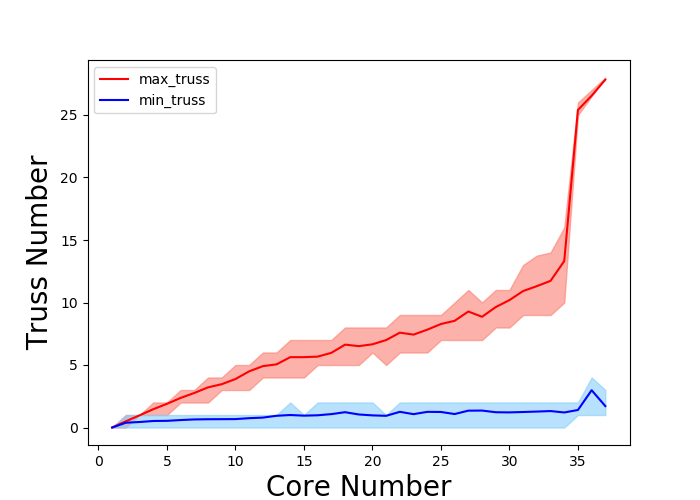}
\vspace{-1.5\baselineskip}
\caption{\tt HepTh}
\end{subfigure}
\hspace{-4ex}
\begin{subfigure}{0.36\textwidth}
\includegraphics[width=\linewidth]{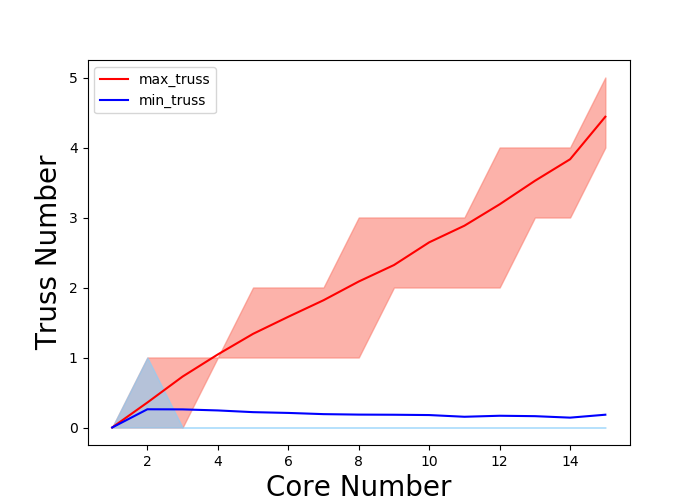}
\vspace{-1.5\baselineskip}
\caption{\tt CiteSeer}
\end{subfigure}
\hspace{-4ex}
\begin{subfigure}{0.36\textwidth}
\includegraphics[width=\linewidth]{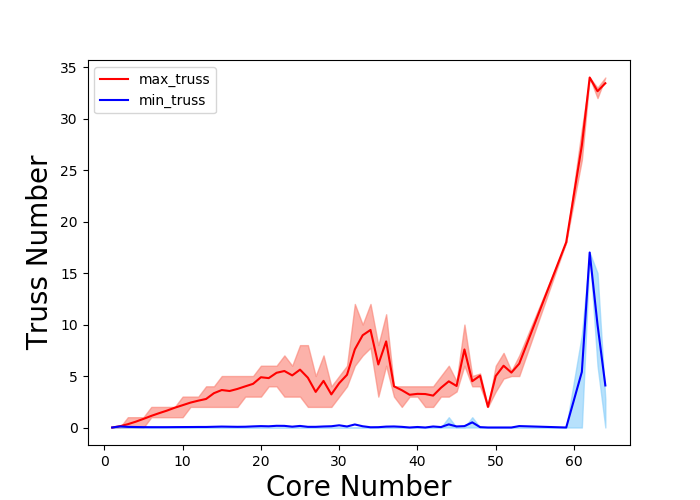}
\vspace{-1.5\baselineskip}
\caption{\tt Patent}
\end{subfigure}
\hspace{-4ex}

\hspace{-2ex}
\begin{subfigure}{0.36\textwidth}
\includegraphics[width=\linewidth]{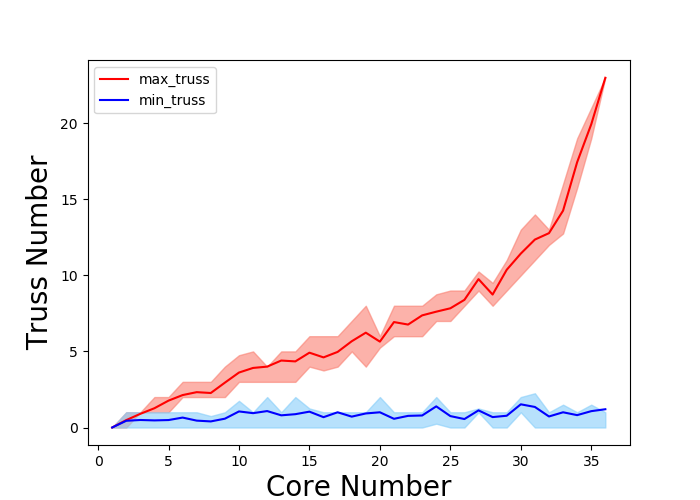}
\vspace{-1.5\baselineskip}
\caption{\tt Blogs}
\end{subfigure}
\hspace{-4ex}
\begin{subfigure}{0.36\textwidth}
\includegraphics[width=\linewidth]{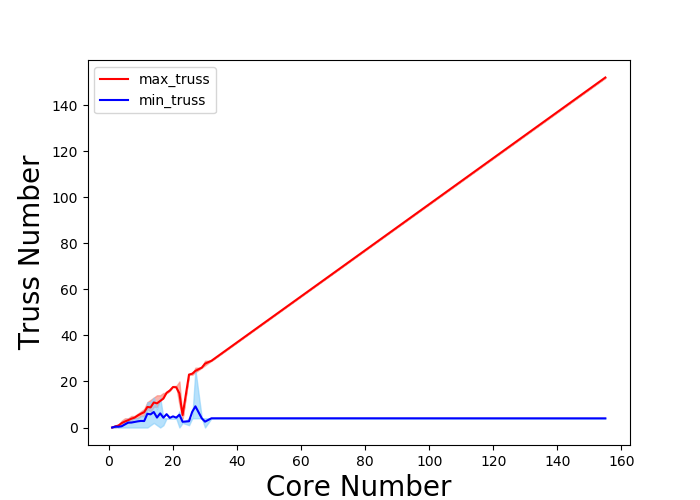}
\vspace{-1.5\baselineskip}
\caption{\tt NotreDame}
\end{subfigure}
\hspace{-4ex}
\vspace{-0.5\baselineskip}
\caption{\bf Vertex interplay (VI) plots for other real-world graphs (listed but not shown in the paper) (Part II).
For each vertex with a particular core number, the maximum and minimum of the truss numbers of surrounding edges are shown.
Average and interquartile ranges are computed over the vertices with same core number.
}
\label{fig:VI_ext_2}
\end{figure*}
\end{center}

~\cref{fig:VI} presents the VI plots for some real-world networks in our dataset (all are available in 
Figures \ref{fig:VI_ext_1} and \ref{fig:VI_ext_2}).
We examine the neighborhood of vertices with a particular core number and consider the maximum and minimum truss numbers of edges adjacent to a given vertex for a fixed core number.
By this, we aim to understand how the core number of a vertex shapes the spectrum of neighborhoods around it.
If there are multiple vertices with the same core number, we report the average and interquartile ranges.
Formally, for each core number $c$, we find the set $S=\{u \in V: K(u)=c\}$ and compute $\frac{1}{|S|}\sum_{u \in S}{min\{T(u, v): v \in N(u)\}}$ (likewise for maximum) along with the interquartile ranges.

One general behavior we observe is that the maximum truss number of adjacent edges is strongly correlated to the core number of the vertex; larger core numbers yield larger truss numbers overall.
Figures~\ref{fig:catsterVI},~\ref{fig:dogsterVI}, and~\ref{fig:flickrVI} show this pattern in the VI plots for social networks: \texttt{Catster}, \texttt{Dogster}, and \texttt{Flickr}.
Regarding the minimum truss numbers, on the other hand, there is a consistent trend for all the vertices regardless of their core numbers.
Every vertex in the network is connected to at least one edge with a very low truss number. 
This is in line with the core-periphery structure~\cite{Borgatti00}; each vertex in the core block is connected to both core and periphery blocks -- truss numbers of the edges between core and periphery are likely to be low.
Note that the variation for vertices with the same core number is also very small, i.e., interquartile ranges are narrow, suggesting a high similarity among those vertices.
We observe this behavior in 8 (of 9) social networks -- \texttt{LiveJournal} exhibits a zigzag trend for the minimum truss numbers, probably because of the large range of core and truss values (it has the largest degeneracy numbers among all the networks).
Autonomous systems exhibit a similar behavior as well; the minimum truss number line has small zigzags in some cases.


Collaboration networks exhibit a different behavior in the VI plots.
Figures~\ref{fig:dmVI},~\ref{fig:dbsVI}, and~\ref{fig:ppVI} present the \texttt{DBLP-dm}, \texttt{DBLP-dbs}, and \texttt{DBLP-pp} networks.
The minimum truss numbers are very close to the maximum ones, as opposed to the consistent trend in the social networks.
Vertices with large core numbers are not connected to any edge with a low truss number.
This implies that cliques of vertices with large core numbers are surrounded by some other cliques with close core numbers.
In some cases the minimum truss number, maximum truss number, and core number are almost equal, indicating both $k$-core and $k$-truss are derived from a clique-like structure (\cref{fig:clique-like}), as annotated in~\cref{fig:dbsVI}. In fact, a collaboration network can be seen as a union of cliques, where authors of each paper form a clique, which in turn yields closed clique-like structures that are not connected to the nodes in the periphery.

We observe yet another distinct behavior in web networks.
Figures~\ref{fig:berkstanVI},~\ref{fig:googleVI}, and~\ref{fig:stanfordVI} show the VI plots for \texttt{BerkStan}, \texttt{Google}, and \texttt{Stanford} networks.
The maximum truss numbers are very small for some vertices with large core numbers, implying that the truss numbers of all adjacent edges are small.
Most of the neighbors are isolated from each other, indicating a sparse neighborhood despite the cohesiveness suggested by the large core number.
Note that a great amount of the neighbors also have a large core number, by definition.
Those vertices serve as structural holes in the network~\cite{Burt09}, as illustrated in~\cref{fig:hub-like} where a node with a large core number is connecting multiple cores isolated from each other.
In~\cref{fig:berkstanVI}, there are 271 vertices with core number 113 but the largest truss number adjacent to those is only 2.
Note that, in the same network, the vertices with core number 111 and 120 are surrounded by some very large truss numbers, indicating a clique-like structure.

\subsection{Random Graphs}

Here we investigate if the observed behaviors in the VI plots are artifacts of some graph characteristics.
For this purpose, we randomize all the real-world networks with the BTER model~\cite{Sesh12}.
As explained in~\cref{sec:back}, the BTER model rewires the edges in a reference network by preserving the clustering coefficient per degree distribution to the best extent.

\cref{fig:randomVI} presents the results for \texttt{Catster}, \texttt{DBLP-dbs}, and \texttt{BerkStan} networks, each represents a different behavior in~\cref{fig:VI}.
Random graphs generated by BTER model capture the common pattern observed in social networks but fail to capture the other interesting behaviors in collaboration and web graphs
The low-maximum and high-minimum behaviors, which are driven by the unusual structures (cliques and structural holes), seem to be distinctive characteristics in real-world graphs.

\begin{figure*}[!t]
\captionsetup[subfigure]{justification=centering}
\begin{subfigure}{0.33\textwidth}
\includegraphics[width=\linewidth]{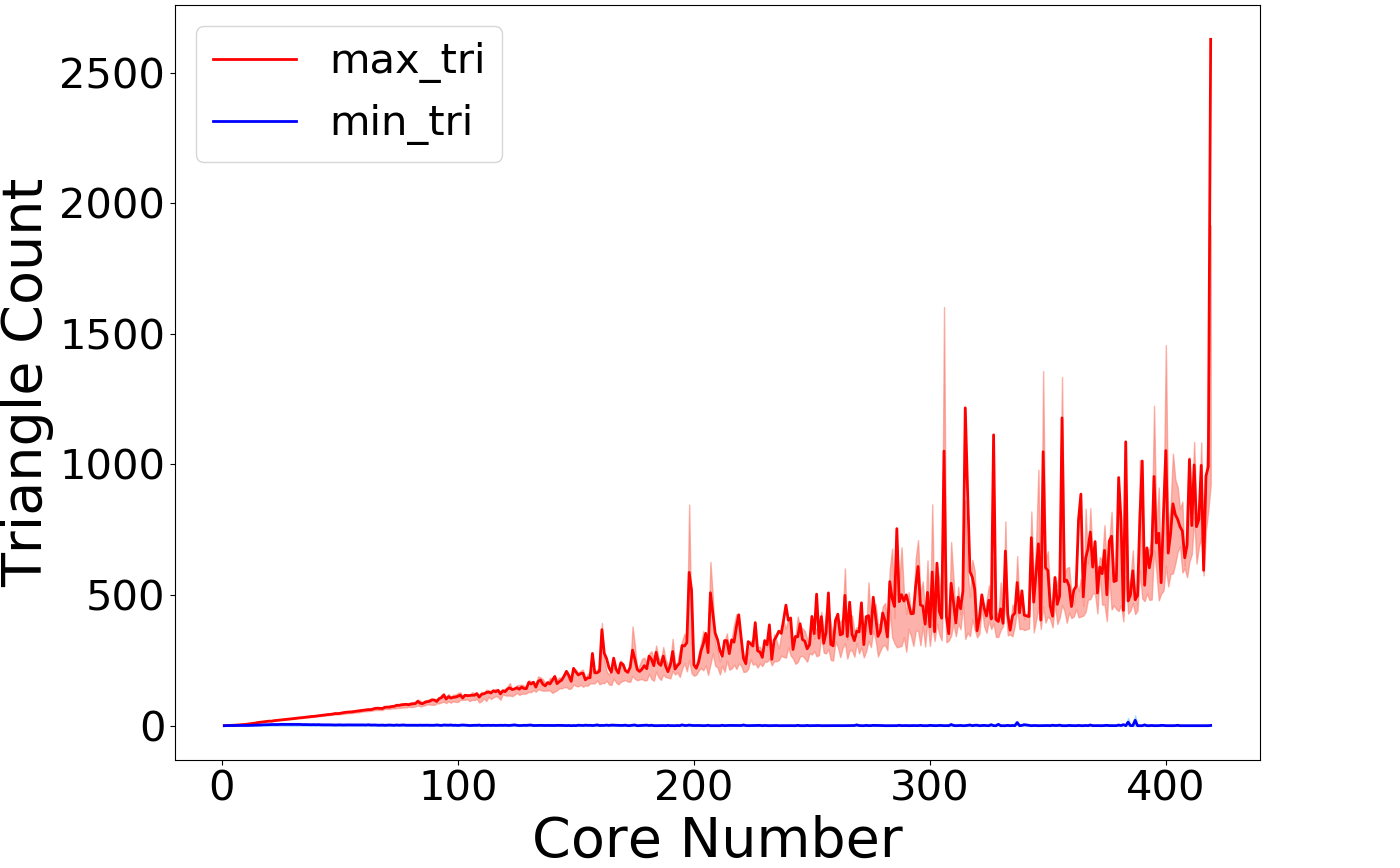}
\caption{\small Core numbers and triangle counts}
\label{fig:catsterALT}
\end{subfigure}
\hspace{1ex}
\begin{subfigure}{0.33\textwidth}
\includegraphics[width=\linewidth]{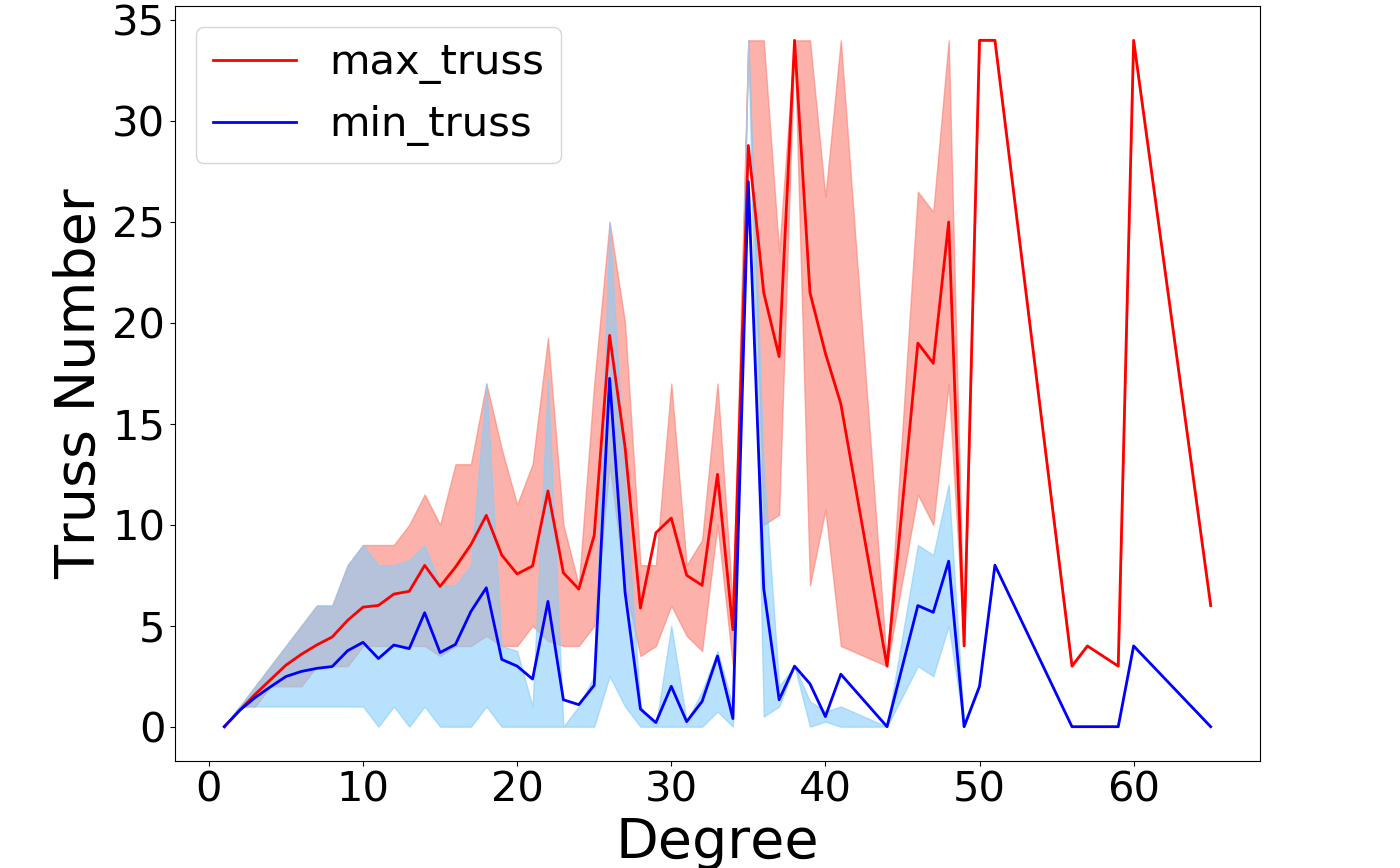}
\caption{\small Degrees and truss number}
\label{fig:dbsALT}
\end{subfigure}
\hspace{1ex}
\begin{subfigure}{0.33\textwidth}
\includegraphics[width=\linewidth]{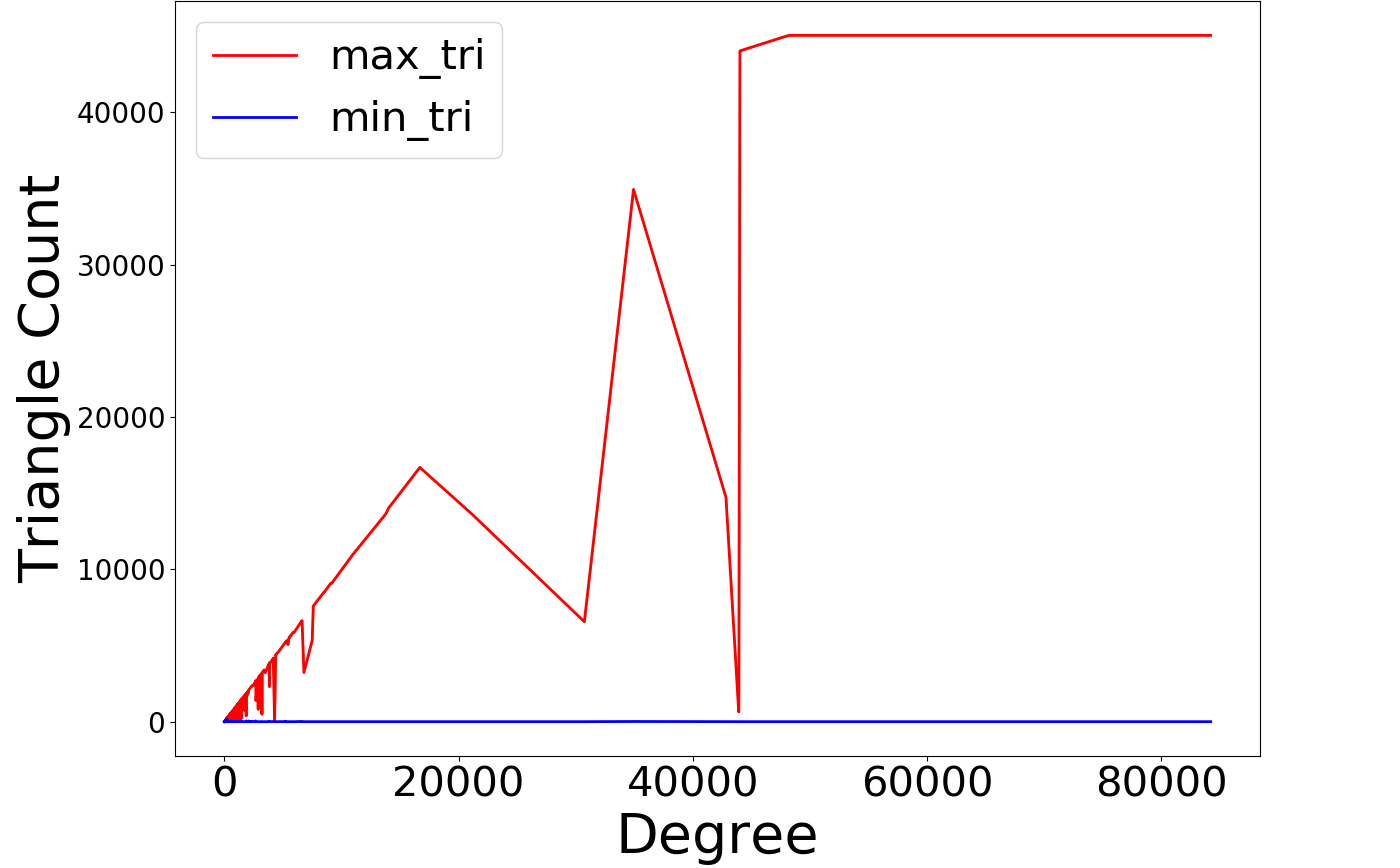}
\caption{\small Degrees and triangle counts }
\label{fig:berkstanALT}
\end{subfigure}
\vspace{-0.5\baselineskip}
\caption{\small \bf Interplays between alternative measures. \cref{fig:catsterALT}, \cref{fig:dbsALT}, and \cref{fig:berkstanALT} show the interplay in \texttt{Catster}, \texttt{DBLP-dbs}, and \texttt{BerkStan}. All fail to generate a consistent and pervasive pattern.
}
\label{fig:otherVIs}
\end{figure*}

\noindent\textbf{Remark.} Given the benefit of analyzing the interplay of core and truss numbers, it is natural to think about using just the degrees of vertices and triangle counts of edges for a similar analysis.
However, the skewed distributions of degrees/triangle counts prevent such analyses; the VI plots become inconsistent for the networks from the same domain and behaviors can be gamed with simple changes in the graph.
\cref{fig:otherVIs} presents a few examples for the networks analyzed above.
In all variations (degree-triangle count, degree-truss, or core-triangle count) the VI plots fail to generate a consistent and pervasive pattern.
\textbf{Core and truss numbers can be considered as the regularized and more robust versions of the vertex degrees and edges' triangle counts.}

\noindent\textbf{Summary.} VI plots present a meaningful graph summary by showing the interplay between core and truss numbers. 
Considering the cohesiveness around vertices (i.e., core numbers) with respect to various neighborhoods they are involved in (i.e., truss numbers) is an effective way to understand the dense regions and structural holes. VI plots of networks belonging to the same domain exhibit consistent behaviors whereas the ones from different domains suggest diverse characteristics. We believe that VI plots would be handy for domain practitioners to analyze the network structure.

\begin{figure*}[!b]
\begin{subfigure}{0.33\textwidth}
\includegraphics[width=\linewidth]{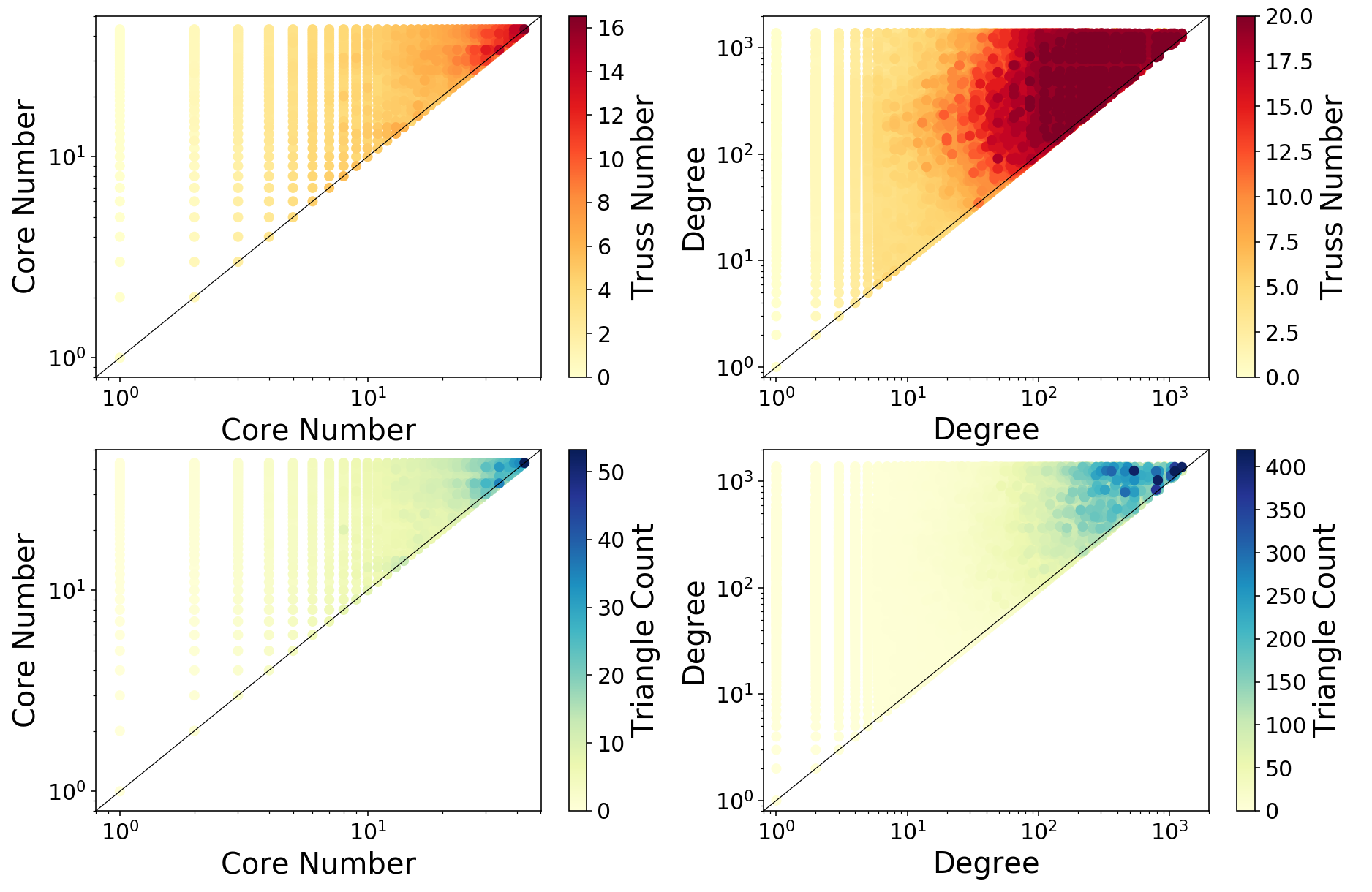}
\caption{\small \texttt{Email}}
\label{fig:email}
\end{subfigure}
\begin{subfigure}{0.33\textwidth}
\includegraphics[width=\linewidth]{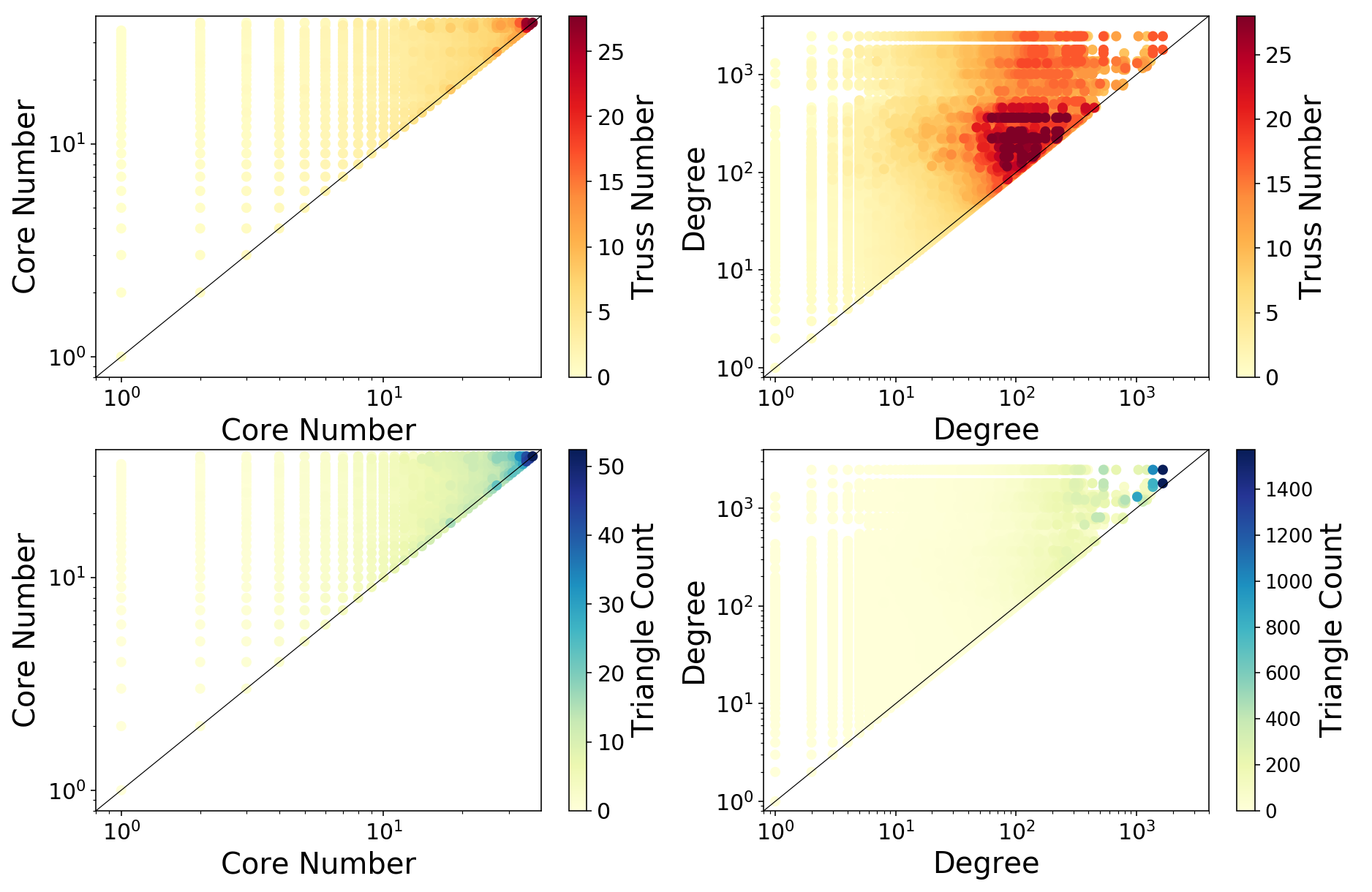}
\caption{\small \texttt{HepTh}}
\label{fig:hepth}
\end{subfigure}
\begin{subfigure}{0.33\textwidth}
\includegraphics[width=\linewidth]{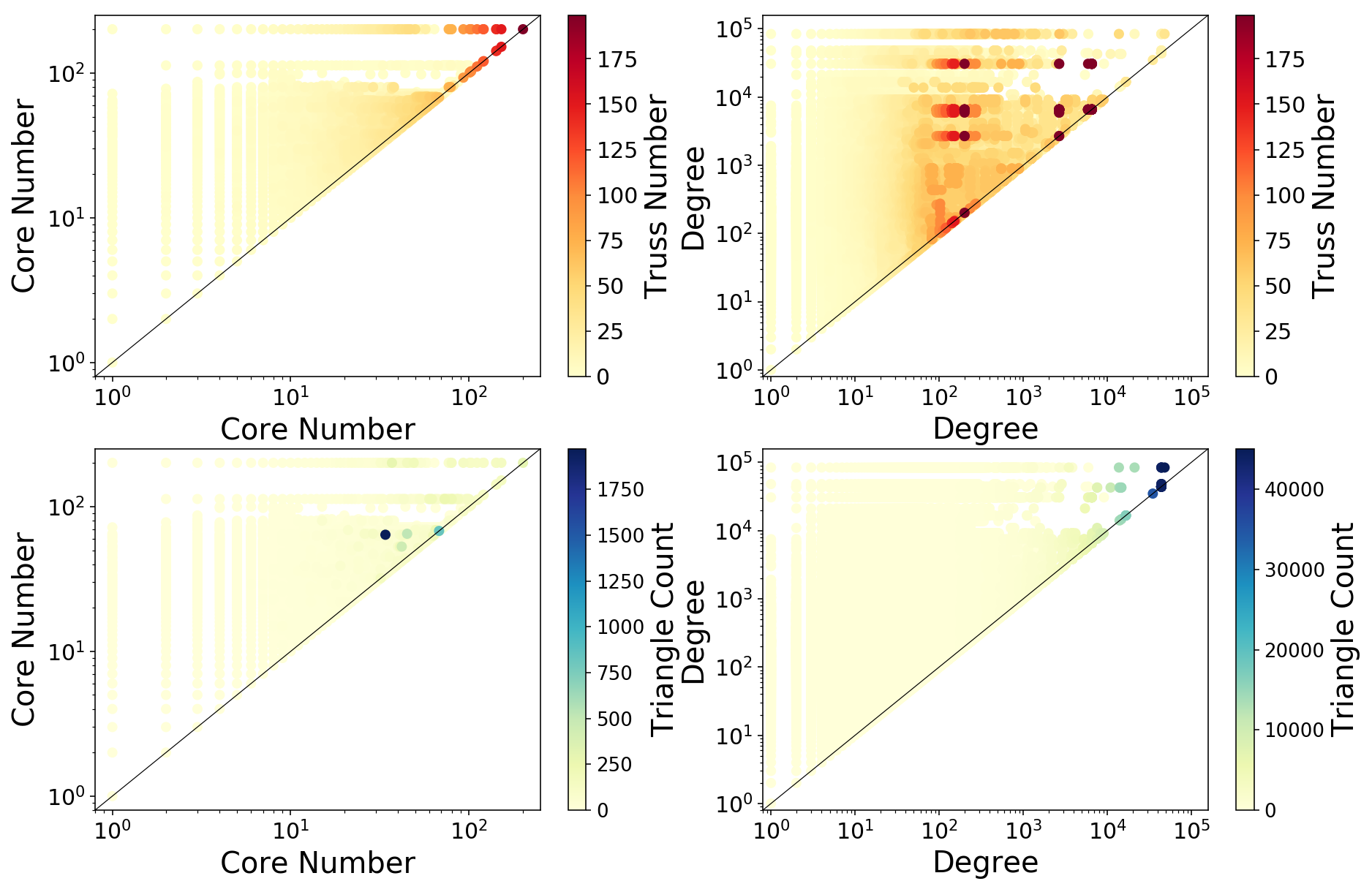}
\caption{\small \texttt{BerkStan}}
\label{fig:berkstan}
\end{subfigure}
\caption{\small \bf Edge interplay (EI) plots for some real-world graphs. For each pair of endpoints with particular core numbers/degrees, the average of the truss numbers/triangle counts of the edges are shown. Social network in Figure \ref{fig:email} exhibits a common and expected behavior, where the truss number/triangle count of edge is larger for two endpoints with large core numbers/degrees. 
Citation network, shown in Figures  \ref{fig:hepth}, presents a different behavior where the edges with large truss numbers are connecting two endpoints with non-maximum degrees.
Web network, shown in Figure \ref{fig:berkstan}, presents another behavior where the two endpoints of the edges with large truss number have very different degrees.
}
\label{EI_real}
\vspace{-3ex}
\end{figure*}

\section{Edge Based Analysis}\label{sec:edge}
In \cref{sec:vertex}, we have explored the neighborhood around a vertex with a specific core number. It is not clear though, how the truss number of an edge is related to the core numbers of its two endpoints. In this section we address the interplay from the perspective of an edge. We introduce the {\bf edge interplay (EI) plot} to capture the truss number of an edge between two endpoints with specific core numbers. We observe consistent behavior in the EI plots of networks from the same domain, and significant difference between the networks from different domains.
Similar to the previous section, we apply our analysis on both real-world graphs and random graphs, and we also consider the degree and triangle count as alternative measures.

\subsection{Real-world Networks}
\begin{figure*}[!t]
\centering
\begin{subfigure}{0.46\textwidth}
\includegraphics[width=\linewidth]{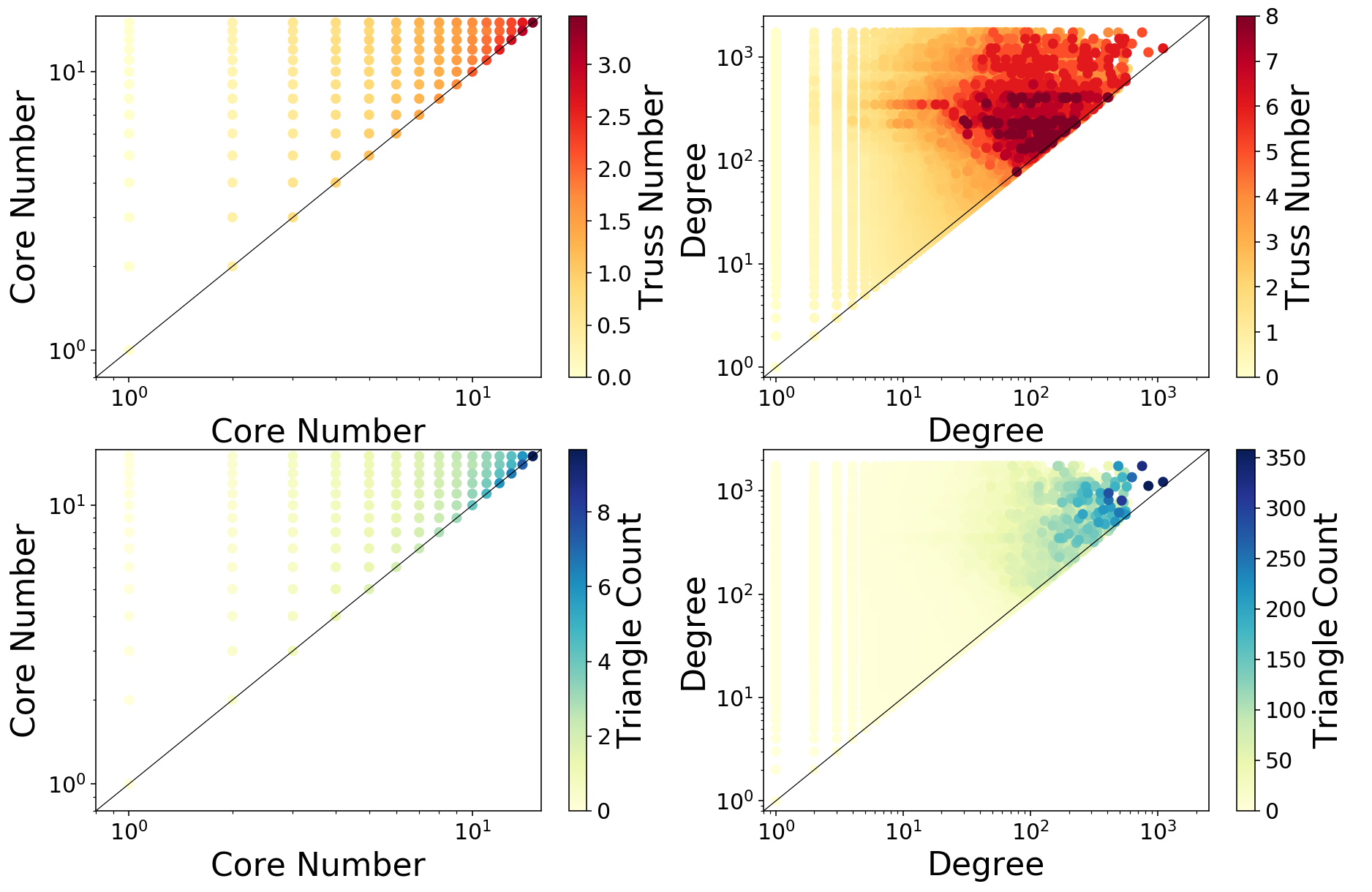}
\caption{\small \texttt{CiteSeer}}
\end{subfigure}
\begin{subfigure}{0.46\textwidth}
\includegraphics[width=\linewidth]{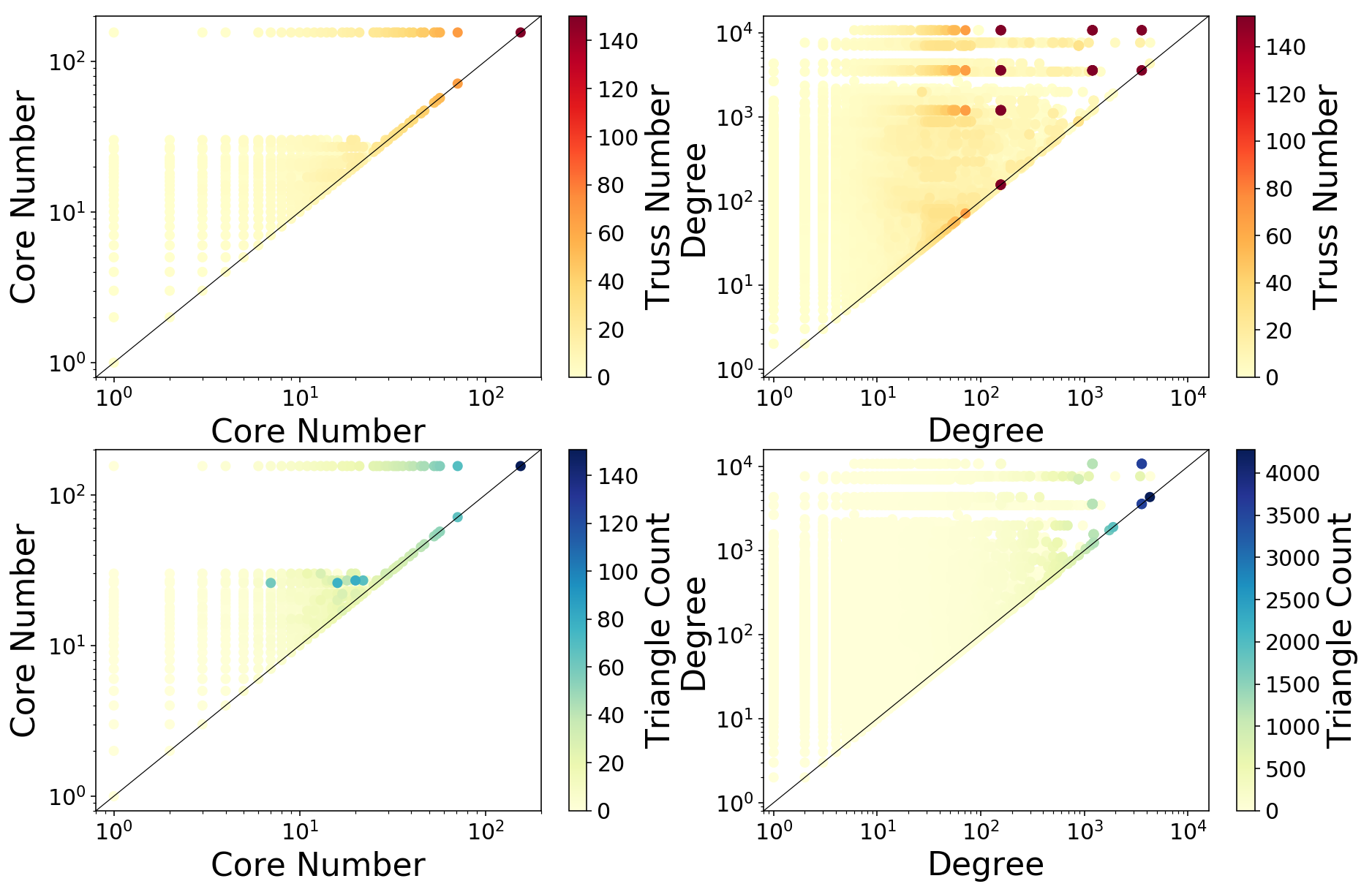}
\caption{\small \texttt{NotreDame}}
\end{subfigure}

\begin{subfigure}{0.46\textwidth}
\includegraphics[width=\linewidth]{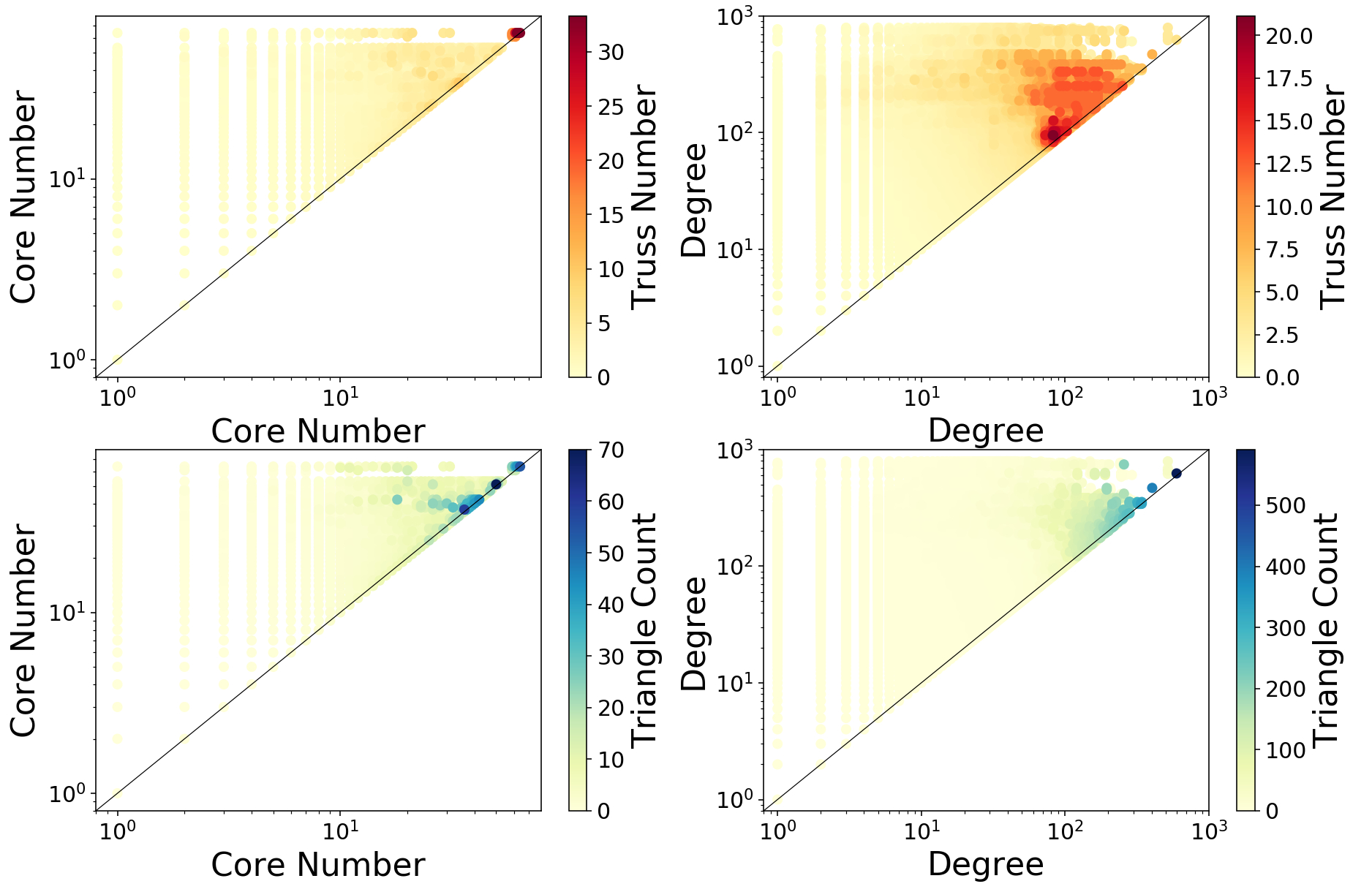}
\caption{\small \texttt{Patent}}
\end{subfigure}
\begin{subfigure}{0.46\textwidth}
\includegraphics[width=\linewidth]{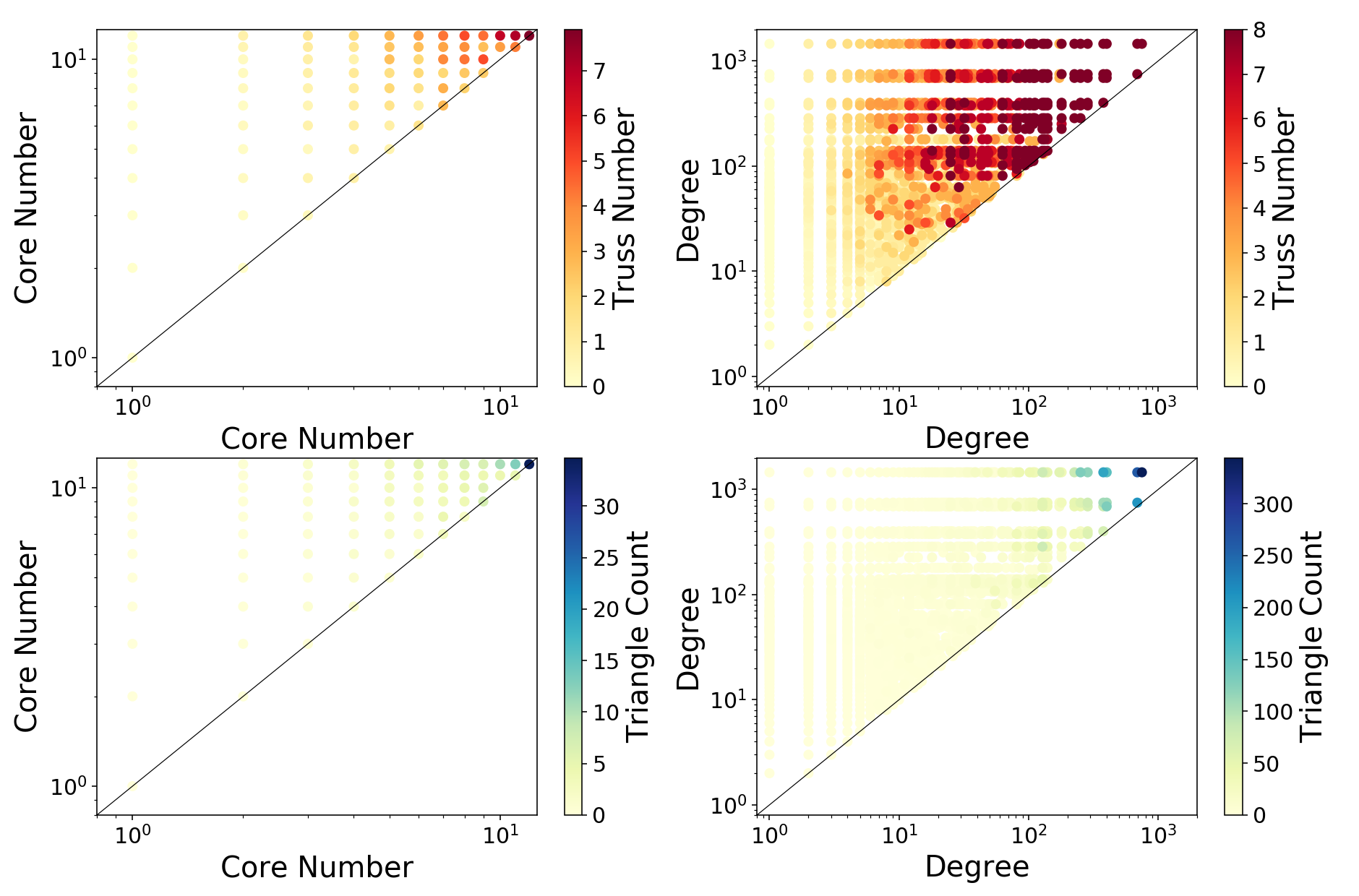}
\caption{\small \texttt{As-733}}
\end{subfigure}

\begin{subfigure}{0.46\textwidth}
\includegraphics[width=\linewidth]{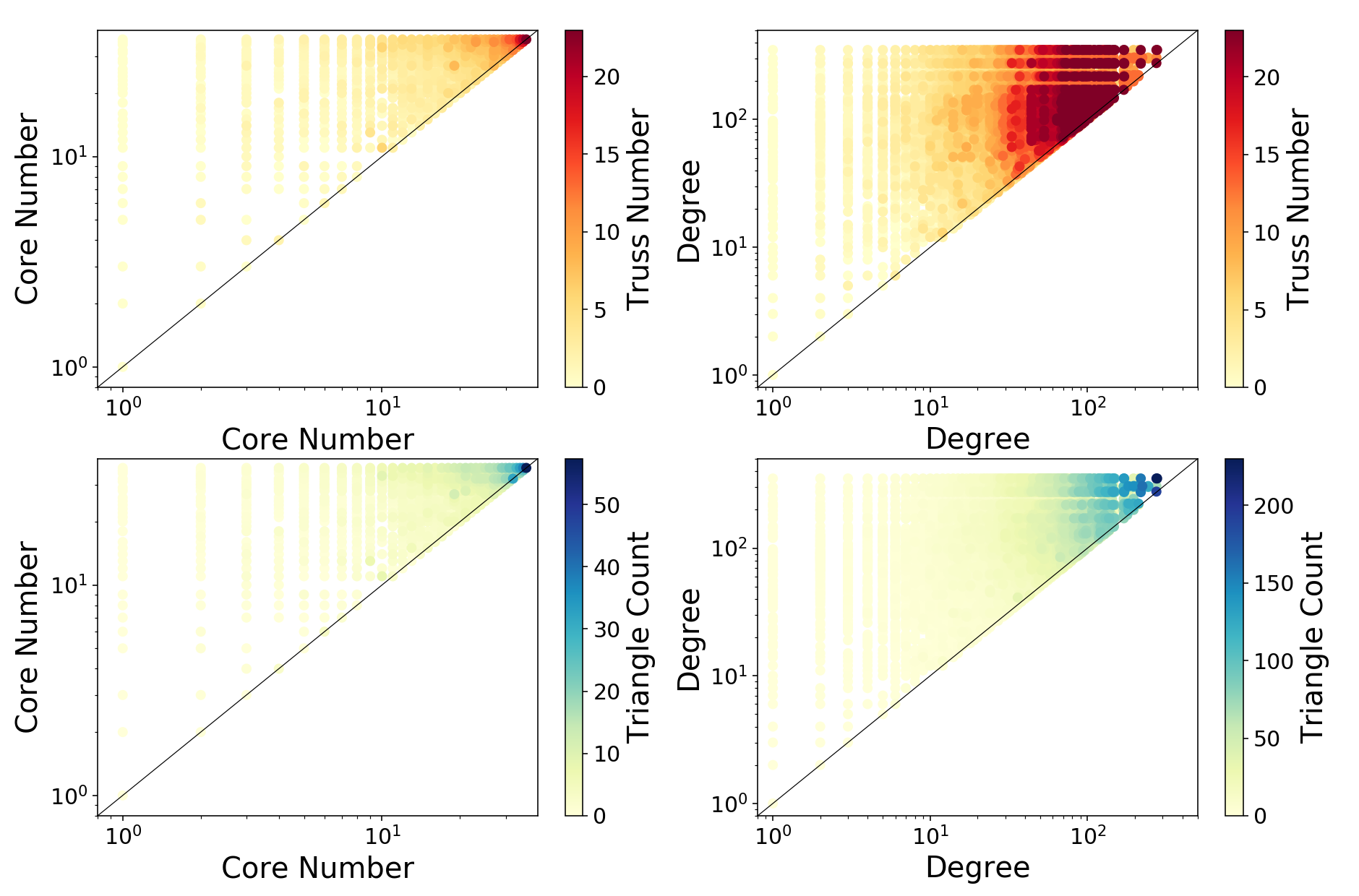}
\caption{\small \texttt{Blogs}}
\end{subfigure}
\begin{subfigure}{0.46\textwidth}
\includegraphics[width=\linewidth]{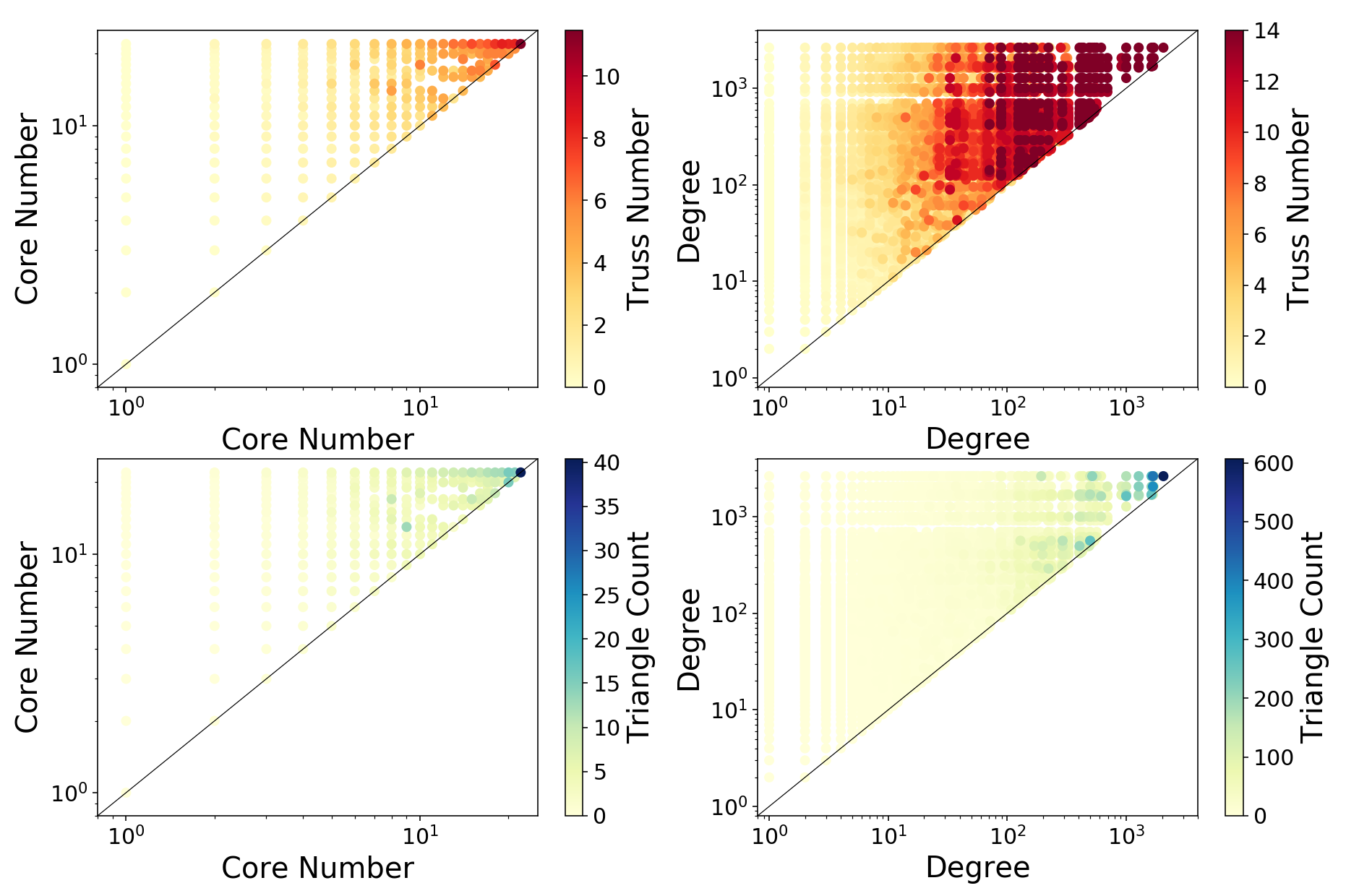}
\caption{\small \texttt{Caida}}
\end{subfigure}

\begin{subfigure}{0.46\textwidth}
\includegraphics[width=\linewidth]{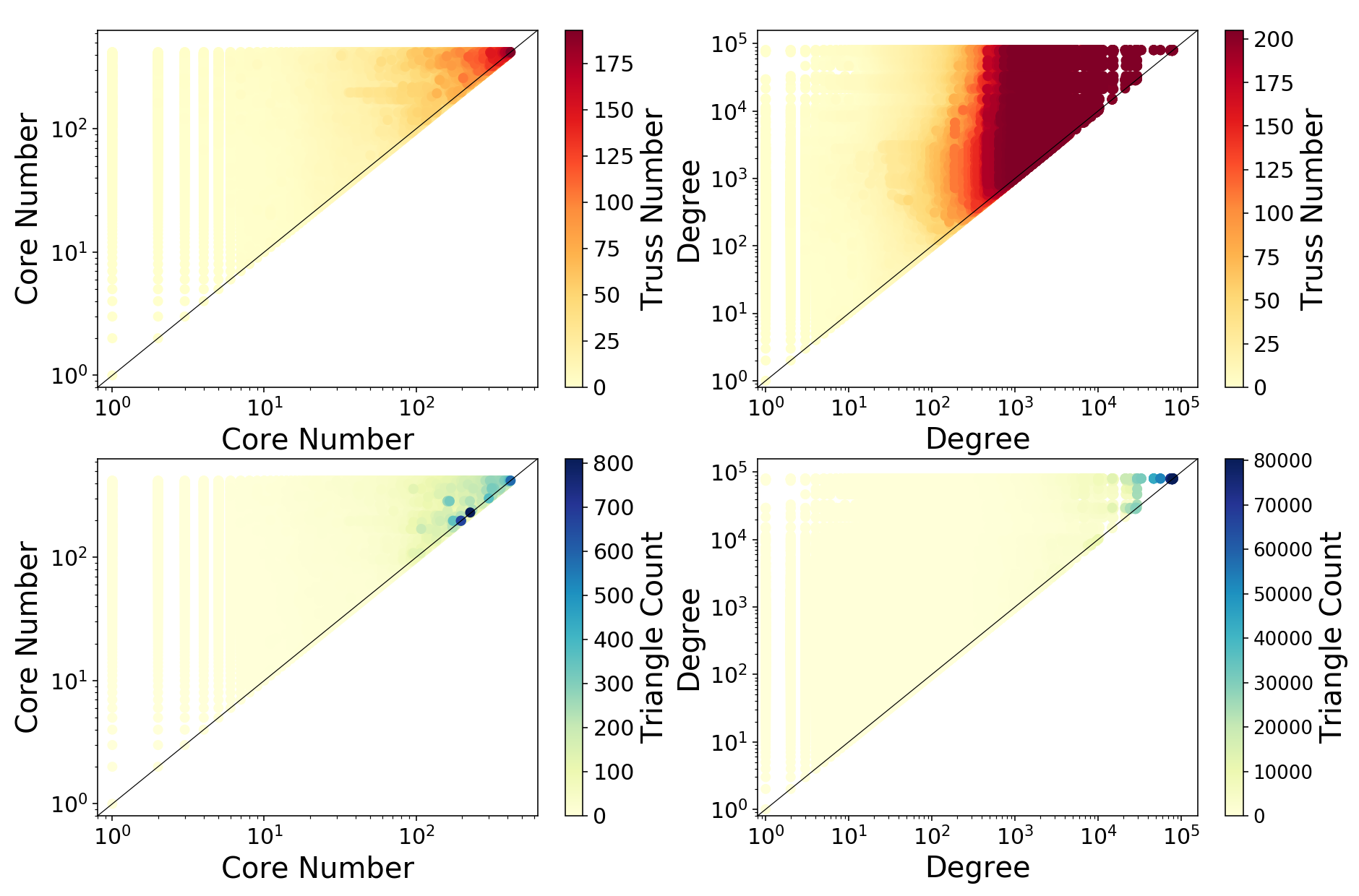}
\caption{\small \texttt{Catster}}
\end{subfigure}
\begin{subfigure}{0.46\textwidth}
\includegraphics[width=\linewidth]{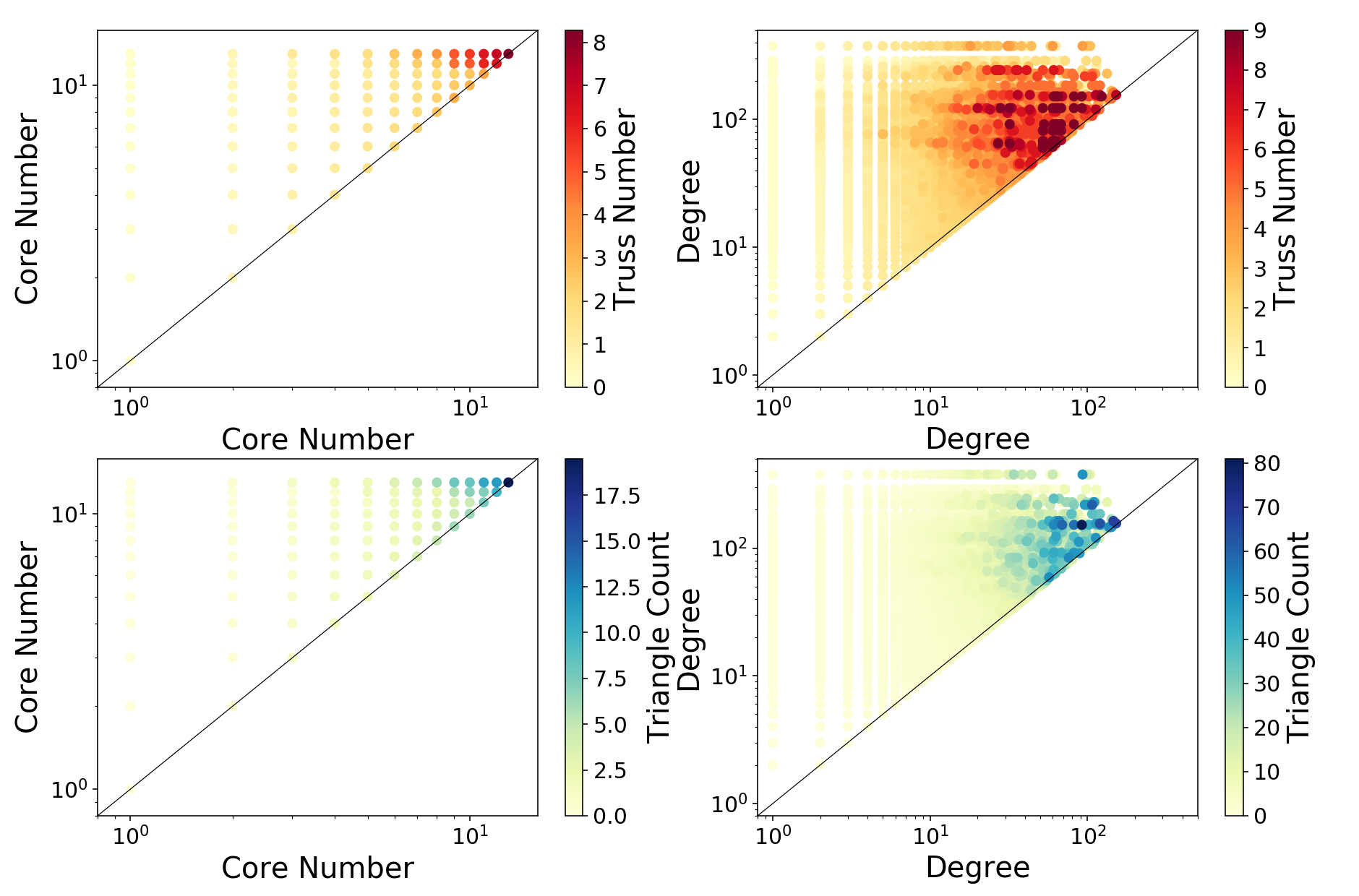}
\caption{\small \texttt{Cora}}
\end{subfigure}

\caption{\small \bf Edge interplay (EI) plots for other real-world graphs (Part I). For each pair of endpoints with particular core numbers/degrees, the average of the truss numbers/triangle counts of the edges are shown.}
\label{fig:EI_real_ext_1}
\end{figure*}

\begin{figure*}[!t]
\centering
\begin{subfigure}{0.46\textwidth}
\includegraphics[width=\linewidth]{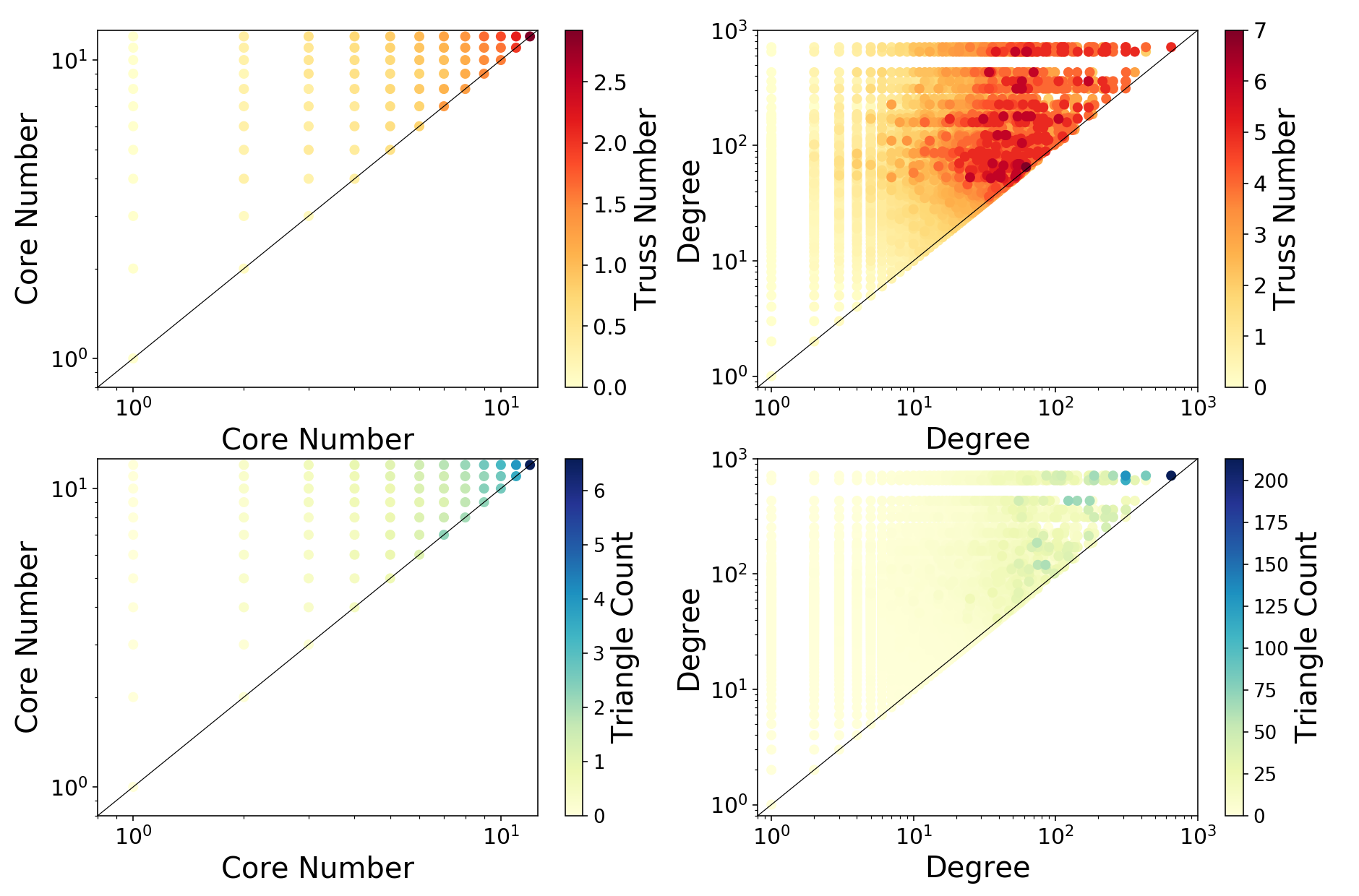}
\caption{\small \texttt{DBLP}}
\end{subfigure}
\begin{subfigure}{0.46\textwidth}
\includegraphics[width=\linewidth]{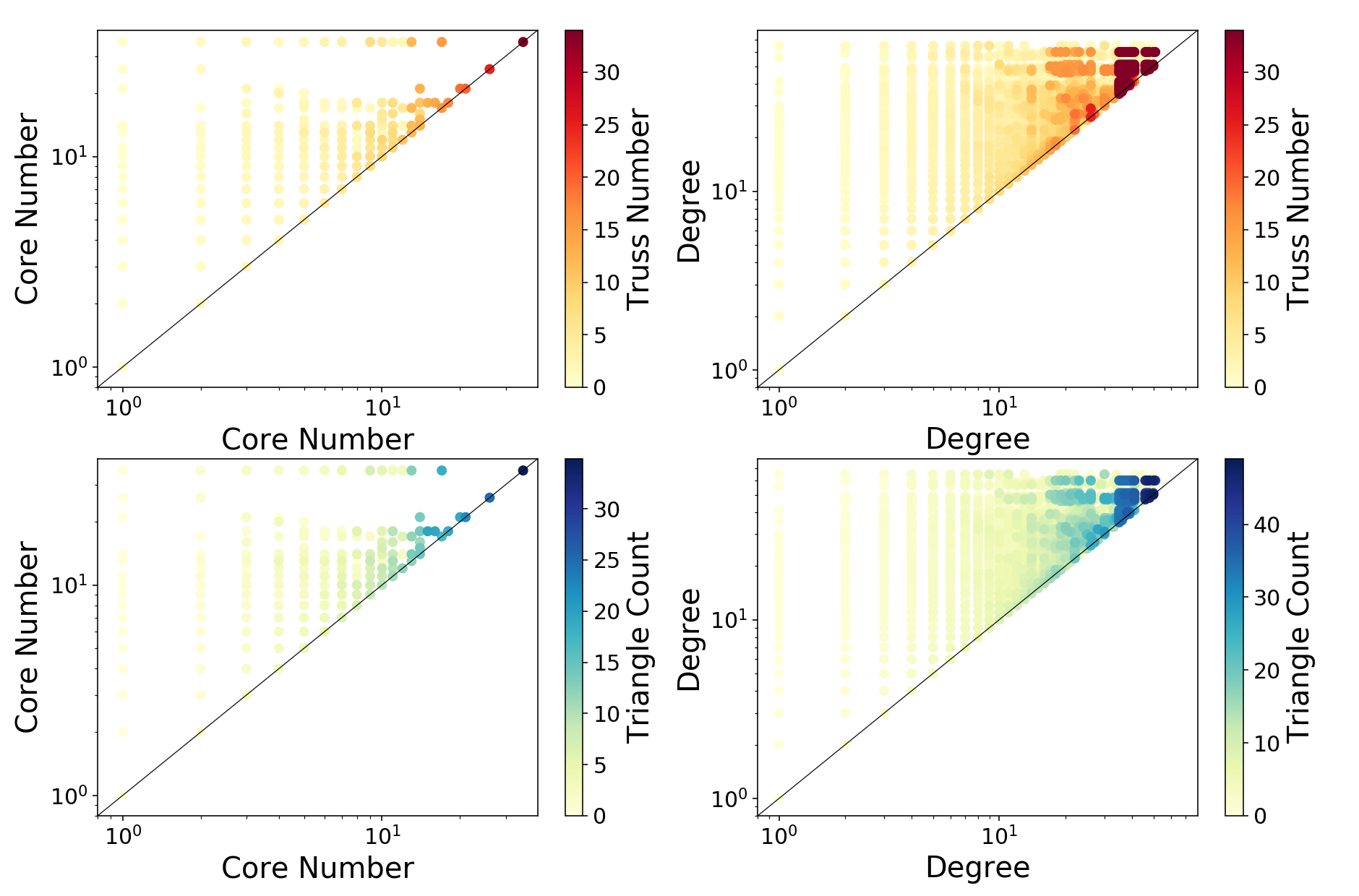}
\caption{\small \texttt{DBLP-dbs}}
\end{subfigure}

\begin{subfigure}{0.46\textwidth}
\includegraphics[width=\linewidth]{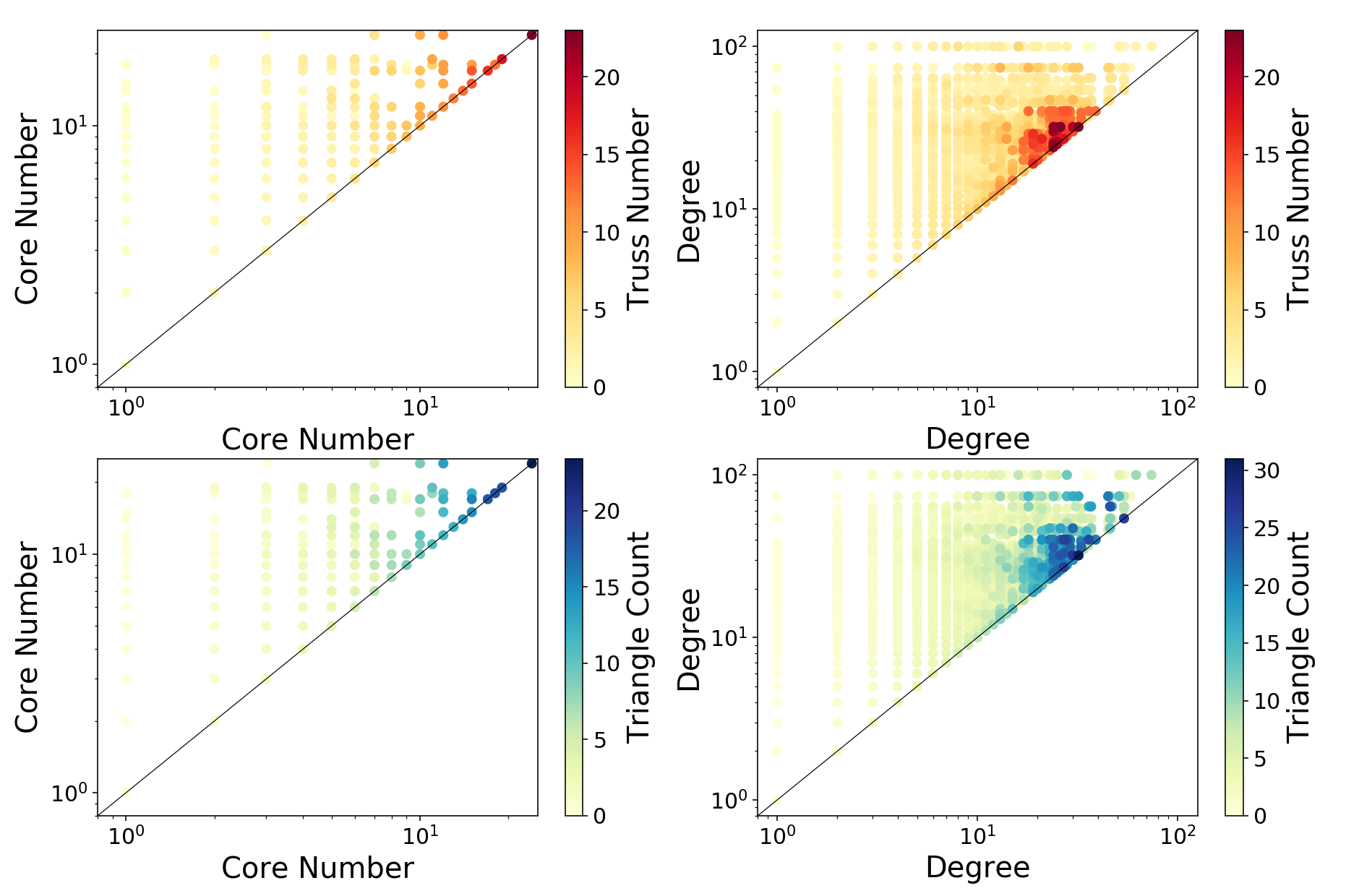}
\caption{\small \texttt{DBLP-dm}}
\end{subfigure}
\begin{subfigure}{0.46\textwidth}
\includegraphics[width=\linewidth]{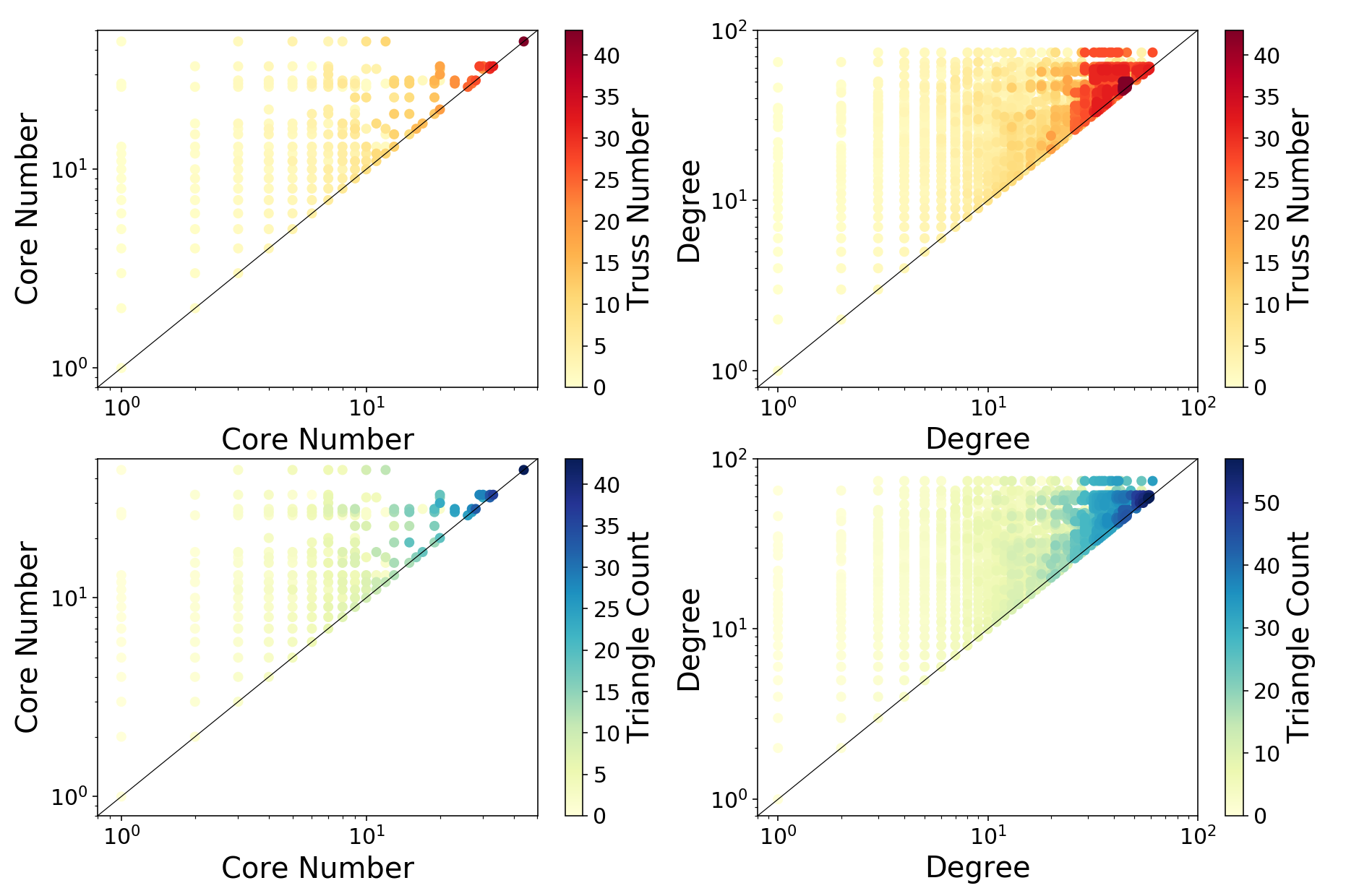}
\caption{\small \texttt{DBLP-pp}}
\end{subfigure}

\begin{subfigure}{0.46\textwidth}
\includegraphics[width=\linewidth]{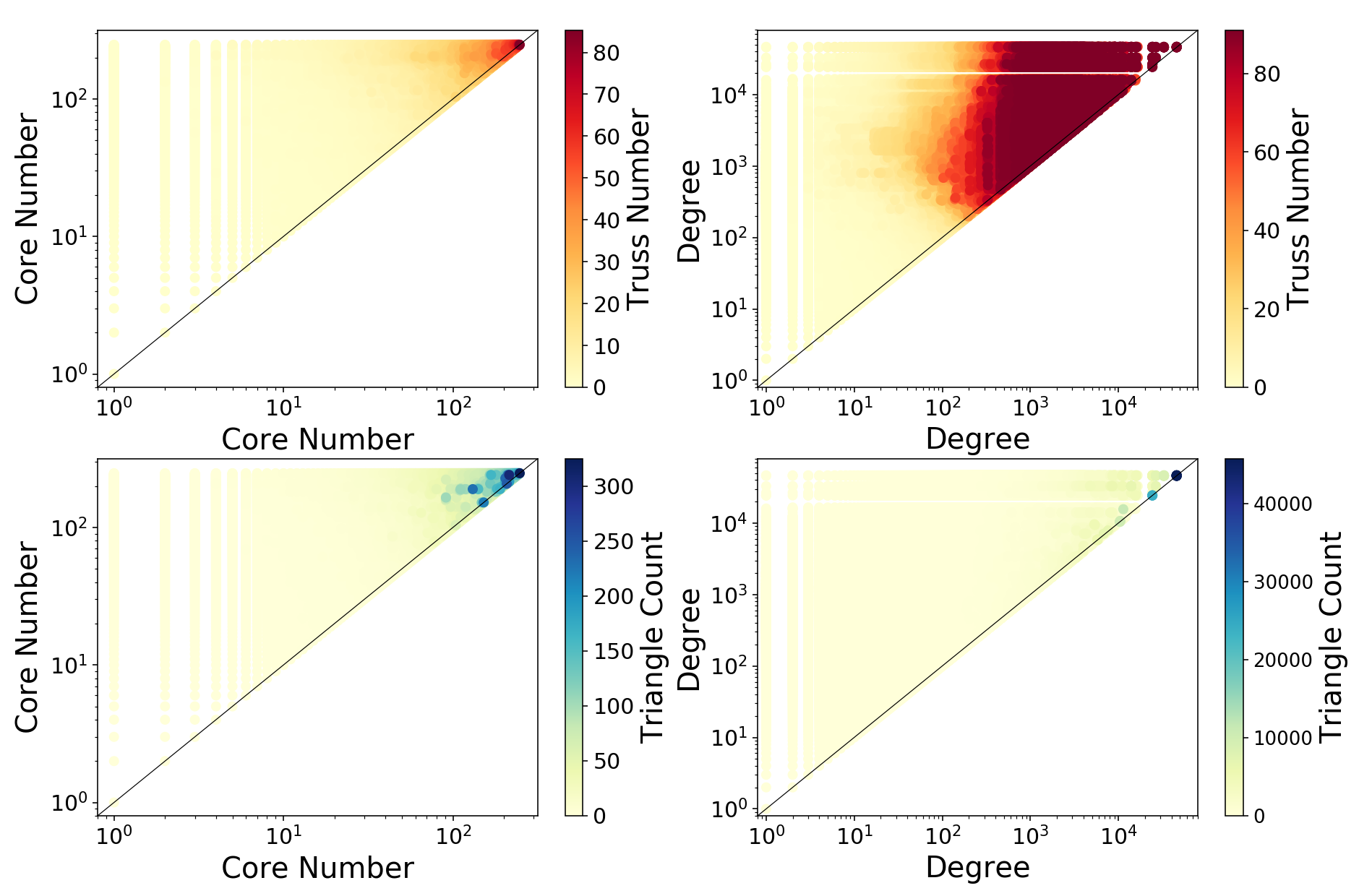}
\caption{\small \texttt{Dogster}}
\end{subfigure}
\begin{subfigure}{0.46\textwidth}
\includegraphics[width=\linewidth]{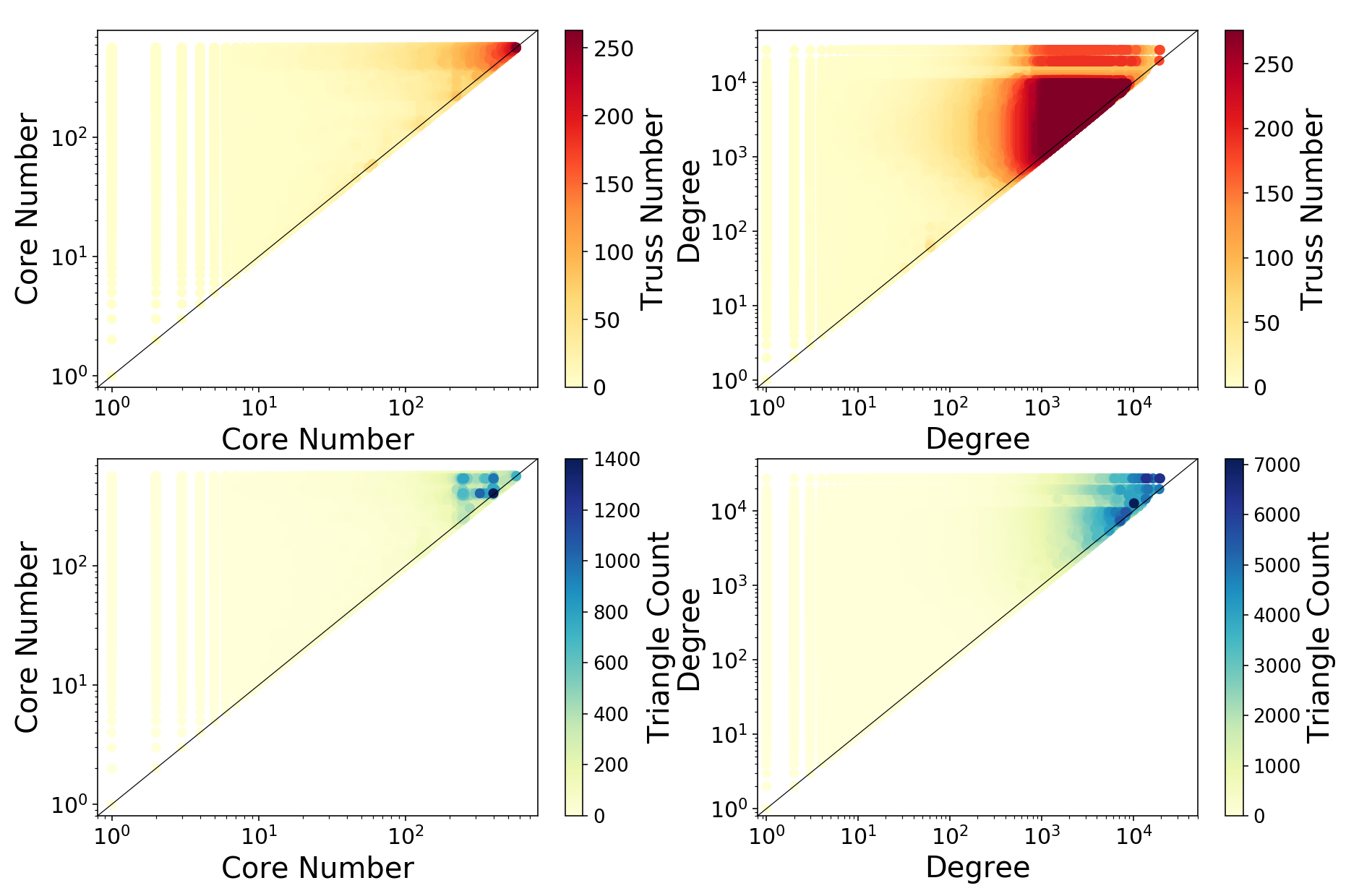}
\caption{\small \texttt{Flickr}}
\end{subfigure}

\begin{subfigure}{0.46\textwidth}
\includegraphics[width=\linewidth]{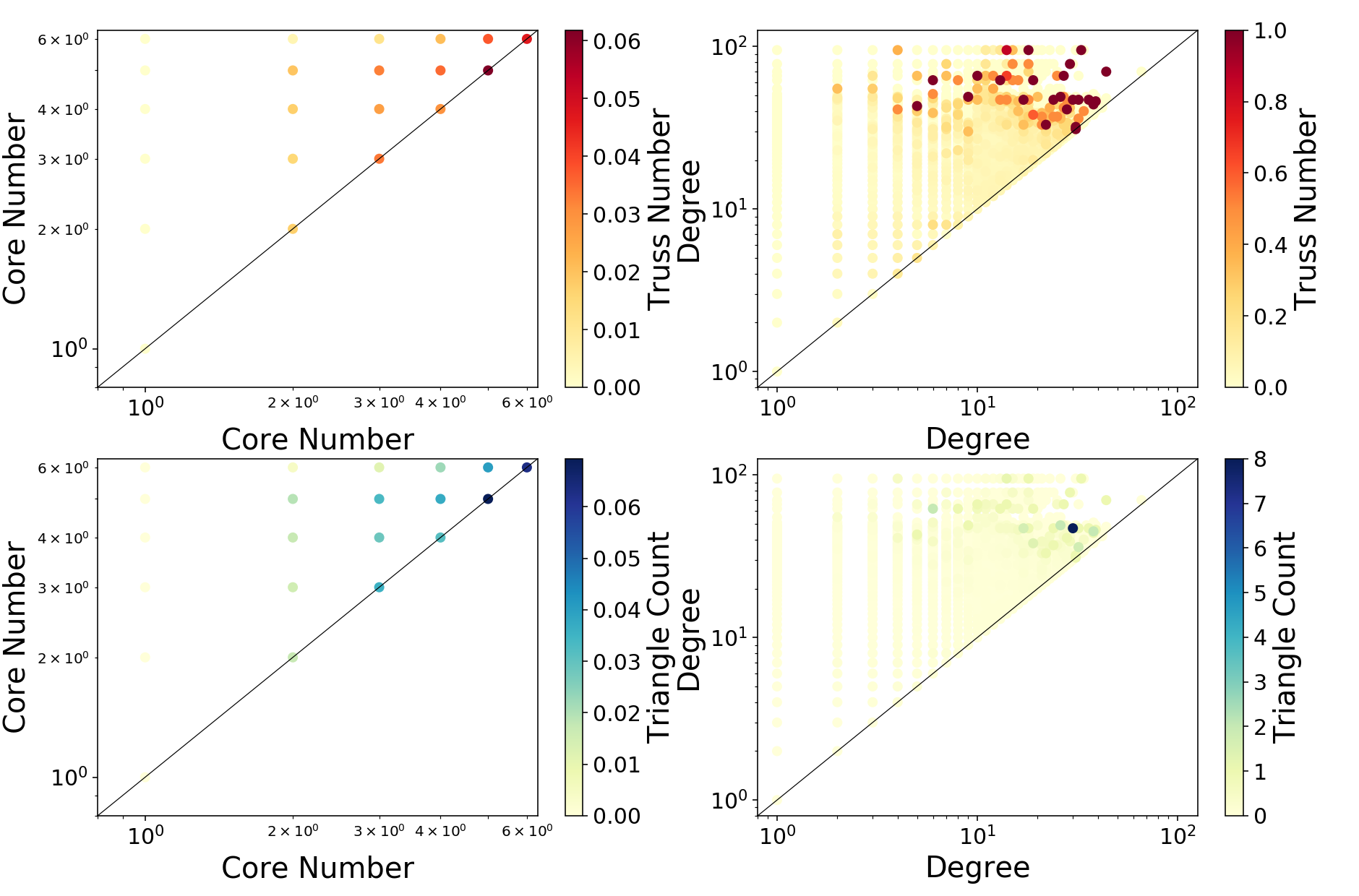}
\caption{\small \texttt{Gnutella}}
\end{subfigure}
\begin{subfigure}{0.46\textwidth}
\includegraphics[width=\linewidth]{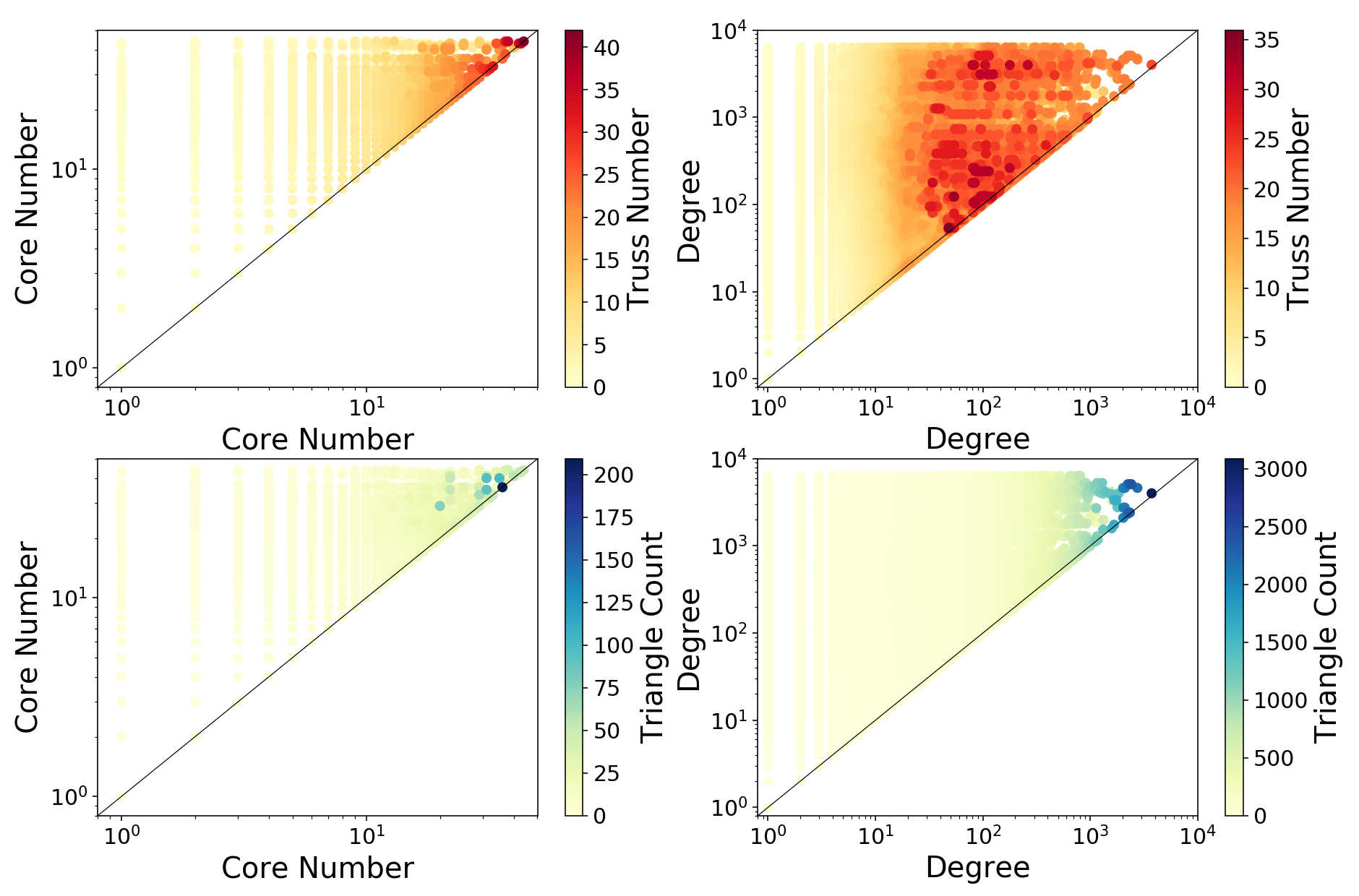}
\caption{\small \texttt{Google}}
\end{subfigure}

\caption{\small \bf Edge interplay (EI) plots for other real-world graphs (Part II). For each pair of endpoints with particular core numbers/degrees, the average of the truss numbers/triangle counts of the edges are shown.}
\label{fig:EI_real_ext_2}
\end{figure*}

\begin{figure*}[!t]
\centering
\begin{subfigure}{0.46\textwidth}
\includegraphics[width=\linewidth]{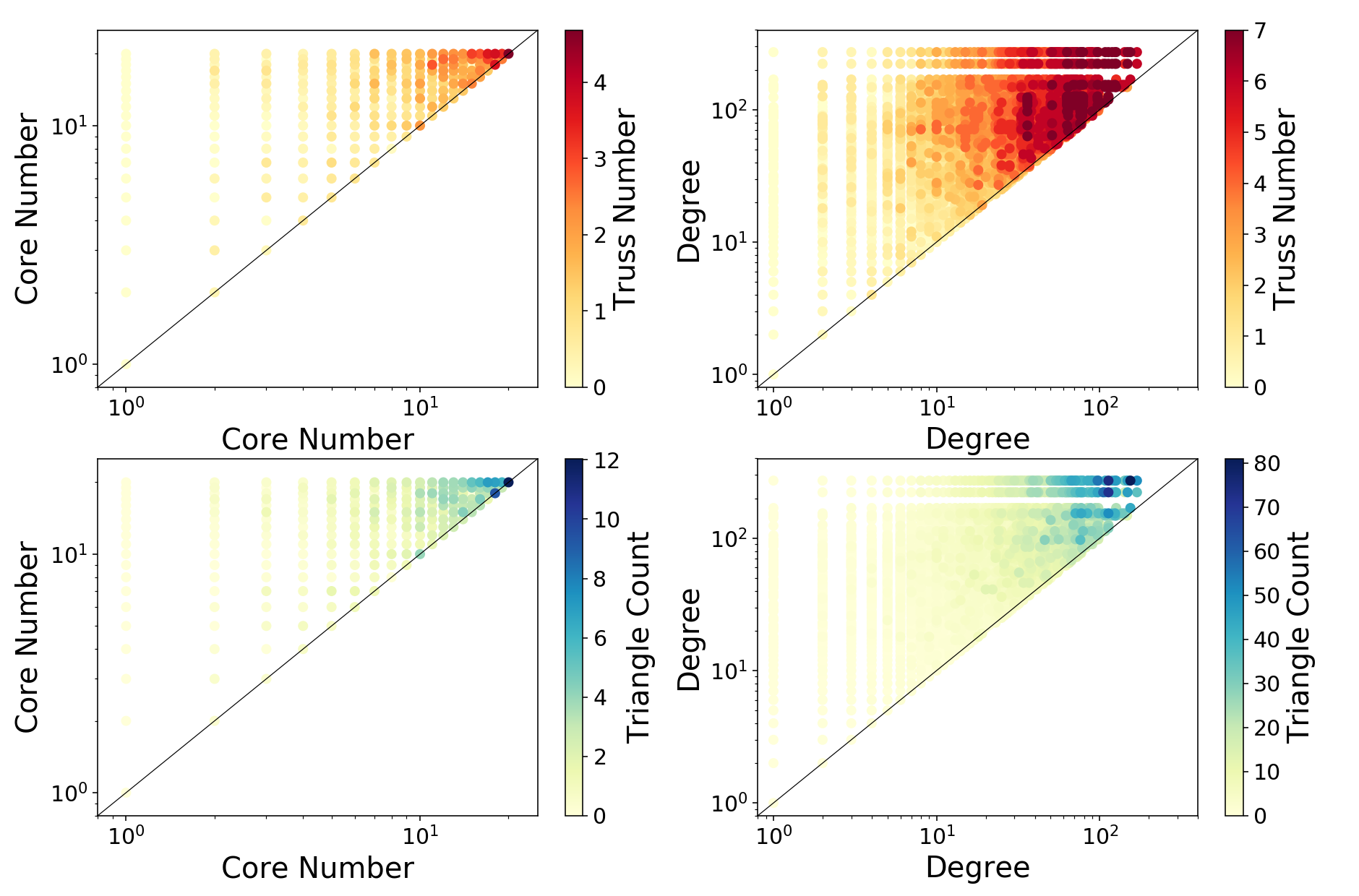}
\caption{\small \texttt{Hamster}}
\end{subfigure}
\begin{subfigure}{0.46\textwidth}
\includegraphics[width=\linewidth]{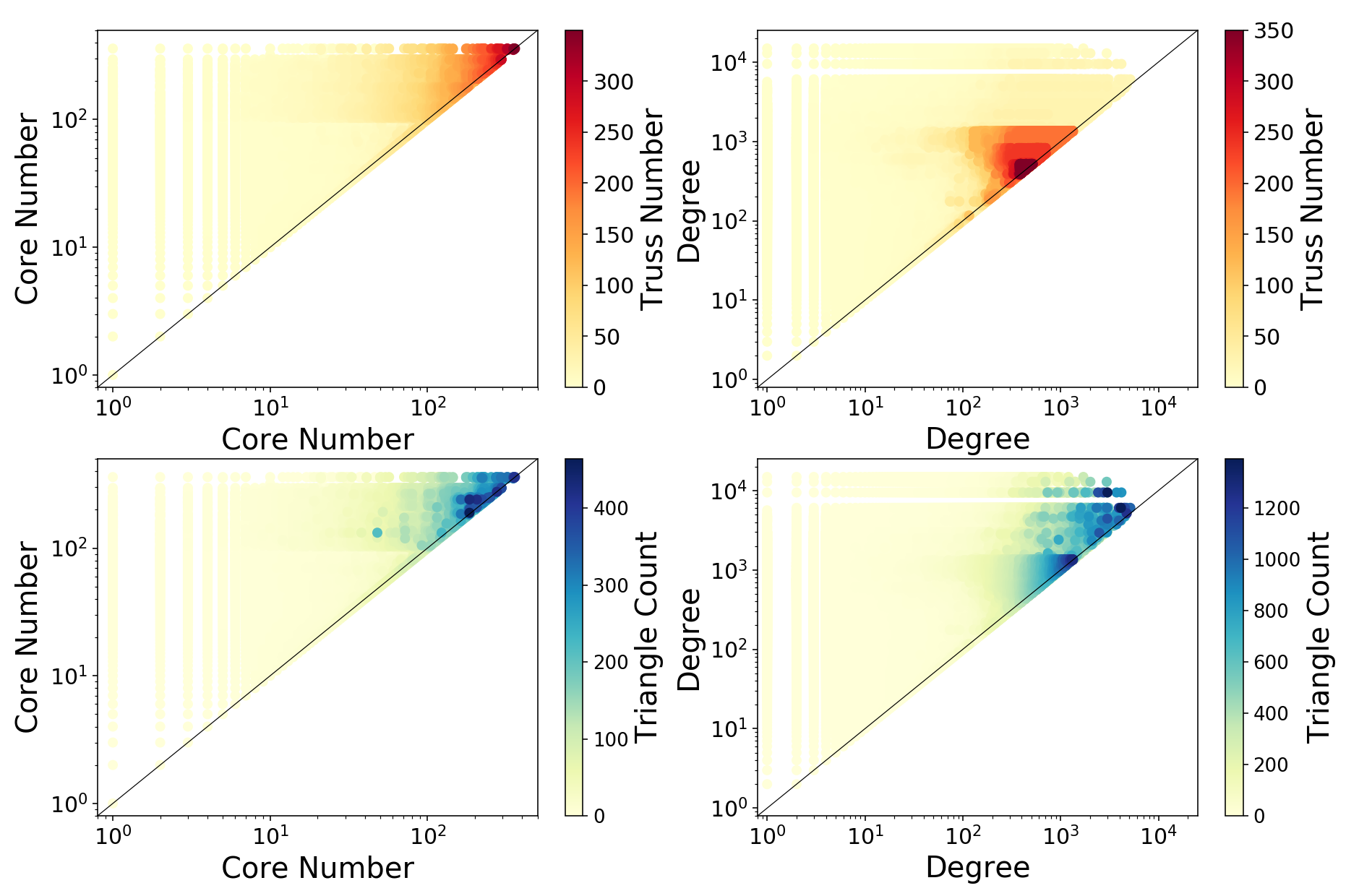}
\caption{\small \texttt{LiveJournal}}
\end{subfigure}

\begin{subfigure}{0.46\textwidth}
\includegraphics[width=\linewidth]{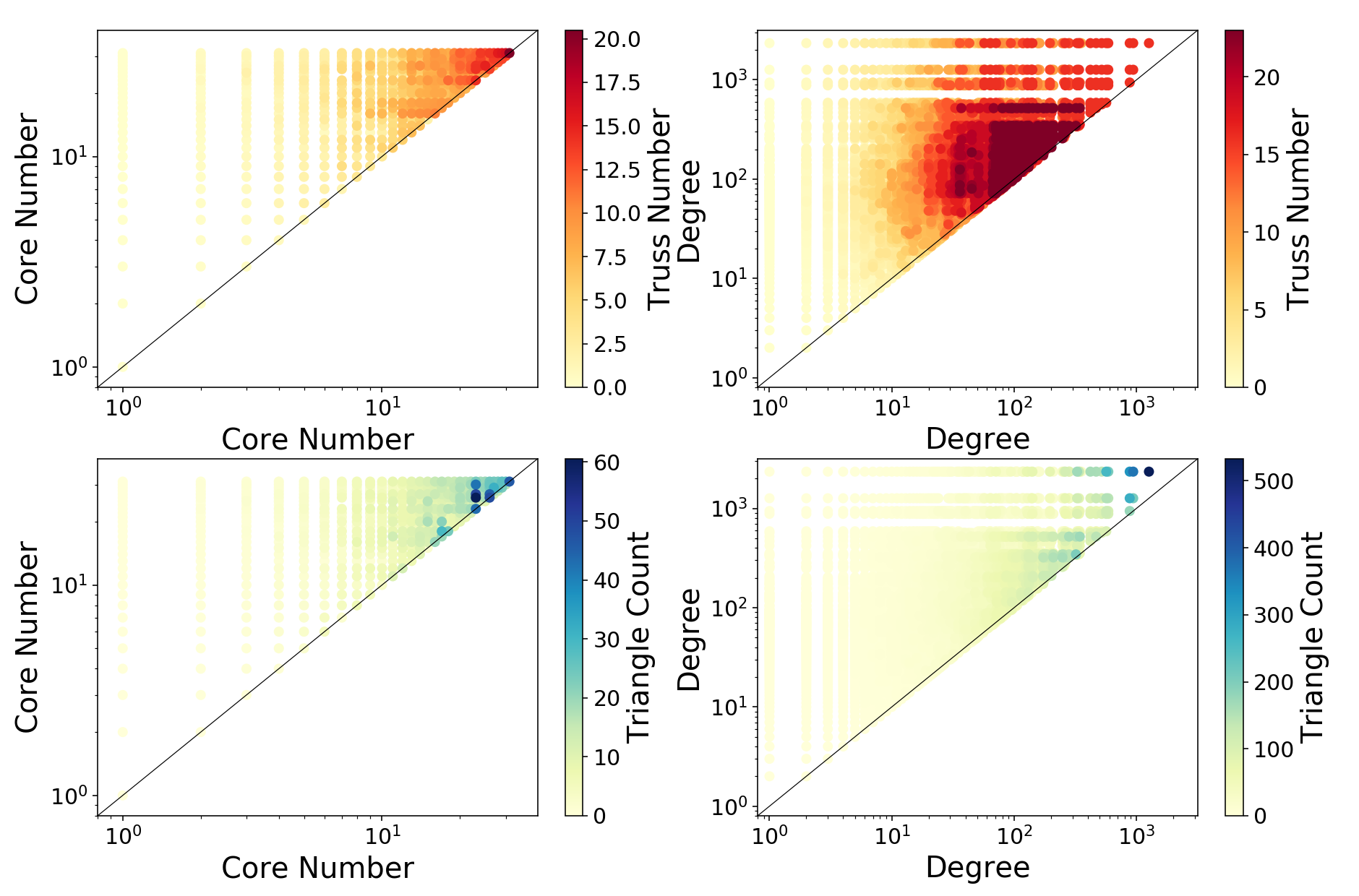}
\caption{\small \texttt{Oregon-2}}
\end{subfigure}
\begin{subfigure}{0.46\textwidth}
\includegraphics[width=\linewidth]{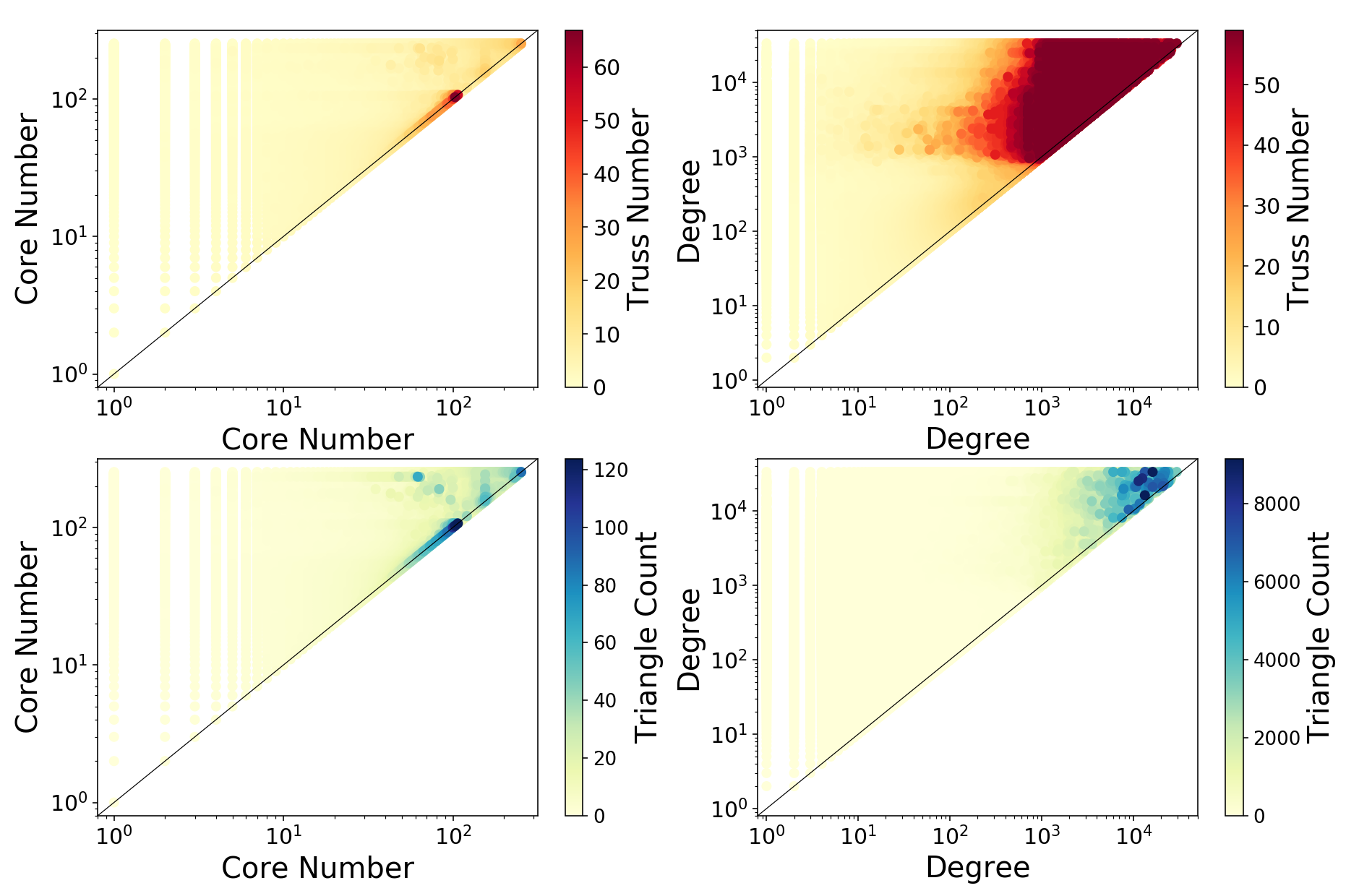}
\caption{\small \texttt{Orkut}}
\end{subfigure}

\begin{subfigure}{0.46\textwidth}
\includegraphics[width=\linewidth]{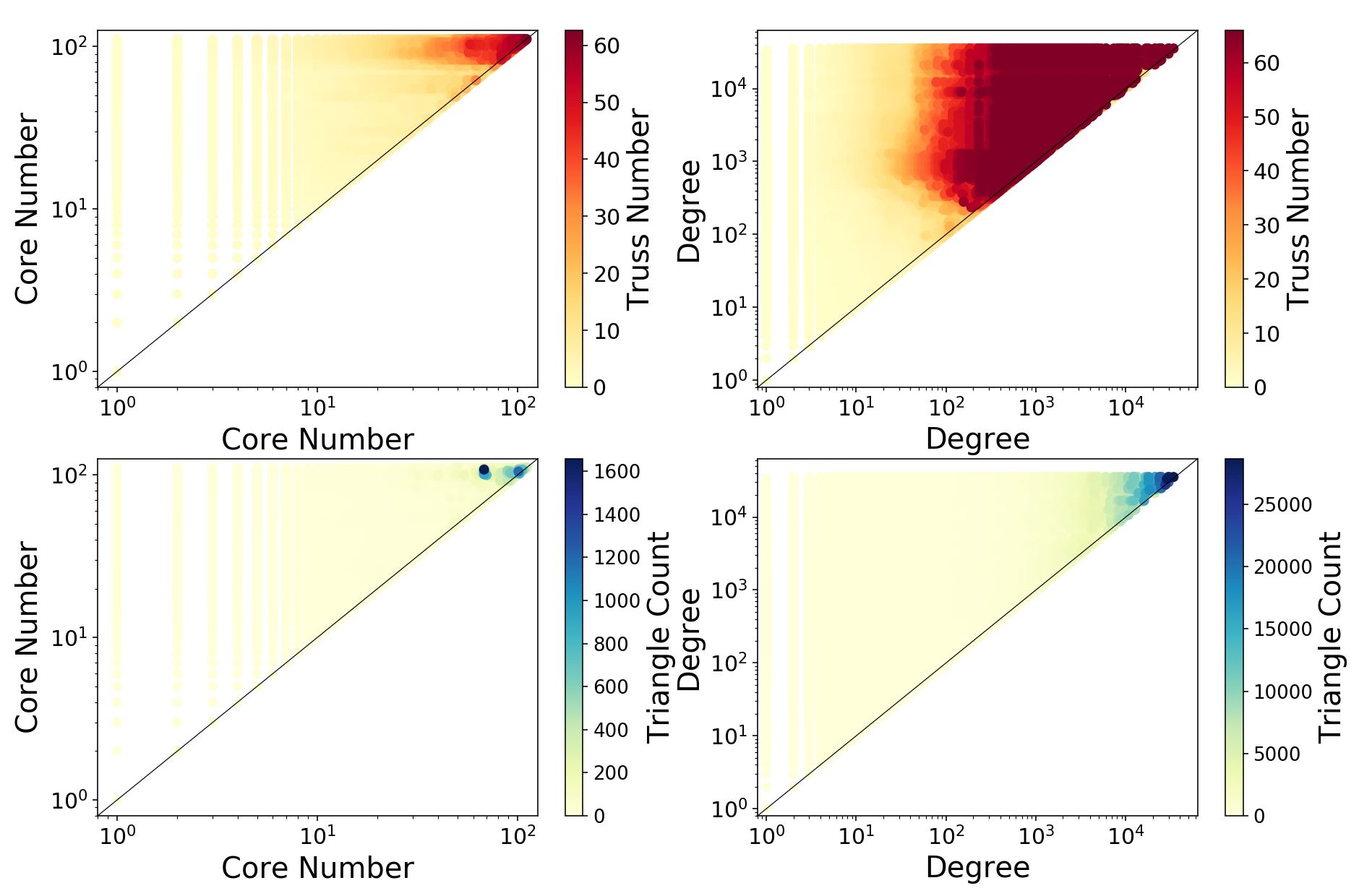}
\caption{\small \texttt{Skitter}}
\end{subfigure}
\begin{subfigure}{0.46\textwidth}
\includegraphics[width=\linewidth]{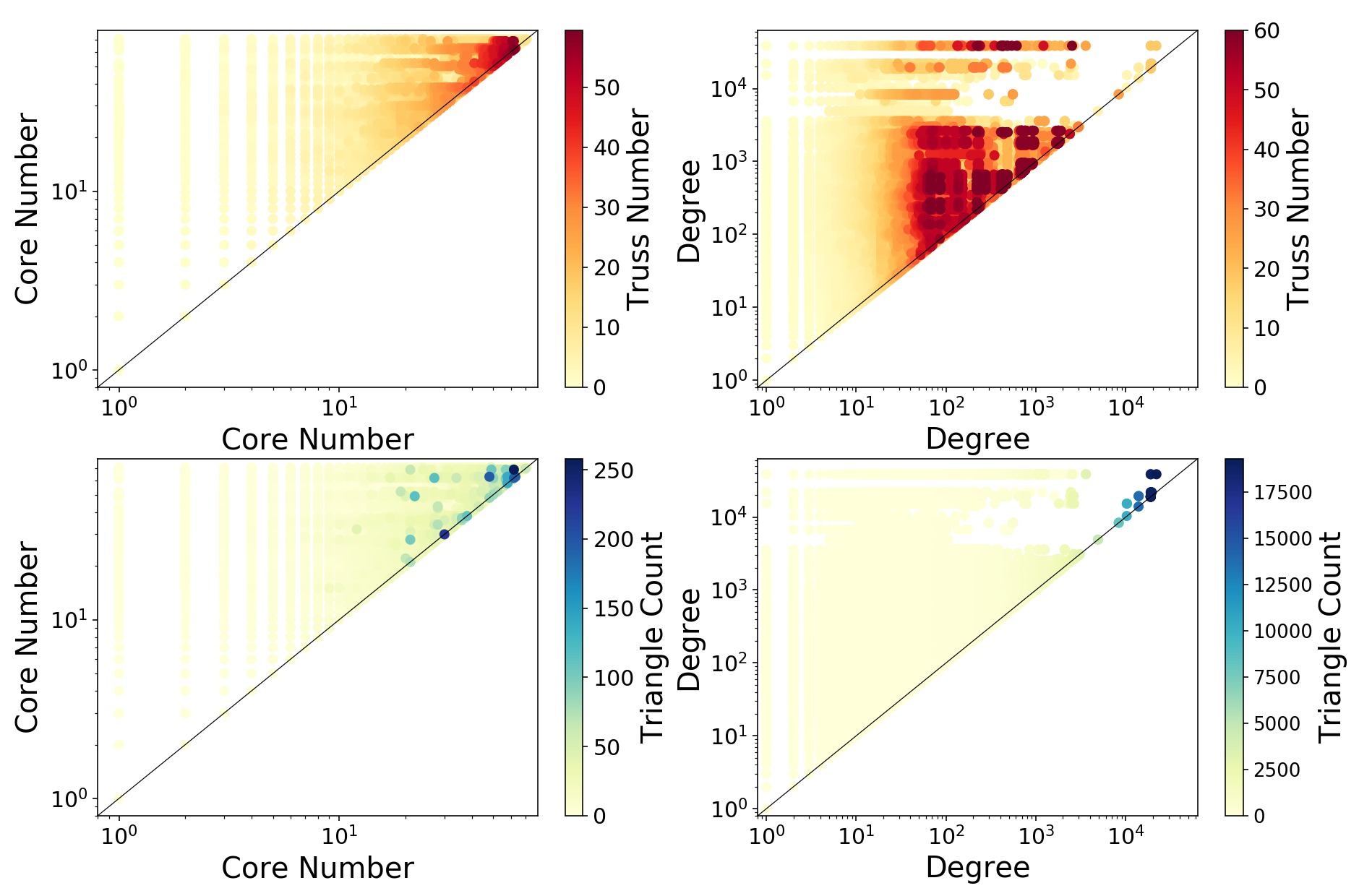}
\caption{\small \texttt{Stanford}}
\end{subfigure}

\begin{subfigure}{0.46\textwidth}
\includegraphics[width=\linewidth]{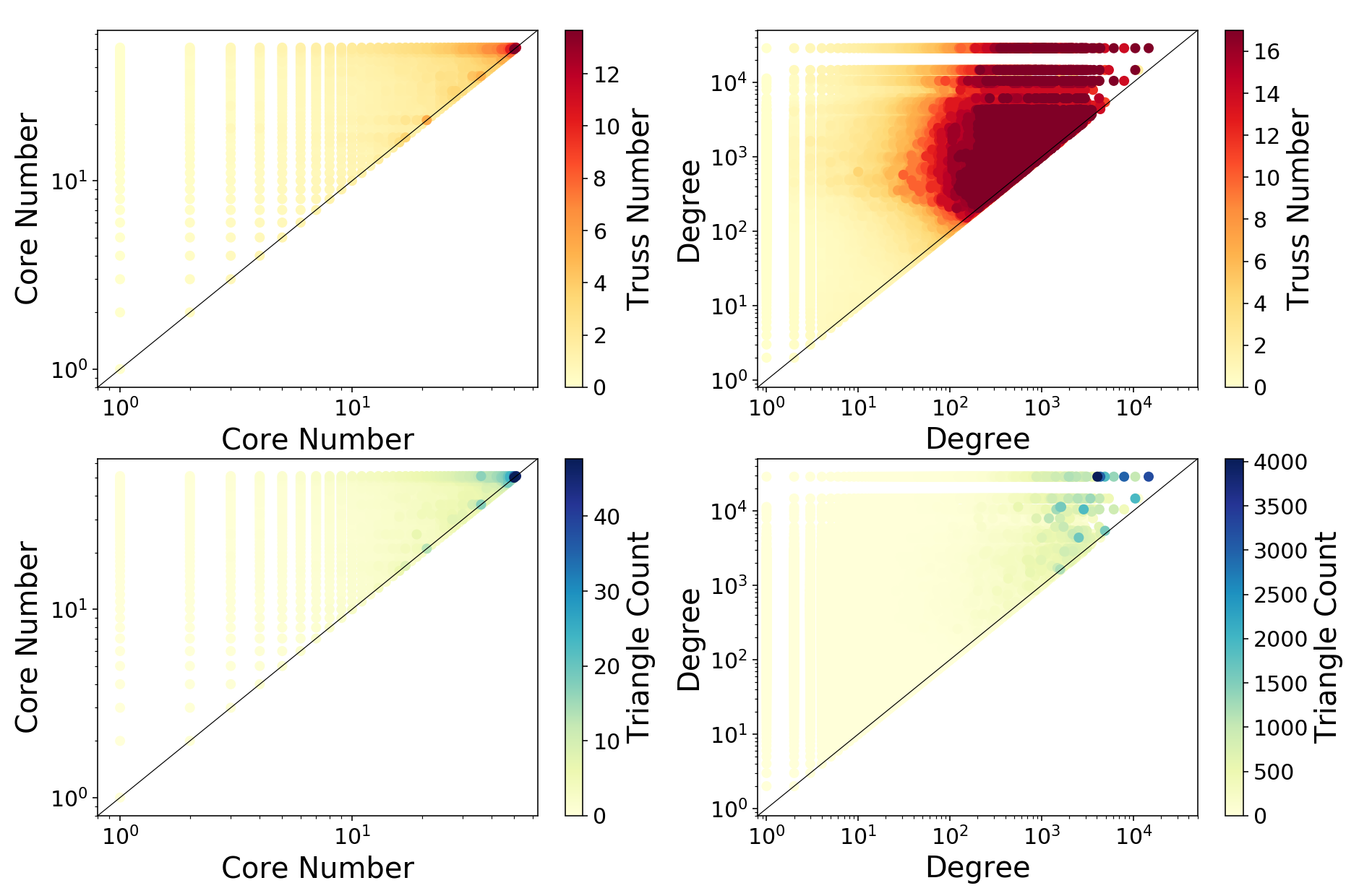}
\caption{\small \texttt{YouTube}}
\end{subfigure}

\caption{\small \bf Edge interplay (EI) plots for other real-world graphs (Part III). For each pair of endpoints with particular core numbers/degrees, the average of the truss numbers/triangle counts of the edges are shown.}.
\label{fig:EI_real_ext_3}
\end{figure*}

Figure \ref{EI_real} presents the EI plots for some real-world networks in our dataset (all are available in 
Figures \ref{fig:EI_real_ext_1}, \ref{fig:EI_real_ext_2}, and \ref{fig:EI_real_ext_3}). 
Here we examine the truss number of edges between two vertices with particular core numbers. 
If there are multiple edges having the same pair of core numbers for their endpoints, we show the mean value of the truss numbers. 
Formally, for each pair of core numbers $c_1 \leqslant c_2$, we find the set $S=\{(u, v)\in E: K(u)=c_1, K(v)=c_2\}$ and compute $\frac{1}{|S|}\sum_{(u,v) \in S}{T(u, v)}$.
In addition to the core-truss interplay, we consider the vertex degrees and edges' triangle counts for a similar analysis. For all EI plot variations (degree-truss, degree-triangle count, and core-triangle), we observe consistent behaviors in networks from the same domains. Those alternative EI plots allow us to discover some interesting behaviors.

One general behavior we observe is that the truss number of an edge is strongly correlated to the core numbers of the two endpoints. This also holds true for all the EI plot variations regarding vertex degrees and edge triangle counts.
Figure \ref{fig:email} shows this pattern in the EI plot of \texttt{Email} network, and we observe this pattern in all social networks, autonomous systems, and collaboration networks.
Note that the dense red regions are small, suggesting a high similarity between the endpoints of edges with large truss numbers.

Citation networks present a different behavior in the EI plots. 
Figure \ref{fig:hepth} presents the EI plot of \texttt{HepTh} network. In the degree-truss interplay, edges with large truss number are connecting two endpoints with non-maximum degrees.
Another interesting behavior appears consistently in web networks. 
Figure \ref{fig:berkstan} presents the EI plot of \texttt{BerkStan} network. Edges with large truss number are connecting two vertices with very different degrees. This implies that there are cohesive structures connecting core and periphery blocks. This structure seems to be artificially constructed for special purposes, e.g., spam link farms.

\subsection{Random Graphs}
\begin{figure*}[!t]
\centering
\vspace{-1ex}
\begin{subfigure}{0.63\textwidth}
\includegraphics[width=\linewidth]{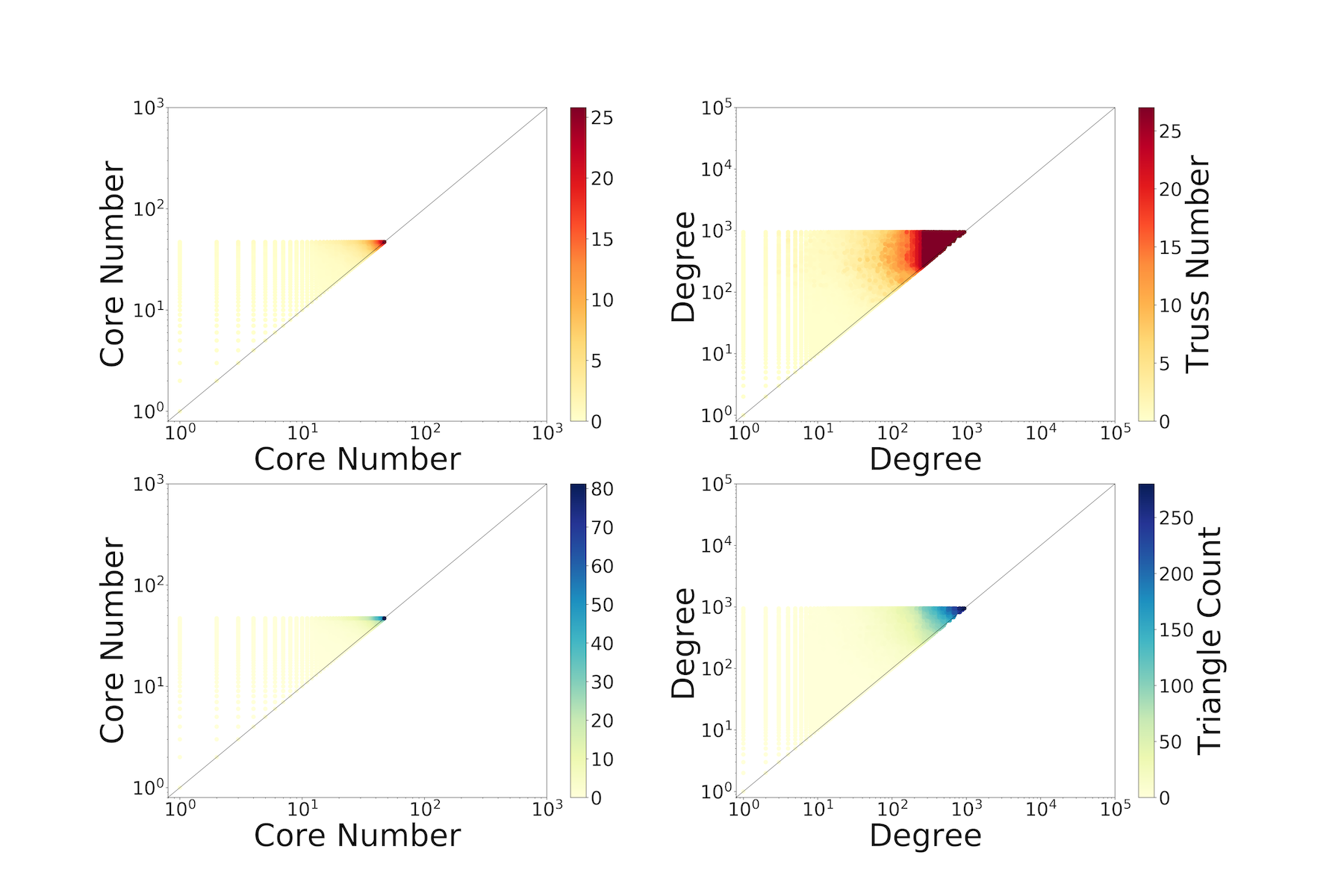}
\caption{\small \texttt{Email} by BTER}
\label{fig:emailBTER}
\end{subfigure}

\begin{subfigure}{0.63\textwidth}
\includegraphics[width=\linewidth]{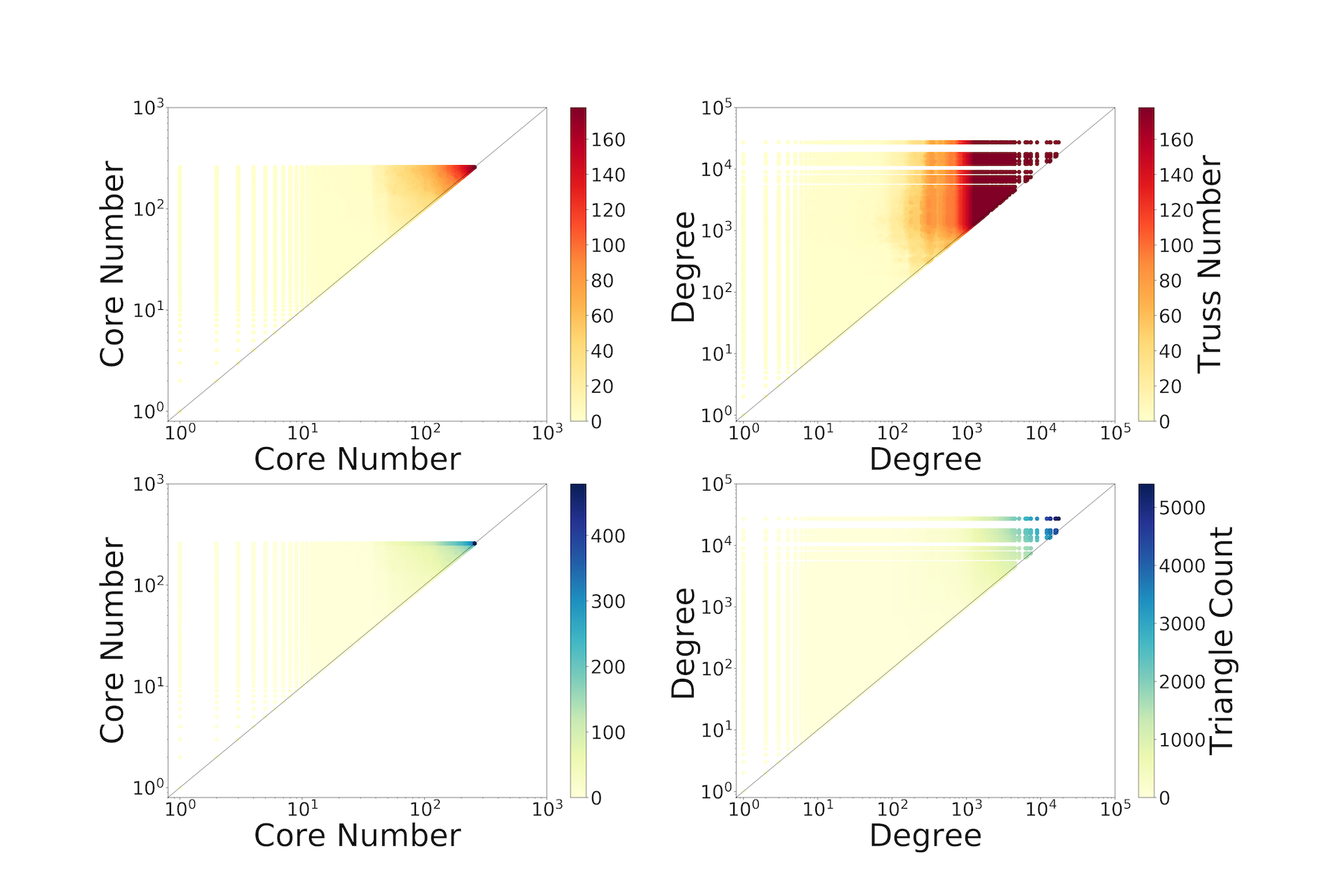}
\caption{\small \texttt{BerkStan} by BTER}
\label{fig:berkstanBTER}
\end{subfigure}
\begin{subfigure}{0.63\textwidth}
\includegraphics[width=\linewidth]{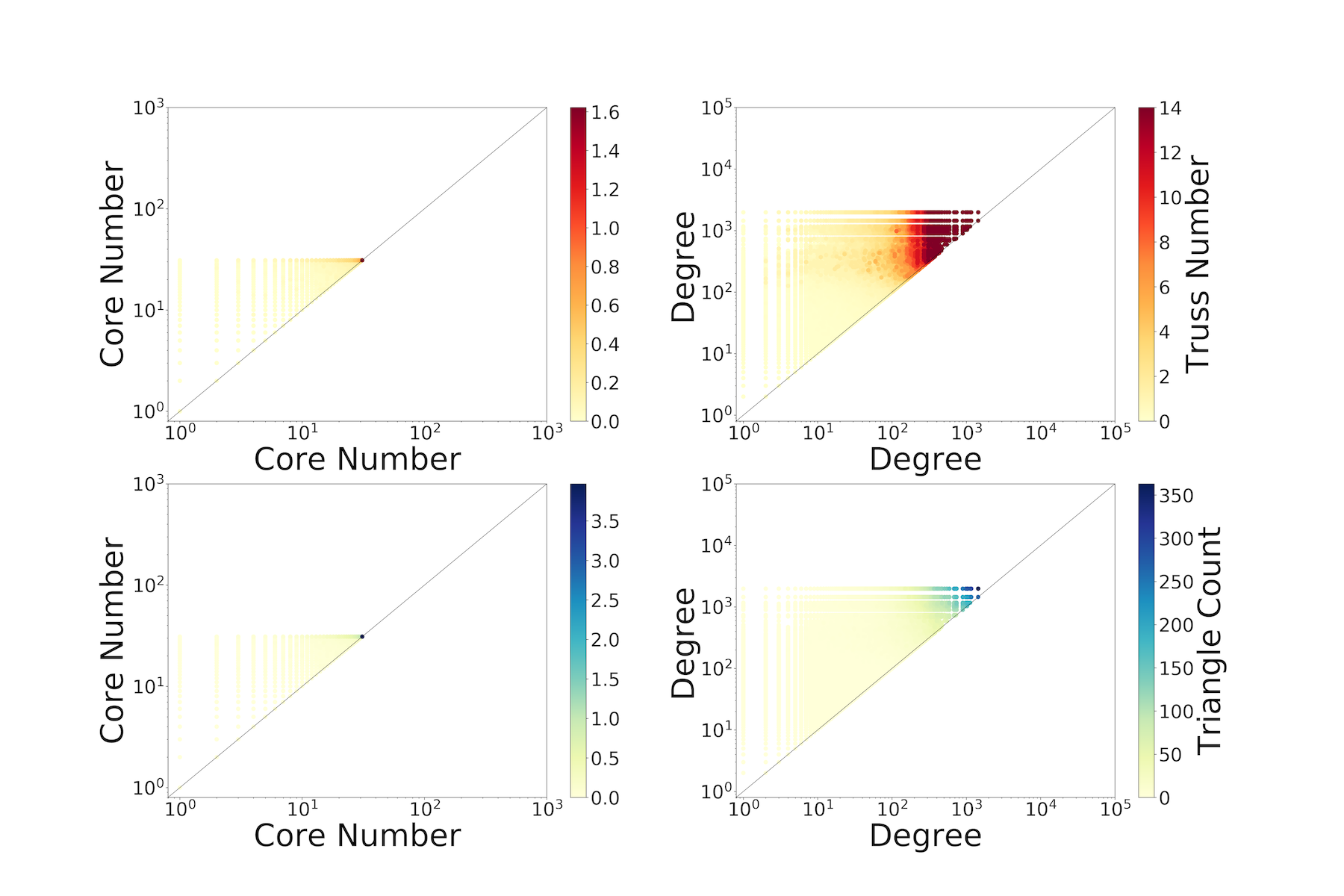}
\caption{\small \texttt{HepTh} by BTER}
\label{fig:hepthBTER}
\end{subfigure}
\caption{\small \bf Edge interplay (EI) plots for random graphs with respect to three  real-world graphs. For each pair of endpoints with particular core numbers/degrees, the average of the truss numbers/triangle counts of the edges are shown.}
\label{fig:EI_random_ext}
\end{figure*}

Here we generate the random graphs with BTER model, by preserving the clustering coefficient distribution per degree.
The EI plots of random graphs are shown in \cref{fig:EI_random_ext}. 
The BTER model captures the general pattern observed in social networks, autonomous systems, and collaboration networks, but fails to capture the interesting behaviors in web networks and citation networks. This indicates the interesting behaviors we observed in EI plot are, again, distinctive characteristics of real-world networks which are driven by specific network structures.

\noindent\textbf{Summary.} Similar to the VI plots, the EI plots present a meaningful graph summary by showing the interplay from the edge perspective. In addition to the core and truss interplay, the EI plots are also capable to capture consistent behavior in the interplay between vertex degrees and edges' triangle counts. This grants us the ability to distinguish diverse characteristics of networks from more domains. 

\section{Anomaly Detection by Core-Truss Discrepancy} \label{sec:alg}
Based on our observations in \cref{sec:vertex} and \cref{sec:edge}, we design the \textsc{Core-Truss discrepancy detection} algorithm (\textsc{Core-TrussDD}) to detect the vertices showing anomalous behaviors of core-truss interplay. We first compute the truss-profile of each vertex, and then cluster the vertices based on their truss-profiles. Within each cluster, we identify the outliers by checking the core number distribution and the Z-scores of the vertices.
We apply our proposed algorithm on \texttt{Email-Eu-core} and \texttt{BerkStan}, and the results show that our algorithm reveals interesting anomalies in real-world network structures.

\subsection{\textsc{Core-TrussDD} algorithm}
\cref{algorithm} provides the pseudocode of \textsc{Core-TrussDD}. 
We first run the core and truss decompositions (\cref{line:core} and \ref{line:truss}) to compute the core numbers of all vertices and the truss numbers of all edges.
Since a vertex can have many edges with different truss numbers, truss numbers cannot be directly used as vertex features. We introduce the vertex truss-profile, which represents the spectrum of truss numbers of all adjacent edges for a vertex. It is the probability distribution of truss numbers around a vertex. This allows us to measure the similarity between two vertices with respect to the truss numbers. We create truss-profiles of vertices in Lines~\ref{line:profile1} to \ref{line:profile2}.

\begin{definition}[Vertex Truss-profile]
Given a graph $G=(V, E)$ and its truss degeneracy $max(T)$, the truss-profile of vertex $v \in V$ is a vector $P_v = [p_0, p_1, \cdots, p_{max(T)}]$, where $p_i = \frac{n(i)}{\sum^{max(T)}_{i=0}n(i)}$ and $n(i)$ is the number of adjacent edges having truss number $i$.
\end{definition}

\begin{algorithm}[!t]
\caption{\textsc{Core-TrussDD}}\label{algorithm}
\hspace*{\algorithmicindent} \textbf{Input:} $G=(V, E)$\\
\hspace*{\algorithmicindent} \textbf{Output:} The set $A$ of anomalous vertices 
\begin{algorithmic}[1]
\State $A \gets \emptyset, \mathcal{P} \gets \emptyset$
\State $K \gets $ \textsc{Core-decomposition}$(G)$ \label{line:core}

\Comment{compute core numbers for all vertices}
\State $T \gets $ \textsc{Truss-decomposition}$(G)$ \label{line:truss}

\Comment{compute truss numbers for all edges}
\State $threshold \gets sort(K)[\frac{|V|}{4}]$ \label{line:threshold1}

\Comment{filter out nodes with low core numbers}
\ForAll{$v \in V$} \label{line:profile1}
\If{$K[v] \geq threshold$} \label{line:threshodl2}
\State $\vec{P_v} \gets$ truss-profile of $v$
\Comment{based on $T$}
\State $\mathcal{P}$.push$(\vec{P_v})$
\EndIf
\EndFor \label{line:profile2}
\State $k \gets$ Elbow-method$(\mathcal{P})$ \label{line:elbow} 
\Comment{optimum number of clusters}
\State $[C_1, C_2, \cdots, C_k] \gets$ k-means$(\mathcal{P}, k)$ \label{line:kmeans}
\Comment{vertex clustering}
\For{$C_i \in [C_1, C_2, \cdots, C_k]$}
\For{$v' \in C_i$}
\State $z[v'] \gets$ Z-score of $K[v']$ in cluster $C_i$ \label{line:z-score}
\If{$|z[v']| > 2$} \label{line:outlier1}
\State $A$.push($v'$) \label{line:outlier2}
\EndIf
\EndFor

\Comment{outliers can also be identified from the histogram}
\EndFor
\State \textbf{return} $A$
\end{algorithmic}
\end{algorithm}

\begin{figure}[!b]
\vspace{-3ex}
\centering

\begin{subfigure}{0.25\textwidth}
\includegraphics[width=\linewidth]{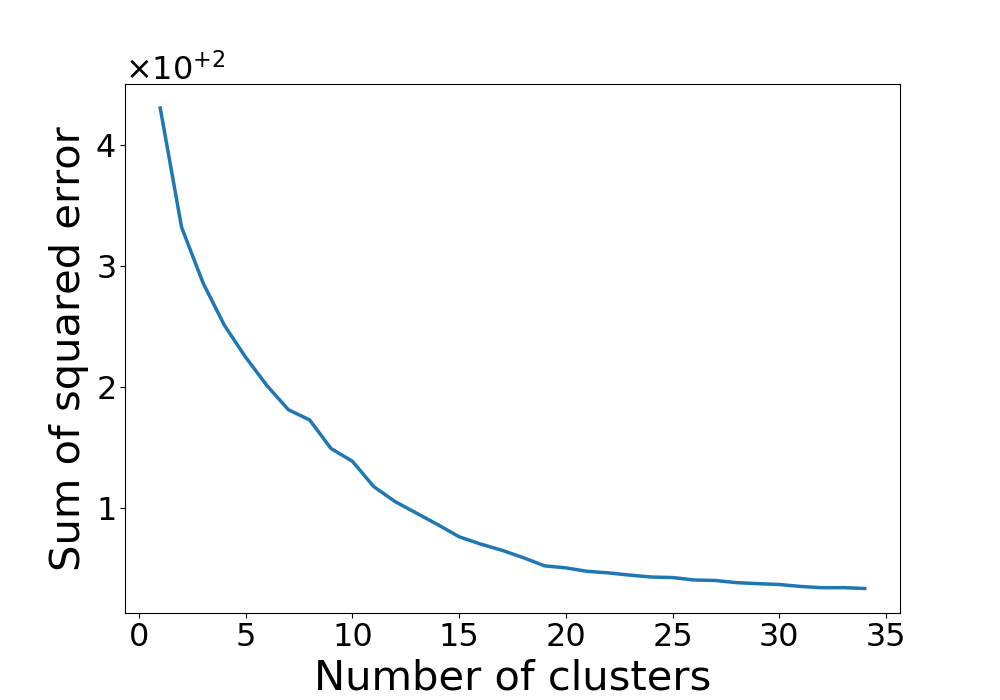}
\caption{\texttt{Email-Eu-core}}
\label{fig:elbow-email}
\end{subfigure}
\hspace{-3ex}
\begin{subfigure}{0.25\textwidth}
\includegraphics[width=\linewidth]{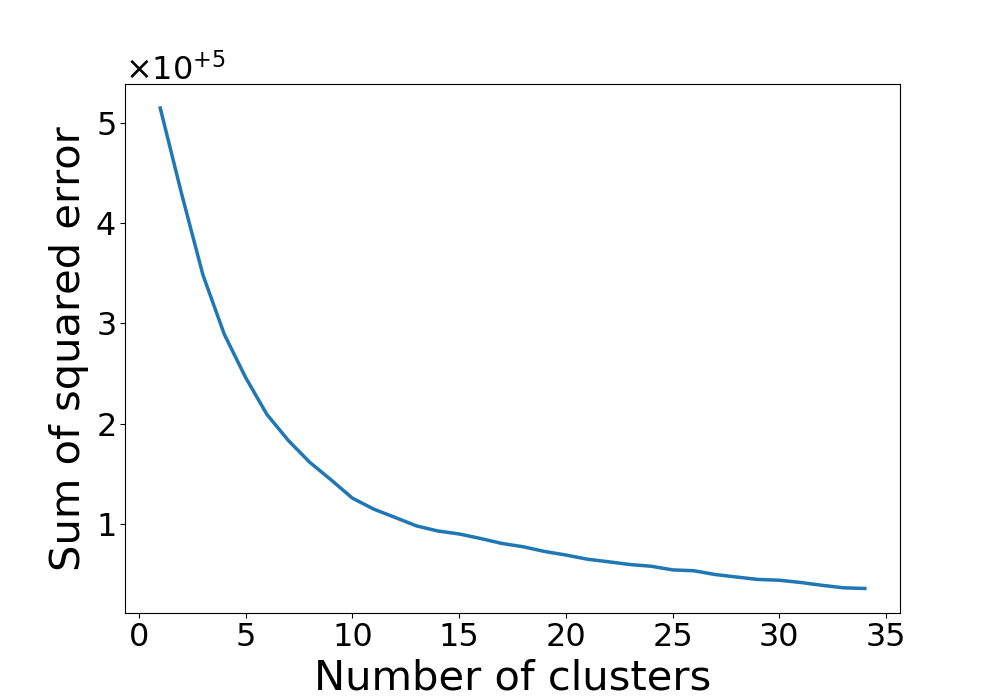}
\caption{\texttt{BerkStan}}
\label{fig:elbow-berkstan}
\end{subfigure}

\caption{\bf \small Exploring the number of clusters using elbow method. Each figure shows the decrease of within-cluster sum of squared error (SSE), with respect to the number of clusters. For each dataset, we select the number of clusters where the SSE stops decreasing. We choose 15 clusters for both \texttt{Email-Eu-core} and \texttt{BerkStan} according to~\cref{fig:elbow-email} and~\cref{fig:elbow-berkstan}. 
}
\label{fig:elbow}
\end{figure}

Real-world networks often present heterogeneous structures. It is hard to capture a general pattern in truss-profiles of all vertices as they can be highly diverse, making anomaly detection challenging. To address this issue, we group the vertices with similar truss-profiles (\cref{line:kmeans}). Here we apply the k-means clustering \cite{k-means}, which minimizes the within-cluster squared distances. 
One important issue of the k-means algorithm is the input parameter $k$, which determines the number of clusters.
We use elbow method \cite{elbow} to select the optimum value of $k$ for each dataset, regarding the within-cluster sum of squares (\cref{line:elbow}). \cref{fig:elbow-email} and \cref{fig:elbow-berkstan} show how the within-cluster sum of squared error (SSE) changes when the number of clusters increases. We use k-means with 15 clusters for both \texttt{Email-Eu-core} and \texttt{BerkStan}, since the SSE is hardly reduced by having more clusters. Another issue of clustering is that each vertex will be assigned to a cluster, including those in the periphery with low importance. This will downgrade the clustering performance and lead to false positives in the anomaly detection. Therefore, we ignore the 25\% of nodes with the lowest core numbers (Lines~\ref{line:threshold1} and \ref{line:threshodl2}).

\begin{figure*}[!t]
\centering
\vspace{-2ex}

\begin{subfigure}{0.33\textwidth}
\includegraphics[width=\linewidth]{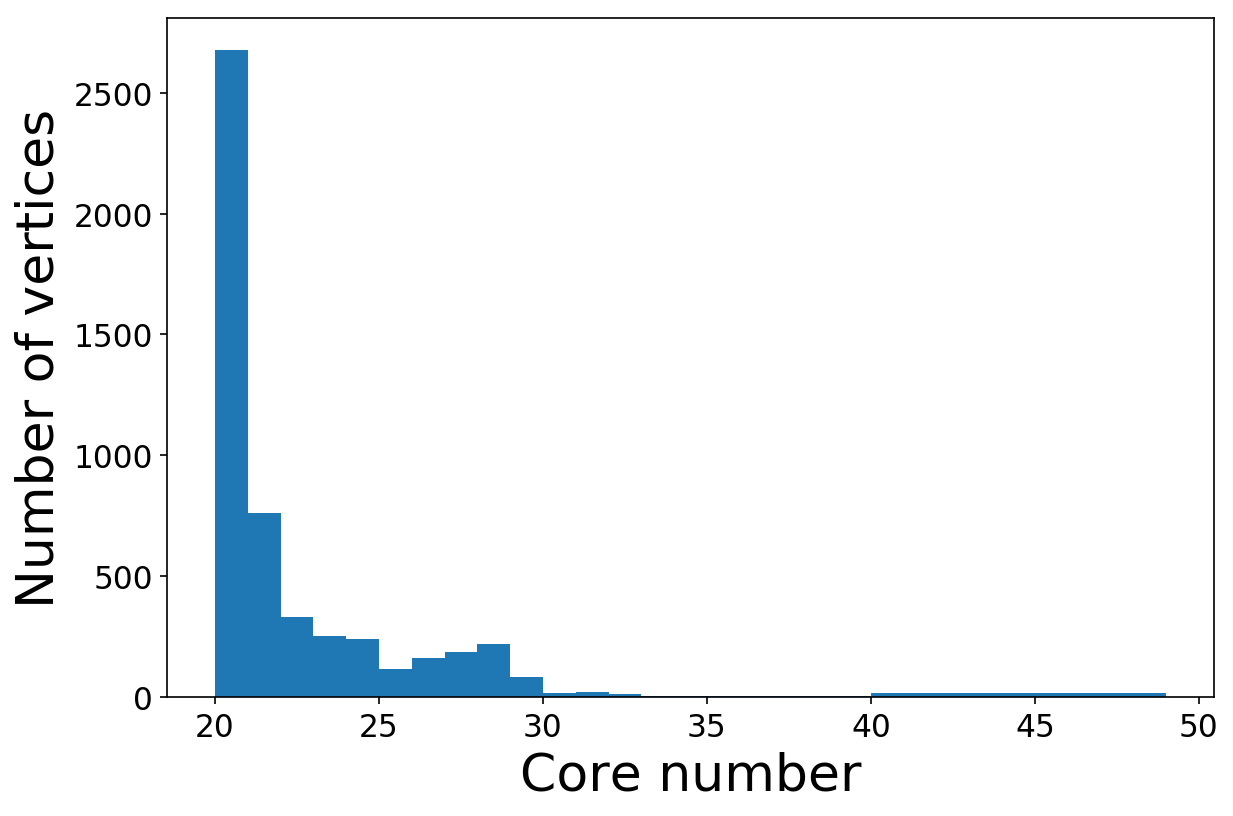}
\caption{Normal cluster}
\label{fig:berkstan-cluster-normal}
\end{subfigure}
\hspace{-2ex}
\begin{subfigure}{0.33\textwidth}
\includegraphics[width=\linewidth]{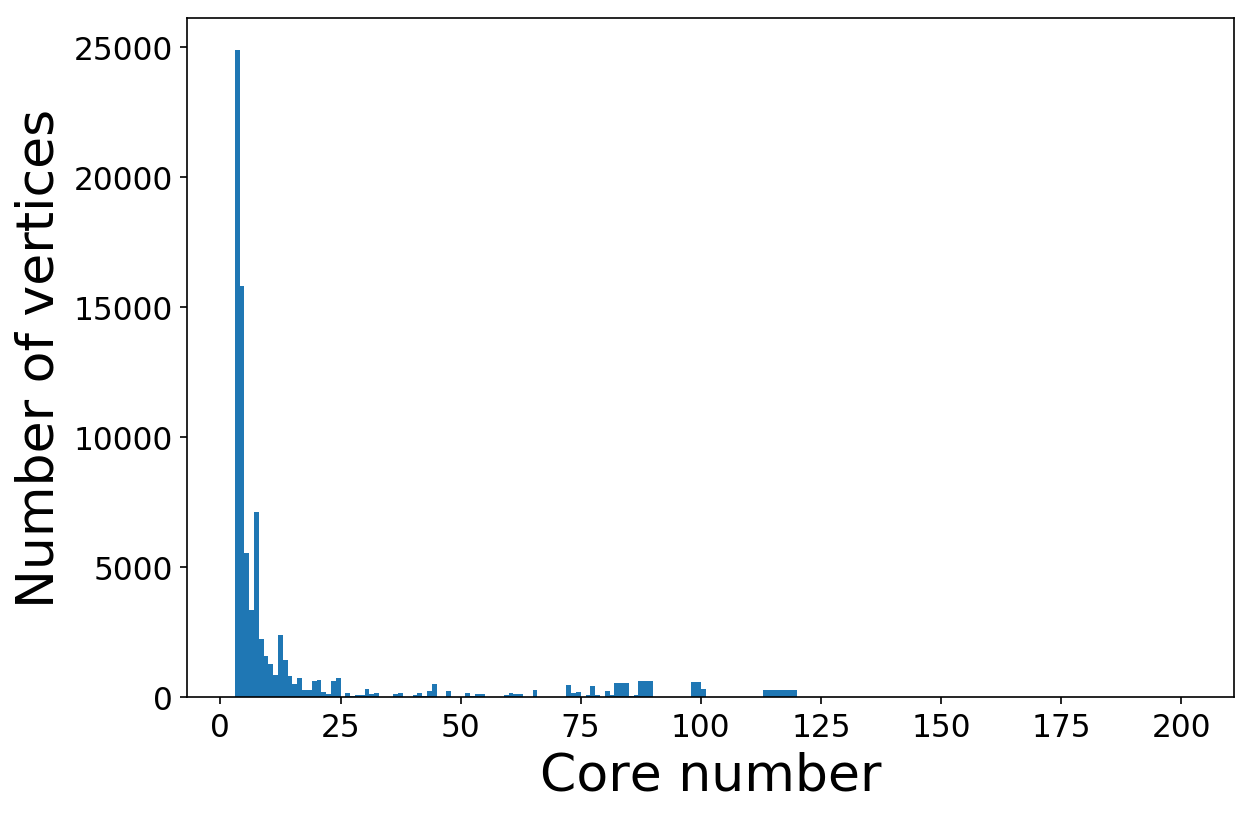}
\caption{Cluster $G$ (gatekeeper anomalies)}
\label{fig:berkstan-cluster-hub}
\end{subfigure}
\hspace{-2ex}
\begin{subfigure}{0.33\textwidth}
\includegraphics[width=\linewidth]{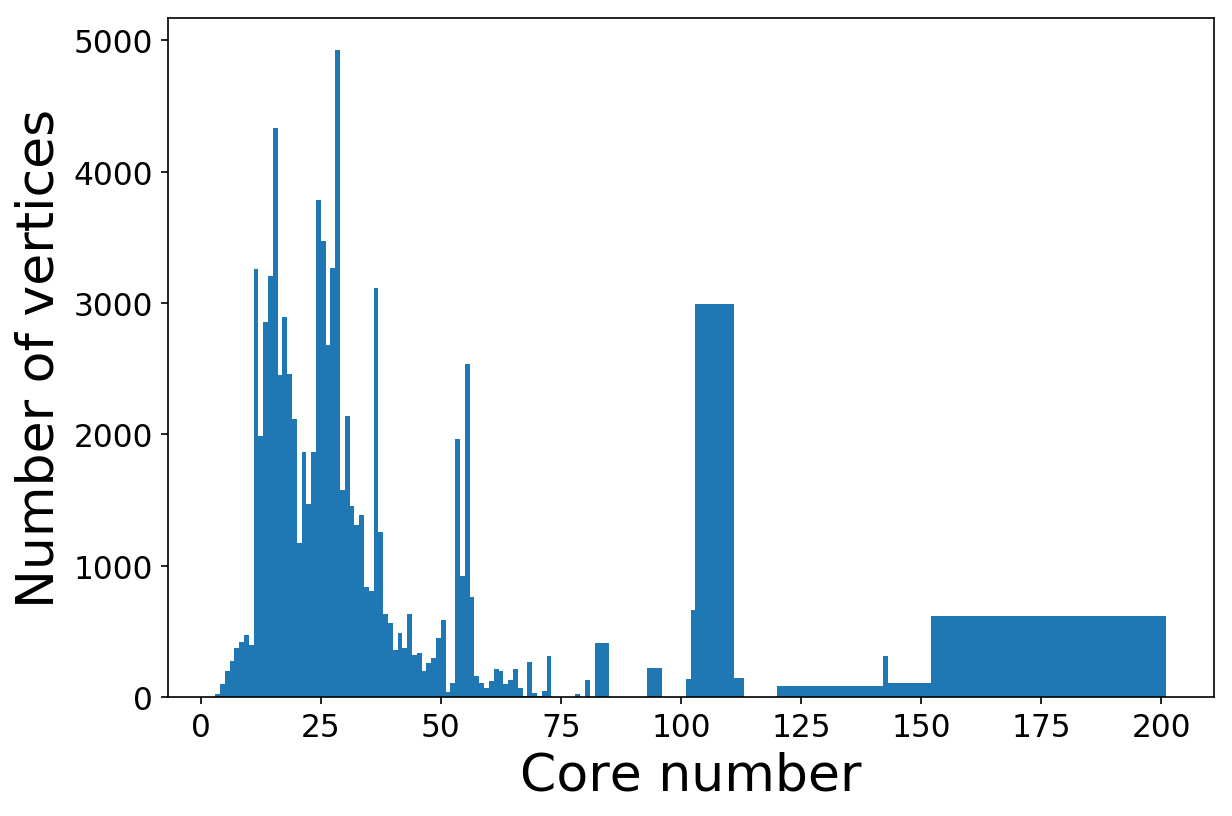}
\caption{Cluster $C$ (clique-like anomalies)}
\label{fig:berkstan-cluster-clique}
\end{subfigure}

\caption{\bf \small Core number distributions in \texttt{BerkStan} clusters. Each figure represents the core number distribution within a cluster. \cref{fig:berkstan-cluster-normal} shows a cluster where all vertices have similar core numbers. This pattern is commonly observed (in 13 out of 15 clusters). \cref{fig:berkstan-cluster-hub} shows another cluster where we observe outliers with core numbers deviated from the major population ($K > 50$). These outliers are found to be the gatekeepers in later analysis. We also observe anomalous vertices in \cref{fig:berkstan-cluster-clique} ($K > 150$), which appear to be clique-like structures.
}
\vspace{-2ex}
\label{fig:berkstan-cluster}
\end{figure*}

\begin{figure}[!b]
\centering
\vspace{-2ex}
\begin{subfigure}{0.24\textwidth}
\includegraphics[width=\linewidth]{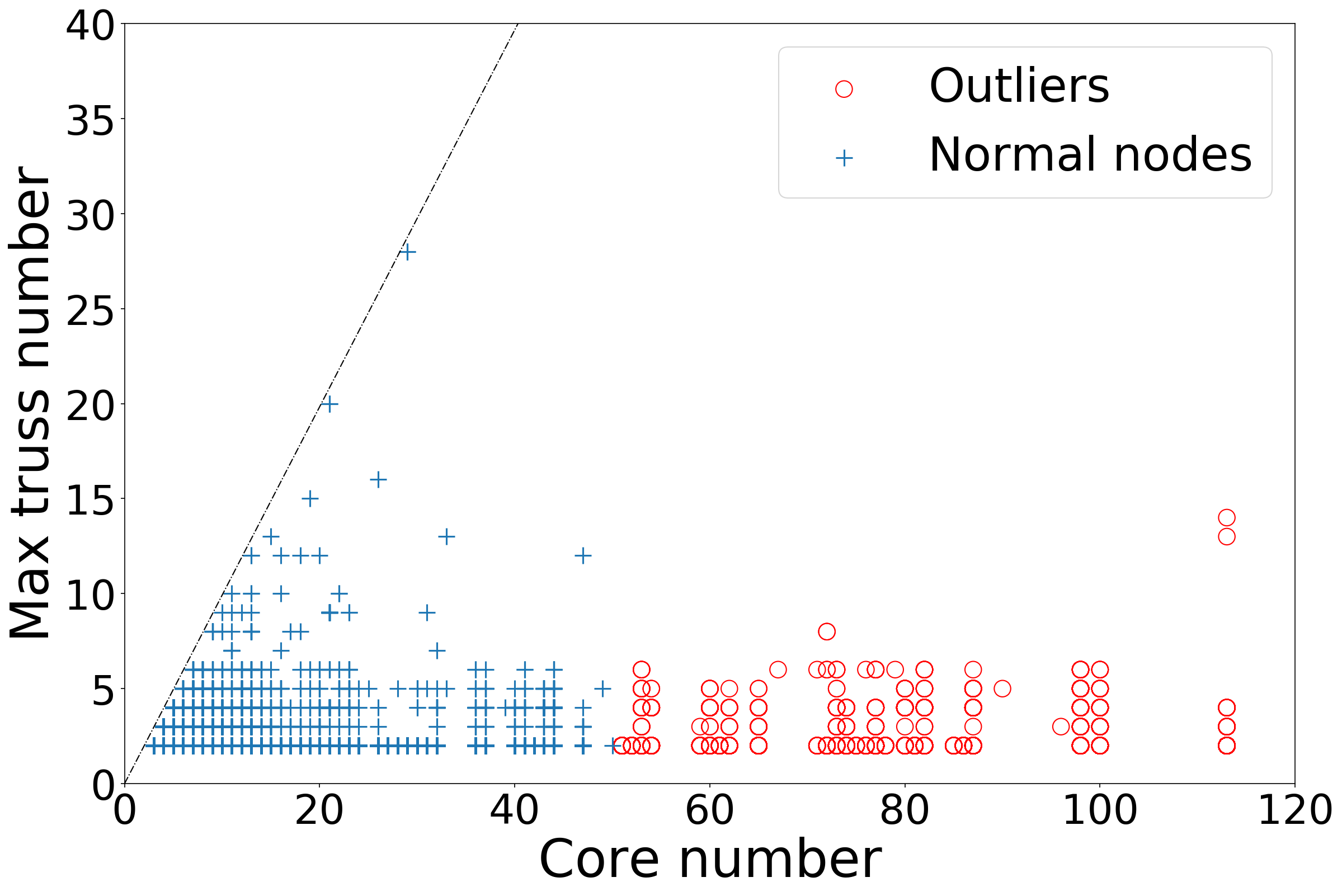}
\caption{\small Cluster $G$ (gatekeepers)}
\label{fig:berkstan-scatter-hub}
\end{subfigure}
\hspace{-1ex}
\begin{subfigure}{0.24\textwidth}
\includegraphics[width=\linewidth]{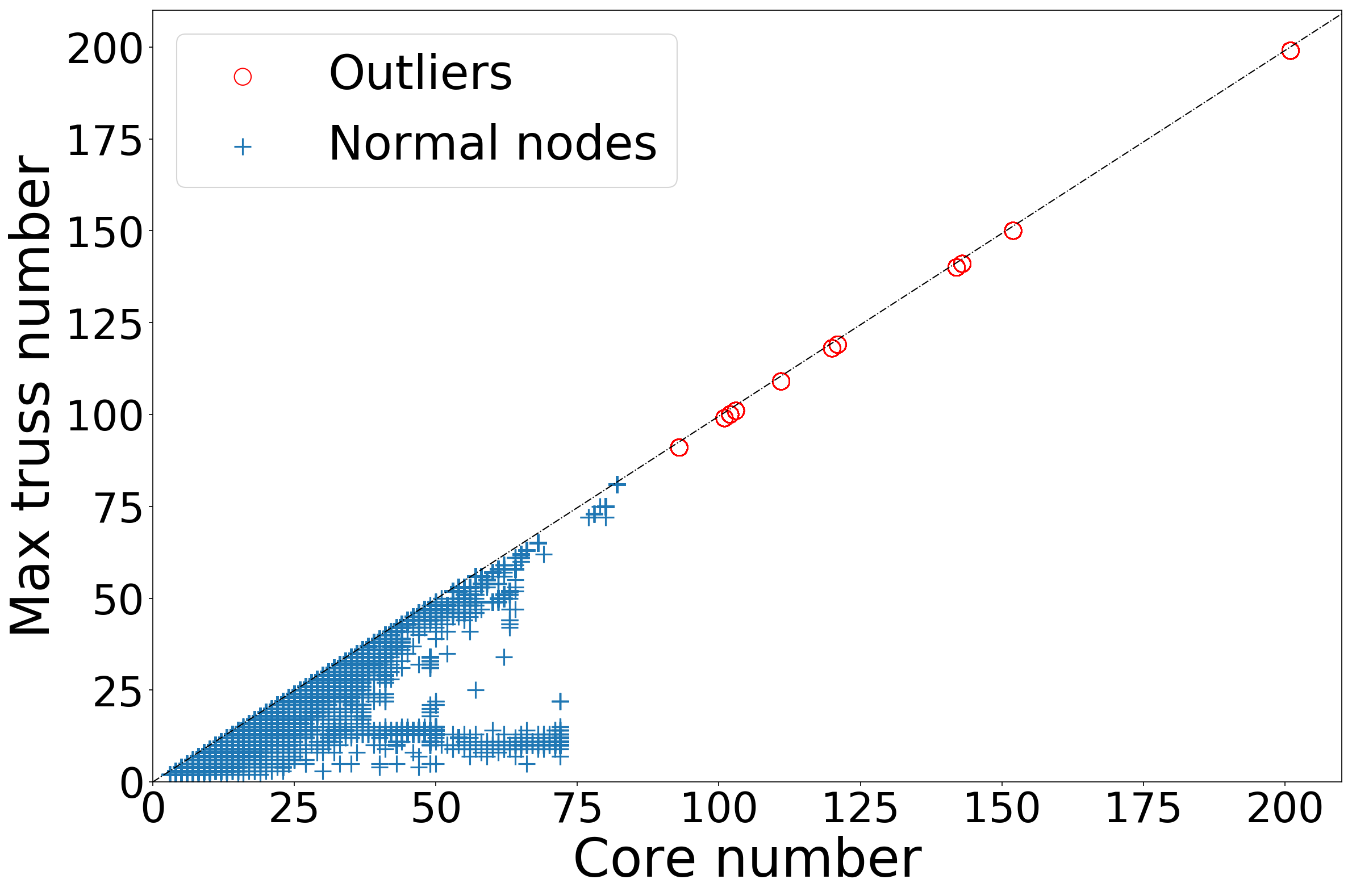}
\caption{\small Cluster $C$ (clique-like outliers)}
\label{fig:berkstan-scatter-clique}
\end{subfigure}

\caption{\bf \small Gatekeeper and clique-like anomalies in \texttt{BerkStan} clusters. In each scatter plot, the blue crosses denote normal nodes and the red circles represent outliers. 
The x-axis represents the core number of the vertex, while the y-axis denotes the maximum truss number of the adjacent edges. The dashed line is $y=x-1$, which indicates the core-truss behavior of cliques.
\cref{fig:berkstan-scatter-hub} shows the patterns of gatekeeper outliers in cluster $G$, where the maximum truss numbers are much less than the core numbers.
\cref{fig:berkstan-scatter-clique} gives the clique-like outliers behaviors in cluster $C$, where the maximum truss numbers are almost equal to the core numbers.
}
\vspace{-1ex}
\label{fig:berkstan-scatter}
\end{figure}

\begin{figure*}[!b]
\centering
\vspace{-3ex}

\begin{subfigure}{0.25\textwidth}
\includegraphics[width=\linewidth]{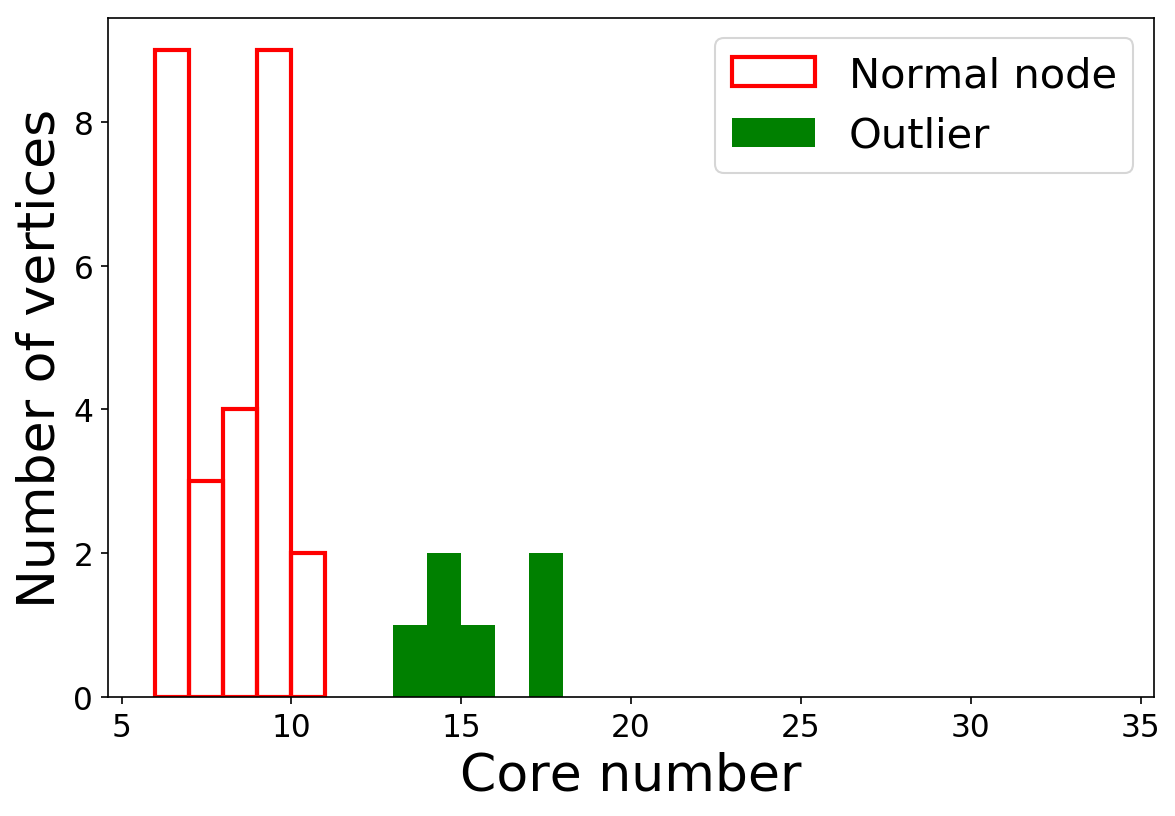}
\end{subfigure}
\hspace{-2ex}
\begin{subfigure}{0.25\textwidth}
\includegraphics[width=\linewidth]{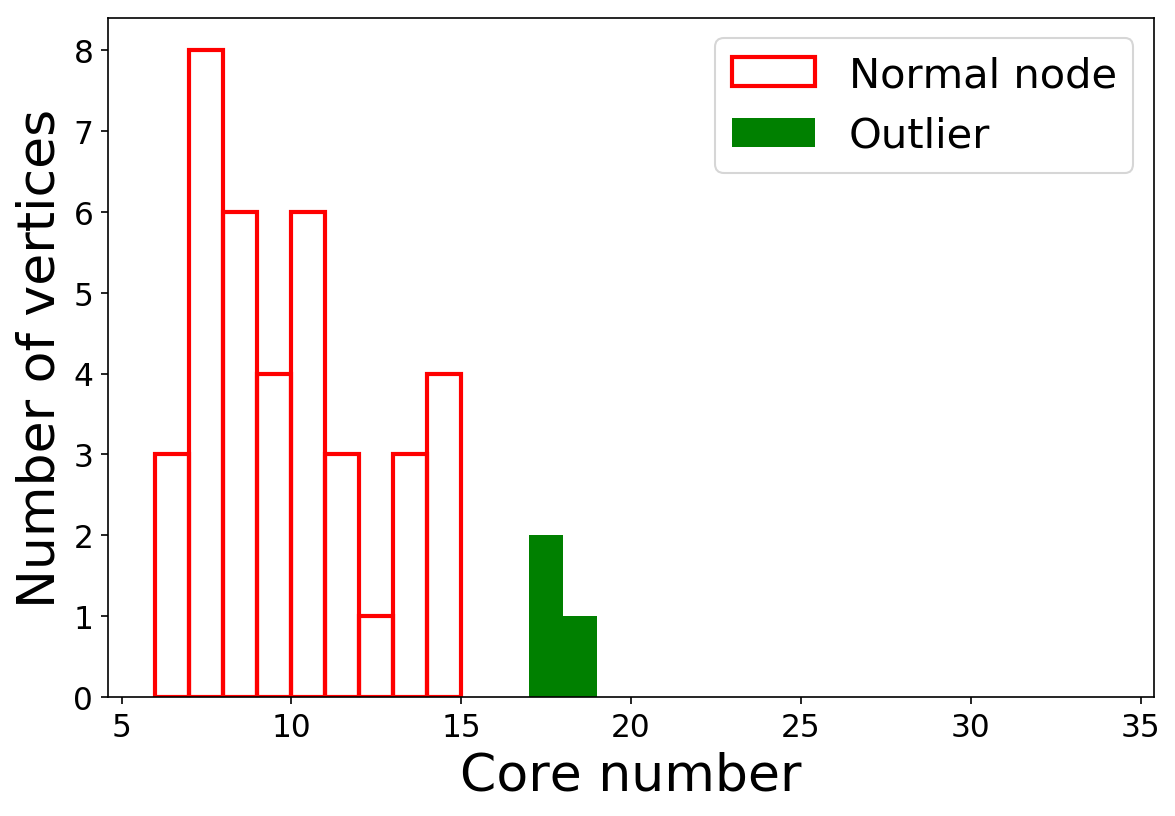}
\end{subfigure}
\hspace{-2ex}
\begin{subfigure}{0.25\textwidth}
\includegraphics[width=\linewidth]{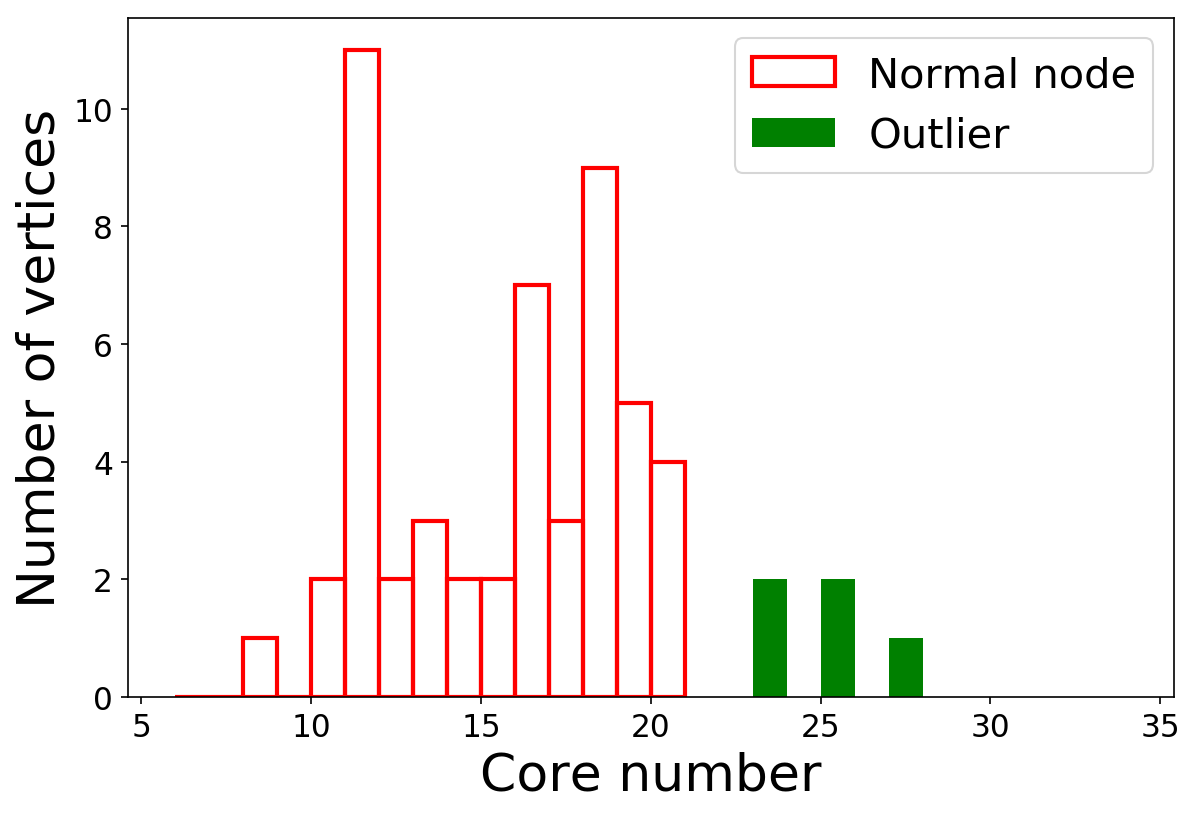}
\end{subfigure}
\hspace{-2ex}
\begin{subfigure}{0.25\textwidth}
\includegraphics[width=\linewidth]{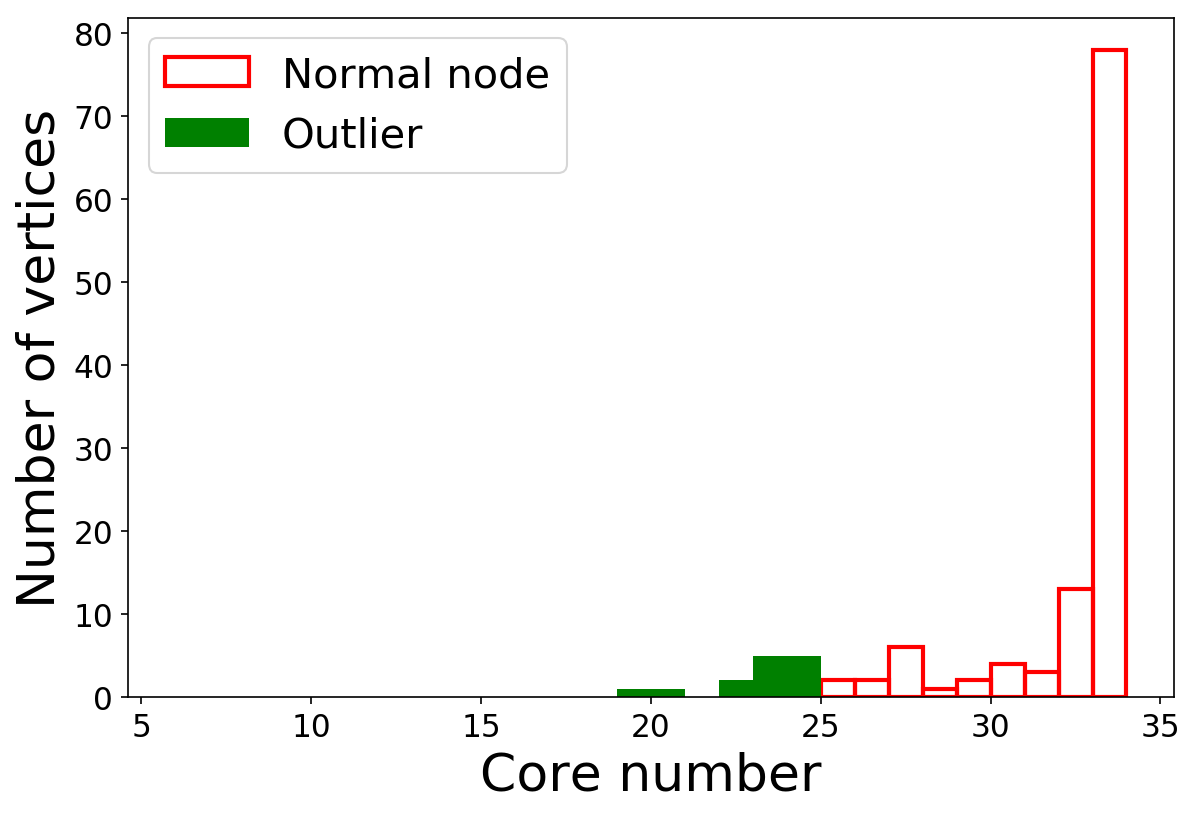}
\end{subfigure}
\vspace{-1ex}

\begin{subfigure}{0.25\textwidth}
\includegraphics[width=\linewidth]{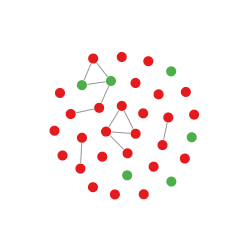}
\caption{Cluster $A$}
\label{fig:email-cluster-1}
\end{subfigure}
\hspace{-2ex}
\begin{subfigure}{0.25\textwidth}
\includegraphics[width=\linewidth]{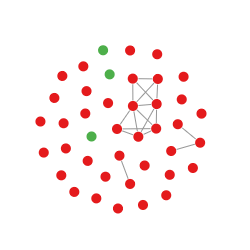}
\caption{Cluster $B$}
\label{fig:email-cluster-4}
\end{subfigure}
\hspace{-2ex}
\begin{subfigure}{0.25\textwidth}
\includegraphics[width=\linewidth]{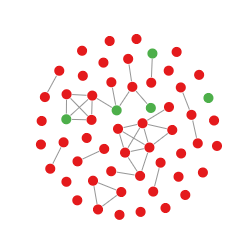}
\caption{Cluster $C$}
\label{fig:email-cluster-9}
\end{subfigure}
\hspace{-2ex}
\begin{subfigure}{0.25\textwidth}
\includegraphics[width=\linewidth]{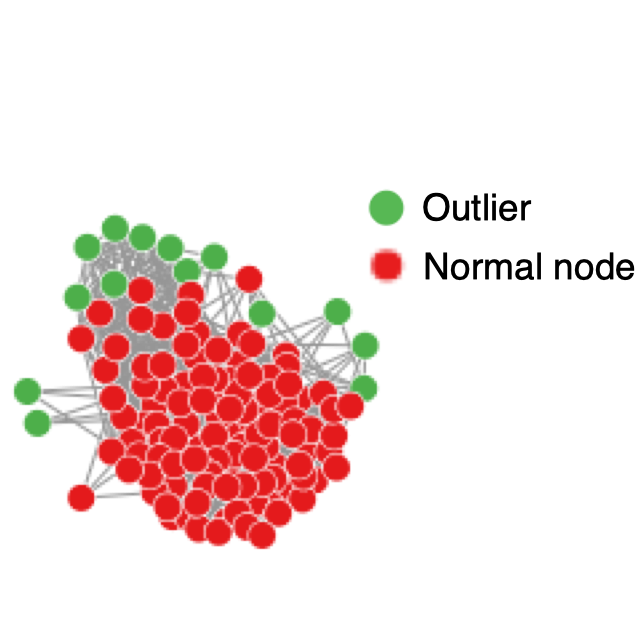}
\caption{Cluster $D$}
\label{fig:email-cluster-14}
\end{subfigure}

\caption{\bf \small Core number distributions and the corresponding subgraphs of \texttt{Email-Eu-core} clusters. The histograms present the core number distributions within the clusters, while the subgraphs at the bottom are induced by the vertices in the clusters. Normal nodes are colored in red and the green circles denote outliers. 
\cref{fig:email-cluster-1}, \cref{fig:email-cluster-4}, and \cref{fig:email-cluster-9} show the pattern of three clusters, where vertices are sparsely connected. \cref{fig:email-cluster-14} presents another cluster which is densely connected. When compared to the normal nodes, the outliers are less likely to be connected with each other, as observed in all clusters.
}
\vspace{-2ex}
\label{fig:email-cluster}
\end{figure*}
After clustering the vertices with respect to the truss-profiles, we investigate the patterns and anomalies of core numbers in each cluster. 
In each cluster, we check the core numbers of the vertices and report the ones that are significantly different than the others.
For this purpose, we compute the Z-score of each vertex in order to measure the deviation of their core numbers with respect to the general patterns (\cref{line:z-score}). For a given vertex $v$ with core number $K_v$, the Z-score is calculated as $z_v = \frac{K_v - \mu}{\sigma}$, where $\mu$ and $\sigma$ are the mean and standard deviation of all core numbers in the cluster. In Lines~\ref{line:outlier1} and \ref{line:outlier2}, we identify the outliers which have the absolute values of Z-scores larger than 2. This corresponds to the 2.28\% of the population in normal distribution. We also plot the distribution of core numbers and identify anomalies which deviate from the general patterns.

\noindent\textbf{Remark.} Note that the k-means algorithm only serves as a subroutine in our anomaly detection algorithm. The purpose of clustering is to partition the vertices into several groups where we can expect common truss-profile behaviors. The clustering performance is not our major concern as we only aim to have a scaled-down problem size for the anomaly detection.

\noindent\textbf{Time complexity:} \textsc{Core-TrussDD} starts with core and truss decomposition, which takes $O(|E|)$ and $O(\sum_{v \in V}{d(v)^2})$ time respectively. The algorithm generates truss-profiles for all vertices in $O(|V| max(T))=O(|V|)$ time, since the truss degeneracy $max(T)$ is $O(1)$ for real-world networks. The k-means clustering can be implemented by the Lloyd's algorithm \cite{lloyd1982least}, which takes $O(k|V|max(T)i)$ time, where $k$ is the number of clusters and $i$ is the number of iterations until convergence. In the worst case the Lloyd's algorithm needs $2^{\sqrt{|V|}}$ iterations. Therefore, the overall time complexity is $O(\sum_{v \in V}{d(v)^2} + k|V|2^{\sqrt{|V|}})$.

\noindent\textbf{Space complexity:} We first construct $K$ and $T$ to store the core numbers of vertices and truss numbers of edges, which take $O(|E|)$ space. In addition, we need $O(|V|max(T))=O(|V|)$ space for the truss-profiles $\mathcal{P}$. The k-means algorithm need $O((|V|+k)max(T)) = O(|V|)$ space. Arrays $A$ and $[C_1, C_2, \cdots, C_k]$ take $O(|V|)$ space, so the total space complexity is $O(|E|)$.

\subsection{Anomalies in real-world networks}

In general, we observe similar core numbers within the clusters.
\cref{fig:berkstan-cluster-normal} gives the core number distribution in one of the clusters in \texttt{BerkStan}. Within the cluster the major population of core numbers lies in a small interval $[20,30]$. Since the vertices are clustered based on their truss-profiles, it indicates that the core and truss decompositions are consistent. We observed similar consistent patterns in 13 out of 15 clusters in \texttt{BerkStan}, as well as in the most clusters in \texttt{Email-Eu-core}. In the remaining two clusters (clusters $G$ and $C$) of \texttt{BerkStan}, we identify significant amounts of outliers (more than 5\%).
\cref{fig:berkstan-cluster-hub} shows the core number distribution in cluster $G$. The majority of vertices have core numbers ranging from 0 to 25, while some outliers have much larger core numbers (note that the core degeneracy of \texttt{BerkStan} is 201).
\cref{fig:berkstan-scatter-hub} shows that the core numbers of these outliers are much greater than the maximum truss numbers of their adjacent edges. This indicates that the anomalies discovered in this cluster are the gatekeepers illustrated in \cref{fig:hub-like}.
In \cref{fig:berkstan-cluster-clique} we observe that most vertices in cluster $C$ have their core numbers lie between 0 and 55. We identify 5352 outlier vertices with core number larger than 92. 
As shown in \cref{fig:berkstan-scatter-clique}, the core numbers are extremely close to the maximum truss numbers of the adjacent edges. 
For all of the 5352 outliers, the difference between their core numbers and the maximum truss numbers is equal to 2. According to our discussion in \cref{sec:vertex}, these are the clique-like structures (\cref{fig:clique-like}) in the network.

Next, we investigate the patterns and anomalies in the \texttt{Email-Eu-core} clusters. Among the 15 clusters four of them present anomalous behaviors. \cref{fig:email-cluster} shows the core number distributions of the four clusters, along with the subgraphs induced by the vertices in the clusters. In cluster $A$ we observe the majority of vertices have core numbers from 6 to 10, while there are six outliers with core numbers larger than 12. Similar anomalies are observed in clusters $B$ and $C$, where the outliers have core numbers significantly larger than the majority of the population in the cluster. Cluster $D$ presents a different picture, where most of the vertices have core numbers of 33 and the outliers have core numbers less than 25.
In order to further analyze these anomalous clusters, we visualize the subgraphs based on the edges within the clusters and display them in \cref{fig:email-cluster}. Note that we are not doing graph clustering but cluster the vertices with respect to the truss-profiles, so they may not be connected at all. We noticed that subgraphs induced by clusters $A$, $B$, and $C$ are structurally sparse.
Vertices with similar truss behavior are often not connected to each other in real-world networks.
Moreover, outliers are less likely to be connected with each other than the normal vertices. 
The subgraph of cluster $D$ shows a very different picture, where the normal vertices form a connected component with relatively high density. This is due to the overall high core numbers in the cluster. Note that the outliers lie in the periphery of the subgraph and few connections exist between them.

\noindent\textbf{Summary.} \textsc{Core-TrussDD} algorithm creates the truss-profile to present the truss number spectrum for a given vertex, which allows us to cluster the vertices with similar truss behavior.  We then analyze the core number distribution and the Z-scores in the cluster to identify the anomalies. These outliers reveal the clique-like and gatekeeper structures in the networks. By further investigating the subgraphs induced from the clusters, we discover distinct interaction patterns between the outliers.

\section{Related Work}\label{sec:related}
The core and truss decompositions, proposed by \cite{Seidman83} and \cite{Cohen08}, is commonly used in dense subgraph discovery. 
One of the works that addresses the relation between core and truss decompositions is \cite{Sariyuce15, Sariyuce-TWEB17}. Sariyuce et al. proposed a network decomposition framework which addresses both core and truss decompositions, and explored the hierarchical and overlap relations among the subgraphs. 
Another related work \cite{Li2018} defined a subgraph structure that combines the definitions of \textit{k}-core and \textit{k}-truss, and proposed the corresponding  decomposition algorithm.

The most related work to our study is introduced by Shin et al~\cite{Shin18}.
They explored the relations between core (truss) degeneracy and the number of triangles.
Note that, our research directly addresses the relationship between the core and truss decompositions based on local perspectives.
We do not limit our analysis to investigate each decomposition on its own.
Instead, we focus on the interplay of core and truss numbers and introduce the VI and EI plots for network analysis. We also propose the \textsc{Core-TrussDD} algorithm to detect the outlier nodes.
Note that, in a prior work we also considered the relationship between clique counts and the maximum core and truss numbers~\cite{liu2019analysis}.

%




\section{Conclusion}\label{sec:conc}
In this paper, we analyzed the interplay between core and truss decompositions in real-world networks.
We introduced VI and EI plots to analyze the network structure by using the core and truss decompositions. 
The VI plot investigates the core-truss interplay from a vertex perspective, by examining the spectrum of edges around vertices with particular core numbers. 
The EI plot explores the interplay from an edge perspective by checking the truss number of an edge and the core numbers of the two endpoints. 
We applied our analysis on real-world networks from various domains, and then validate our findings by evaluating the random graphs generated by the BTER model. The VI and EI plots reveal consistent pattern in social networks and autonomous systems, and identify some interesting behaviors in collaboration networks, citation networks, and web networks.
Inspired by our observation in VI and EI plots, we proposed the \textsc{Core-TrussDD} algorithm to identify anomalies in the networks by utilizing the core-turss interplay . We analyzed the characteristics of the outliers indentified by our algorithm, and the results support our findings in the VI plots. We believe our study would be handy for domain practitioners to analyze the dense subgraph structure of networks, and provide important insights for anomaly detection.
 

%

\noindent {\bf Acknowledgments.} This research was supported by NSF-1910063 award, JP Morgan Chase and Company Faculty Research Award, and Center for Computational Research at the University at Buffalo~\cite{CCR}.
\bibliographystyle{IEEEtran}
\bibliography{IEEEabrv,paper}



\vfill
\vfill


\end{document}